%% file: all.tex
\documentclass[a4paper,11pt,twoside]{book}
\pdfoutput=1

\input{usepackage}

\input{newcommand}

\makeatletter %
\renewcommand{\@makeschapterhead}[1]{%
  \indent {\Huge #1}
  \vspace{10pt}
  \hrule 
  \vspace{40pt}
}
\makeatother %

\newcites{pub}{List of Publications}

\makeglossary

\begin{document}
\pagenumbering{roman}
\pagestyle{empty}

\input{symbols.tex}

\hyphenation{Fesh-bach Op-pen-hei-mer func-tion wave-func-tion}

\begin{center}
\hrule\par
\vspace{1em}\par
{\bf \huge Universality in Ultra-Cold}\par
\vspace{0.5em}\par
{\bf \huge Few- and Many-Boson Systems }\par
\vspace{1em}\par
\hrule\par
\vspace{4em}
{\bf \LARGE Martin Th{\o}gersen}\par
\vspace{1em}\par
Department of Physics and Astronomy\\
Faculty of Science\\
{\AA}rhus University, Denmark\par
\strut\vfill\par
\includegraphics[width=8cm]{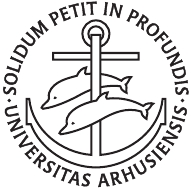}
\strut\vfill\par
{\bf \Large Dissertation for the degree of\\
  Doctor of Philosophy}\par
\vspace{0.5em}\par
{\Large July 2009}
\end{center}

\newpage
\phantom{e}
\vfill
\noindent \textcopyright{} 2009 Martin Th{\o}gersen\\
Department of Physics and Astronomy\\
Aarhus University\\
Ny Munkegade, Bldg. 1520\\ 
DK-8000 Aarhus C\\
Denmark\\
Phone: +45 26 70 22 43\\
Email: martint@phys.au.dk\\

\vspace{1em} \noindent 
1st edition, July 2009.\\
Online version and bibliography: \url{http://www.martint.dk/thesis}\\

\vspace{1em} \noindent Cover image: Illustration of a Bose-Einstein
condensate, bringing together the few-body concepts of Efimov states and
Borromean binding.\\
Central image: NIST/JILA/CU-Boulder, 1995 (public domain). 
Left image: B. D. Esry and C. H. Greene, Nature, \textbf{440} (2006) 289.
Top image: Public domain, GPL.

\vspace{2em}
\noindent Layout and typography chosen and implemented by the writer.\\
Font: Computer Modern, 11pt.\\
Typesetting done with \LaTeXe{}, Bib\TeX{} and the report class.\\
Used packages: \input{autopackage.tex}\\
Figures are made with Gnuplot 4.2 and TikZ/PGF.\\
Printed by SUN-Tryk, Aarhus University.

\newpage
\begin{center}
\framebox{
  \parbox{0.95\textwidth}{
    \noindent This dissertation has been submitted
    to the Faculty of Science at Aarhus University, Denmark, in
    partial fulfillment of the requirements for the PhD degree in
    physics.  The work presented has been performed under the
    supervision of Ass. Prof. Dmitri Fedorov and Sen. Ass. Prof. Aksel
    S. Jensen. The work was carried out at the Department of Physics
    and Astronomy in Aarhus during the period August 2005 to July
    2009.  The group of Brett D. Esry at the Physics Department,
    Kansas State University, Manhattan, KS, USA, is also acknowledged
    for its hospitality during the spring of 2008.%
  }  
}
\end{center}
\vspace{30em}
\begin{flushright}
\emph{A physicist is just an atom's way of looking at itself.}\\
Niels Bohr (1885 - 1962)
\end{flushright}

\pagestyle{plain}
\tableofcontents

\chapter*{Outline}
\addcontentsline{toc}{chapter}{Outline} 

This thesis describes theoretical investigations of universality
and finite-range corrections in few- and many-boson systems. The major
part of this work concerns ultra-cold trapped atomic gases, but some
of the results and methods may be applicable to small molecules, and
nuclei too.

In the introductory chapter~\ref{chap:intro} we briefly describe some
of the basic concepts and phenomenology used in this dissertation such
as universality, Bose-Einstein condensation, Efimov physics, and
Borromean binding. The typical experimental systems of interest are
also described.

Chapter~\ref{chap:theory} reviews the relevant theoretical background
and methods for few-body systems as well as mean-field models for
condensates. The numerical procedures we implement are also described.
This gives a common background for the rest of the dissertation. The
main chapters~\ref{chap:efimov}--\ref{chap:tf}
can be read almost independently, but they often refer back to
chapter~\ref{chap:theory}.

In chapter~\ref{chap:efimov} we investigate universal finite-range
corrections to Efimov physics in three-boson systems. Connections to
Borromean binding are made. We also describe the effects of putting
the system in a finite trap.
Chapter~\ref{chap:efimov2} continues this line by investigating the
conditions for Efimov physics, in particular for large effective
range.
The content of chapter~\ref{chap:efimov} and \ref{chap:efimov2} was
published in \cite{pub-thogersen08-3,pub-thogersen09a}, but new
material is also presented. (See page~\pageref{chap:publications} for
a list of publications.)

In chapter~\ref{chap:N-efimov} we show the existence of a many-body
Efimov effect based on two-body correlations. The features of the
effect and experimental signatures are discussed.
This chapter is based on \cite{pub-thogersen08-2, pub-fedorov08-1} as
well as some new results.

In chapter~\ref{chap:correlations} we numerically consider trapped
few-boson systems of order 10--30 particles including two-body
correlations. Condensate-type states are identified. We investigate
energies and correlations for these condensate states and discuss
possible experimental signatures. Appendix~\ref{chap:appendix-obdm}
and \ref{chap:appendix-corr} contains some technical details used in
this chapter.
The content of this chapter is composed of selected material from
\cite{pub-thogersen07, pub-thogersen08-1, pub-fedorov08-1,
  pub-jensen07}.

Chapter~\ref{chap:mean-field} approaches the question of effective
range corrections in condensates from the mean-field point of view. A
modified Gross-Pitaevskii equation is combined with a higher-order
Feshbach resonance model. Effects on stability and decay mechanisms
are predicted. 
In chapter~\ref{chap:tf} we continue the mean-field analysis of
condensates. We present the Thomas-Fermi approximation in the case of
higher-order interactions. Some details are found in
appendix~\ref{chap:appendix-tf}.
Chapter~\ref{chap:mean-field} and \ref{chap:tf} are based
on \cite{pub-thogersen09b, pub-zinner09a, pub-zinner09b}.

Conclusions and outlook are given in chapter~\ref{chap:summary}. The
back-matter contains appendices
and the bibliography.

The results presented in \cite{pub-fedorov09-1} do not fit in the
line of this thesis and is omitted.

Online information: This dissertation is available online at the
address \url{http://www.martint.dk/thesis}, as well as on the arXiv
preprint server. The bibliography for the thesis is also available on
the above address, including abstracts and direct (doi) links to all
referenced articles. This will hopefully be useful for the reader.

\chapter*{Acknowledgements}
\addcontentsline{toc}{chapter}{Acknowledgements} 

First of all I would like to thank my two cheerful supervisors, Dmitri
V. Fedorov and Aksel S. Jensen, for all their support.  They have
always been available when I needed help, ranging from technical
details to conceptual understanding of physical phenomena and overall
strategies.  Besides teaching me a lot of physics they have also
taught me the main principles of independent research, how to write
papers that could hopefully be interesting to others, and how to
maneuver through the political maze of scientific publication. Having
two supervisors has been a true privilege and I will take the best
with me from each of them.

Special thanks goes to my close colleague and friend Nikolaj
T. Zinner. He has always been interested in discussing a wide range of
my questions and given feedback on my research. Especially during the
past year after Nikolaj went abroad our cooperation was extended and
resulted in some nice papers. Nikolaj has also been kind to proofread
major parts of this thesis.

I acknowledge the cooperation and help from many of my colleagues at
{\AA}rhus University. In particular Nicolai Nygaard and Uffe
V. Poulsen have been very kind to answer my questions and to comment
on my newest ideas when I occasionally have dropped by their office.
Also Thomas Kj{\ae}rgaard and David C. Hansen for carrying out a few
projects with me.

During my longer visit as a research scholar in USA, prof. Brett
D. Esry and his group at Kansas State University were very kind to host
me.  I appreciate the work we did together during the stay and
afterwards. Also, I am grateful for the hospitality and openness shown
by all my friends in Kansas.  I must also thank all the people I have
met around the world at conferences and workshops during the last four
years.  Especially the people in the NordForsk network on coherent
quantum gases.

I am pleased to have been a part of the subatomic group at {\AA}rhus
University. Thanks goes to previous and current members of this group,
in particular Karsten Riisager, Hans Fynbo, Christian Diget, Hans
Henrik Knudsen, Solveig Hyldegaard, Oliver Kirsebom, Jacob Johansen,
and Raquel Alvarez.  Our coffee and lunch breaks have always been
amusing experiences.

A big thank goes out to all my friends at {\AA}rhus University. This
includes people in the MatFys2001 group, TK, FFB, and too many others
to list here.  I will miss you all, but hope to keep in touch with
many of you.

I am grateful to my parents, sister and the rest of my family for
giving me their full support and for bearing with me during periods of
physical and mental absence. Most of all I am indebted to my beloved
wife Susanne. I would not have been able to complete this major
project without her.

\cleardoublepage
\addcontentsline{toc}{chapter}{List of Publications} 
{
\input{pub.bbl}

}
\label{chap:publications}
\nocitepub{pub-thogersen08-3,pub-thogersen09a,pub-thogersen08-2,
  pub-fedorov08-1,pub-fedorov09-1,pub-thogersen07, pub-thogersen08-1,
  pub-fedorov08-1, pub-jensen07, pub-thogersen09b, pub-zinner09a, pub-zinner09b}

\cleardoublepage
\pagestyle{fancy}
\pagenumbering{arabic}
\symb{*}
\chapter{Introduction}
\label{chap:intro}
\input{intro.tex}

\chapter{Theoretical and Numerical Background}
\label{chap:theory}
\input{theory.tex}

\chapter{Efimov Physics: Finite Range and Trap Effects}
\label{chap:efimov}
\input{efimov.tex}
\chapter{Conditions for Efimov Physics}
\label{chap:efimov2}
\input{efimov2.tex}

\chapter{$N$-Body Efimov Effect}
\label{chap:N-efimov}
\input{N-efimov.tex}
\chapter{Two-Body Correlations in Condensates}
\label{chap:correlations}
\input{correlations.tex}

\chapter{Mean-Field BEC with Higher-Order Interactions}
\label{chap:mean-field}
\input{mean-field.tex}

\chapter{Higher-Order Thomas-Fermi Approximation}
\label{chap:tf}

\input{tf.tex}
\chapter{Summary and Outlook}
\label{chap:summary}
\input{summary.tex}

\appendix
\chapter{Calculation of Condensate Fractions}
\label{chap:appendix-obdm}

\input{appendix-obdm.tex}
\chapter{Examples of Strongly Correlated Wave-Functions}
\label{chap:appendix-corr}

\input{appendix-corr.tex}

\chapter{TF Boundary Conditions and the Virial Equation}
\label{chap:appendix-tf}

\input{appendix-tf.tex}

%
\cleardoublepage
\addcontentsline{toc}{chapter}{Bibliography} 
\nocite{thogersen07,thogersen08-1,thogersen08-2,thogersen08-3,jensen07,fedorov08-1,fedorov09-1,thogersen09a,thogersen09b,zinner09b,zinner09a}
{%
\input{all.bbl}

}

\end{document}

%% file: usepackage.tex
\usepackage[english]{babel}
\usepackage{graphicx}
\usepackage{amssymb,amsmath}
\usepackage{fancyhdr}
\usepackage[final]{fixme} %
\usepackage{tikz,xxcolor}
\usepackage{multibib}
\usepackage[it,small]{caption}
\usepackage[toc,style=super,cols=2]{glossary}
\usepackage{threeparttable}
\usepackage{url}
\usepackage{rotating}

%% file: newcommand.tex
\newcommand{\bm}[1]{ \mbox{\boldmath $#1$}  }
\newcommand{\abs}[1]{\lvert #1 \rvert}
\newcommand{\pitwo}{ \frac{\pi}{2} }

\newcommand{\pisix}{ \frac{\pi}{6} }
\newcommand{\ud}{\,\mathrm{d}}

\newcommand{\rot}{{\cal R}}

\newcommand{\symb}{\useglosentry}

%% file: symbols.tex
\storeglosentry{a}{name={$a$},sort=a,%
  description=$s$-wave scattering length}

\storeglosentry{Re}{name={$R_e$},sort=Re,%
  description=Effective range}

%% file: autopackage.tex
%
babel, graphicx, amssymb, amsmath, fancyhdr, fixme, tikz, xxcolor, multibib, caption, glossary, threeparttable, url, rotating

%% file: pub.bbl
\newcommand{\etalchar}[1]{$^{#1}$}

%% file: intro.tex
Universality or model-independence is desirable in theoretical and
experimental physics because the scope becomes broader and the
applications more transparent and flexible. Universality generally
refers to systems determined by only a few large-distance parameters,
implying that they become independent of the specific short-range
interactions and structure.  Thus, once theoretical formulations are
established they can be applied to a wide range of fields, e.g.
nuclear, atomic, and molecular physics. Much effort has therefore
been devoted to extract universal features in many branches of
physics.

\section{Ultra-Cold Atomic Gases}

One system in which universality can be found is Bose-Einstein
condensates (BEC) where cold massive bosons occupy the same
quantum-mechanical state. The BEC concept was proposed almost a
century ago \cite{einstein24}, but remained an elusive goal for atomic
gas experiments for many years.  The quest for this cold grail was
motivated by several factors. The idea of a macroscopic number of
particles in the same single-particle ground state is simple, elegant,
and easy to grasp. Also, it is direct evidence of quantum mechanics
and particle-wave duality on a macroscopic scale.

Since the experimental realization of the first BEC atomic $^{87}$Rb
two decades ago \cite{anderson95} the field of ultra-cold atomic gases
has picked up speed.  Bose-Einstein condensation has now been achieved
for most of the alkali gases in numerous experimental groups,
typically with $^7$Li, $^{23}$Na, $^{39}$K, $^{41}$K, $^{85}$Rb,
$^{87}$Rb, and $^{133}$Cs \cite{pethick02}.  In the wake of the BEC
milestone \cite{anderson95} a tremendous variety of experiments has
been realized \cite{bloch08,giorgini08}.  This includes the
interference between two condensates proving full coherence
\cite{andrews97}, structured arrays of quantized vortices in rotating
condensates \cite{abo-shaeer01}, and propagating solitons and
sound-like waves \cite{pitaevskii03}.  Besides standard (harmonic)
traps, numerous experiments have been performed with different external
confinement, including one- and two-dimensional gases and optical
lattices with e.g. Mott-insulating phases \cite{bloch08}.  Experiments
with ultra-cold fermions have also given rise to new physics, e.g.
BEC-BCS crossover and super-fluidity have been realized
\cite{giorgini08}.  Moreover, it is possible to produce cold gases
with a wide range of mixed bosonic and fermionic species
\cite{stan04,inouye04}.  Production of bosonic gases with a small
definite number of atoms is also in progress \cite{dudarev07}.

Currently, one of the most important experimental tools for ultra-cold
gases is the use of Feshbach resonances \cite{kohler06,chin09}, which
are found in virtually any atomic system. This allows for arbitrary
tuning of atomic scattering lengths (i.e.  low-energy interaction
strengths) by varying an external magnetic field.  Tunable scattering
lengths have been a key ingredient for controlling the production,
macroscopic stability, and decay of condensates \cite{dalfovo99}. The
Feshbach resonance method also allows for production of weakly bound
diatomic molecules (shallow dimers) \cite{kohler06,chin09}.

\begin{figure}[htb]
  \centering
  \includegraphics[width=6.2cm, height=5cm]{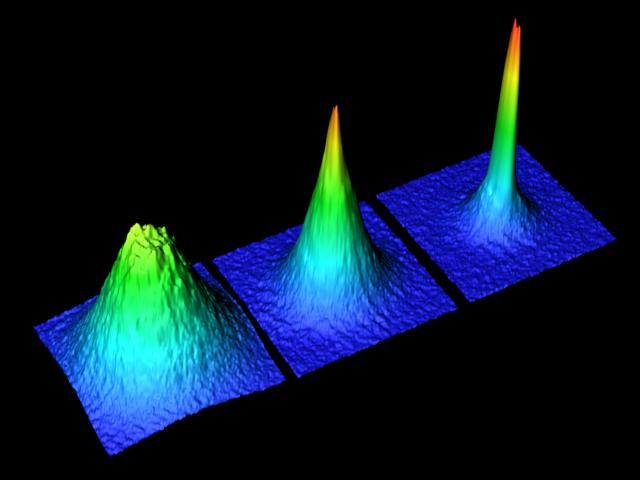}%
  \includegraphics[width=6.2cm, height=5cm]{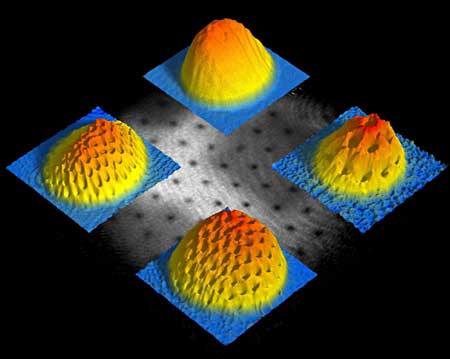}
  \caption{Experimental realization of Bose-Einstein condensation in
    trapped alkali gases. Left: The characteristic ``spike'' of
    condensed $^{87}$Rb atoms emerges from the thermal cloud as the
    temperature is lowered below a critical value. Right: A rotating
    $^{23}$Na BEC with vortices arranged in a highly ordered
    structure.  From ETH Zurich (2002) and MIT (2001)
    \cite{abo-shaeer01}.}
  \label{fig:exp-bec}
\end{figure}

From the perspective of universality the scattering length is the
decisive parameter for ultra-cold gases. Most features of the above
experiments can be understood in terms of this parameter alone. This
simplifies the theoretical descriptions immensely and makes the
interpretations of the physical effects extremely sharp and clear.
Most of the theoretical models for condensates are based on some sort
of mean-field approximation with a contact interaction proportional to
the scattering length. This means that all high-energy effects are
integrated out and short-range model-dependent correlations are
neglected.  Thus it gives simple models which by definition are
strictly universal.

When such universal models are used it is of course essential to know the
precise ``window'' where the results can be applied to experiments. In
many realistic cases this window is in fact very narrow. It is
therefore important to know the qualitative significance of
higher-order effects at the border of this window. The higher-order
effects may ultimately be expressed in terms new model-independent
parameters, but can also be strictly non-universal.

\section{Few-Body Systems and the Efimov Effect}

Another long-standing prediction of universality is the Efimov effect,
which was presented in the early 1970s
\cite{efimov70,efimov71,amado71,amado72,efimov73} and has been
discussed thoroughly since
\cite{fonseca79,fedorov93,fedorov94a,amorim97,nielsen99a,fedorov01,nielsen01,braaten06}.
It is an anomalous effect which occurs in neutral three-body
systems when at least two of the underlying subsystems have resonant
$s$-wave interactions, i.e. a diverging scattering length (or at least
much larger than the range of the physical force).  This corresponds
to the subsystems being close to the threshold for binding. Under
these conditions an exceptional sequence of infinitely many
geometrically spaced three-body bound states occurs, with an
accumulation point at zero energy. Moreover, the ratio between the
energies is a universal constant depending only on the mass ratios and
the particle statistics. For three identical bosons the famous ratio
is \cite{nielsen01}
\begin{equation}
  \label{eq:scaling-factor}
  \frac{E_T^{(n)}}{E_T^{(n+1)}}=\textrm{e}^{2\pi/s_0}\simeq 515.0\;.
\end{equation}
We emphasize that this specific scaling factor is independent of the
mass, and that the same effect in principle can be found at all
physical scales. The Efimov states have large sizes which are also
geometrically spaced with ratios $515.0^{1/2}=22.7$, and lie outside
the classically allowed region.

The effect was initially proposed by V. Efimov within the field of
nuclear physics and has been searched for over many years in nuclei --
unfortunately this search has been futile.  The only possible
three-body candidate without Coulomb interactions is two neutrons and
a (positively charged) core. The possible resonant $s$-wave
interactions should then be between each of neutrons and the
core\footnote{Two neutrons with a spin-singlet function can be in a
  relative $s$-wave state. However, the singlet $n$--$n$ scattering
  length is $\sim 20$fm, which is not large enough.}.  Such three-nucleon
systems, called two-neutron halos, do in fact exist. Common examples
are $^6$He ($\alpha+n+n$) and $^{11}$Li ($^9$Li$+n+n$) \cite{jensen04}
where the essential degrees of freedom correspond to the core$+n+n$
structure.  Unfortunately the heavy-light-light mass ratio gives an
extremely large scaling factor (as compared to the already large
factor in eq.~\eqref{eq:scaling-factor}) \cite{jensen03}.  Thus one
can only hope for ``accidental fine-tuning'' of the neutron-core
scattering length to see the Efimov effect.  One of the few candidates
is $^{20}$C \cite{amorim97,braaten06}. The Efimov effect has also been
searched for in atomic systems. There is a general theoretical
consensus~\cite{lim77,esry96,esry06,nielsen98} that the first excited
state of the atomic helium trimer $^4$He$_3$ is an Efimov state,
although the experimental observation so far proved
elusive~\cite{bruhl05}.

\begin{figure}[htb]
  \centering
  \begin{minipage}[c]{0.47\textwidth}
    \centering
    \includegraphics[width=4cm, height=4cm]{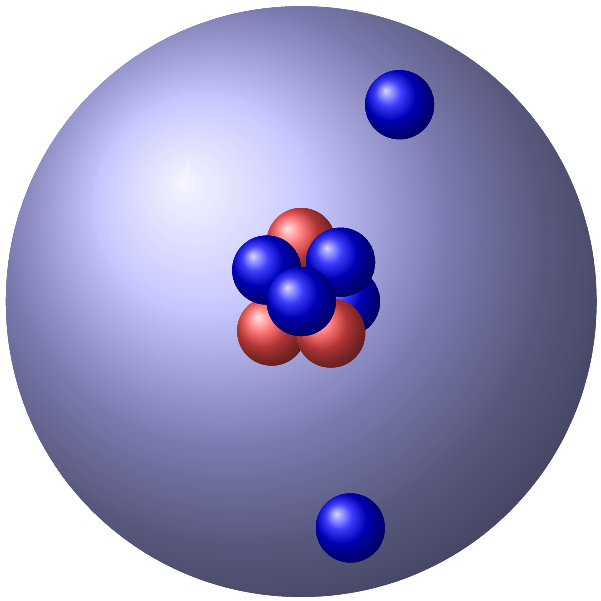}
  \end{minipage}%
  \hfill
  \begin{minipage}[c]{0.47\textwidth}
    \centering
    \includegraphics[width=3.5cm, height=3.5cm]{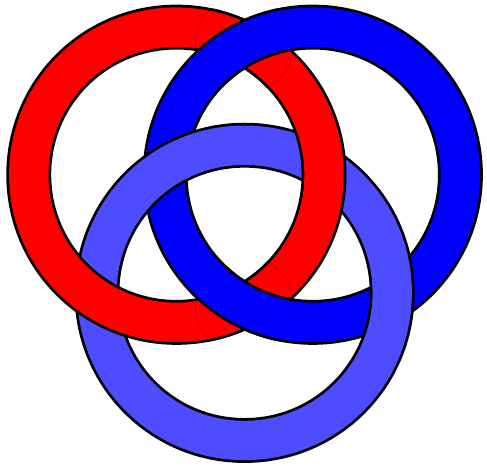}
  \end{minipage}
  \caption{Left: Schematic two-neutron halo nucleus with the neutrons
    weakly bound to the core. Right: Interlocked Borromean rings: If
    one ring is removed the others fall apart. The name ``Borromean''
    is historical and comes from the aristocratic Italian Borromeo
    family who used the rings in their coat of arms.}
  \label{fig:halo+borro}
\end{figure}

Another concept closely related to the Efimov effect is Borromean
binding. The phrase refers to atypical bound few-body systems where
none of the individual subsystems are bound. This is often symbolized
by the interlocked Borromean rings, see fig.~\ref{fig:halo+borro}.
Borromean systems are typically weakly bound with large spatial
probabilities in classically forbidden regions. This implies that the
structure is mainly determined by the binding energy alone, and are
thus good universal candidates.  The Efimov effect can be seen as a
critical limit of Borromean binding: Here the two-body subsystems are
exactly at the threshold but infinitely many bound three-body states
exists.  Examples of Borromean systems are the two-neutron halos
$^6$He and $^{11}$Li discussed above as well as $^9$Be ($\alpha+\alpha+n$)
\cite{jensen04}.

\section{Efimov Physics in Atomic Gases}

The obstacles for observing the Efimov effect in nuclei are three-fold:
i) nuclei are electrically charged, ii) the are no methods for tuning
the interactions externally so one must rely on accidental tuning, and
iii) the remaining candidates have very unfortunate mass ratios.
Instead, the approach of observing the Efimov effect indirectly via
three-body recombination losses in ultra-cold atomic gases attract
much attention currently. As compared to nuclei, atomic gases are very
promising for several reasons: i) atoms are electrically neutral, ii)
scattering lengths can be tuned by magnetic Feshbach resonances, and
iii) by using different species with appropriate mass ratios
(heavy-heavy-light) the universal scaling factor can be reduced.
Concerning the last part, specific proposals for ultra-cold atomic gas
mixtures were given in \cite{dincao06}, for example the realized
boson-fermion mixtures $^{23}$Na-$^{6}$Li \cite{stan04} or
$^{87}$Rb-$^{40}$K \cite{inouye04}, or even better $^{133}$Cs-$^{6}$Li
or $^{87}$Rb-$^{6}$Li.

The basic idea for observing Efimov states using Feshbach resonances
is to measure the bound state spectrum indirectly. When the
scattering length is tuned around large values, a series of resonance
and interference effects occurs in the three-body recombination rate
\cite{esry06,braaten06}. These features are located at critical
scattering lengths with the same scaling factor as the Efimov
sequence, i.e. 22.7 for identical bosons.
The first experimental indications of such features has been found
recently in cold $^{133}$Cs gases \cite{kraemer06,knopp09}.  Also, a
universal trimer state is reported in a three-component $^6$Li Fermi
gas \cite{wenz09}.  Recent theoretical \cite{dincao09,vonstecher09}
and experimental \cite{ferlaino09} work indicate the existence of a
universal four-body effect where an atom is weakly bound to an Efimov
trimer.

The phrase ``Efimov physics'' was introduced recently in the context
of atomic gases and Feshbach resonances \cite{dincao05}, but is now
used more and more frequently.  We will adopt this phrase to the
extent possible: By ``Efimov effect'' we refer to the infinite
accumulation of trimers states at the threshold, while by ``Efimov
physics'' we mean the broader concept in cold gases using Feshbach
resonances and recombinations loss for observing the Efimov effect and
related features.

The experiments performed until now strongly indicate that individual
features of Efimov states have been found. However, the results are
yet not conclusive \cite{lee07}. Definite evidence requires two, or
even better a whole sequence, of the states with the correct relative
scaling to be observed.  The universal window is still quite small and
experimental control is not at its final level yet. Theoretical
physics can help at this point by predicting quantitatively the
regions where universality occurs, as well as the corrections at the
border of these regions.

The subject of universality in atomic gases will face many new
challenges during the next decades.  In the most recent review on
universality in few-body systems \cite{braaten06}, three subjects were
marked as the frontiers of this field: The $N$-body problem for
$N\ge4$, effective range corrections, and large $p$-wave scattering
length.  These issues are clearly within reach in ultra-cold gases and
indicate a promising future.

\section{Angle of this Thesis}
In this dissertation we carry out theoretical investigations of
universality and its limits in few- and many-boson systems.  
We focus on ultra-cold trapped atomic gases, but strive to present the
results in universal terms and via model-independent parameters. Thus,
much of the work may hopefully be used or continued in other areas of
physics.

We will investigate effective range corrections and trap effects to
Efimov physics, as well as the limits of the universal predictions.
Next we investigate the possibility of an Efimov effect in $N$-body
systems.  We also consider trapped BEC-like states from the few-body
level, where we will be concerned with model-independence and two-body
correlations. The approach to answer these questions will be from the
pure few-body level, keeping in mind only to include the degrees of
freedom needed for physical relevance. Finally, we approach
Bose-Einstein condensates from the angle of standard mean-field theory,
but with higher-order interaction terms included.

\vspace{1cm}
\paragraph{Note added.}
The observation of an Efimov spectrum in an ultra-cold gas has just
been published \cite{zaccanti09}, giving the first definite proof of
the universal scaling factors.  This happened at the very final stage
of writing this dissertation (I, the author, only became aware of the
results a few days before handing in the thesis). To keep the
chronological order, no contents or conclusions have been modified at
all. A more elaborate note is found in the summary chapter.

%% file: theory.tex
In this chapter we consider some of the theoretical background for
few-body systems. For the two-body systems we discuss basic universal
behavior, as well as some results on Feshbach resonances, atomic
interactions, and model potentials. Specifically we discuss the large
effective range.  For the three-body systems we derive the Efimov
effect in the hyper-spherical adiabatic approximation and explain
experimental consequences for recombination rates. 
We also briefly review some mean-field concepts. Finally we describe
the stochastic variational method which is one of the main numerical
tools for our work.

\section{Two-Body Systems}
\subsection{Low-Energy Scattering}
Let us briefly review some relevant low-energy concepts for two-body
scattering and bound states.  We will be concerned with the limit
where where only the lowest partial waves, $l=0$, contribute
\cite{bransden03}. Here the scattering can be described by the radial
Schr{\"o}dinger equation
\begin{equation}
  \label{eq:radial-eq}
  \left(\frac{\partial^2}{\partial r^2}-U(r)+k^2\right)u(r)=0,
\end{equation}
where $r=|\bm r_1-\bm r_2|$ is the inter-particle distance,
$U(r)=2mV(r)/\hbar^2$ is the two-body interaction, $m$ the particle
mass, $u(r)=R(r)r$ the $s$-wave radial function, $k=\sqrt{2mE}/\hbar$
the wave number, and $E$ the relative energy.  The asymptotic behavior
\begin{equation}
  u\sim \sin(kr+\delta(k))
\end{equation}
defines the $s$-wave phase shift $\delta(k)$ as function of the
incident energy. The scattering amplitude $f(k)$, defined as the ratio
of outgoing spherical waves and ingoing plane waves, is in this case
spherically symmetric and related to the phase shift by
\begin{equation}
  f(k)=\frac{1}{k\cot\delta(k)-ik}.
\end{equation}
The total cross-section is then given by $\sigma=8\pi|f(k)|^2$ for
identical bosons. For small energies one traditionally employ the
effective range expansion for the phase shift,
\begin{equation}
  \label{eq:eff-range-exp}
  k\cot\delta(k)=-\frac{1}{a}+\frac{R_e}{2}k^2+\dots,
\end{equation}%
where $a$\symb{a} is the scattering length\footnote{We use this sign
  convention for $a$ throughout this dissertation, i.e. a weakly bound
  state corresponds to $a>0$.} and $R_e$\symb{Re} is the effective
range.

The poles of the scattering amplitude $f(k)$ defines the bound states,
virtual states, and resonances for the two-body problem.  The bound
states are located on the positive imaginary axis in the complex
$k$-plane, i.e. $k=i\kappa$, $\kappa>0$. The binding wave number
$\kappa$ is then determined by the equation,
\begin{equation}
  \label{eq:poles}
  i\kappa \cot\delta(i\kappa)+\kappa=0.
\end{equation}
The corresponding energy is given by\footnote{the subscript $D$
  referring to ``dimer'', i.e. a two-body bound state.}
\begin{equation}
  E_D=\frac{\hbar^2 k^2}{m}=-\frac{\hbar^2 \kappa^2}{m},
\end{equation}
where the binding energy is $B_D=|E_D|$.
Contrary to the bound states, the virtual states (or anti-bound
states) are located on the negative imaginary $k$-axis, i.e.
$k=i\kappa$, $\kappa<0$. The corresponding energy is also real and
negative,
\begin{equation}
  E_V=\frac{\hbar^2 k^2}{m}=-\frac{\hbar^2 \kappa^2}{m},
\end{equation}
however, the pole is located on the second Riemann sheet, and can thus
only be measured indirectly via analytic continuation of the cross
section.

\subsection{Universal Bound States}

Using only the lowest scattering length term in the effective range
expansion of eq.~\eqref{eq:eff-range-exp} and inserting in
eq.~\eqref{eq:poles}, we find a single pole with wave number
$\kappa=1/a$. For $a>0$ this gives a bound state with energy
\begin{equation}
  \label{eq:ED-scat}
  E_D=-\frac{\hbar^2}{ma^2},
\end{equation}
which holds when the scattering length is much larger than the
inter-atomic potential range, $r_0$. In atomic physics such a weakly
bound two-body state is often referred to as a shallow dimer.  This
type of state is universal, or model-independent, in the sense that it
depends only on the scattering length and not the details of the
short-range potential: The major part of the wave-function is an
exponential tail located outside $r_0$, which is an essential example of
universality.
The state can also be described in a zero-range model where the usual
boundary condition $u(0)=0$ is replaced by
\begin{equation}
  \left.\frac{1}{u}\frac{\partial u}{\partial r}\right|_{r=0}
    =\left.k\cot\delta(k)\right|_{k\to0}
    =-\frac{1}{a}.
\end{equation}
This leads to the wave number $\kappa=1/a$ and exponentially decreasing
wave-function $u\sim\exp(-r/a)$.

If we also include the effective range in
eq.~\eqref{eq:eff-range-exp}, the binding wave number is determined by
$R_e\kappa^2/2-\kappa+1/a=0$. This gives
\begin{equation}
  \label{eq:kappa-wavenumber}
  \kappa=\frac{1}{R_e}\left(1-\sqrt{1-2R_e/a}\right),
\end{equation}
where we have chosen the negative root. The state corresponding to the
positive root is unphysical, since it is strongly bound with energy
$E\sim\hbar^2/(mR_e^2)$ comparable to the potential depth
\cite{braaten06}.
Thus for $a>0$ we have a bound dimer with energy
\begin{equation}
  \label{eq:ED-reff}
  E_D = -\frac{\hbar^2}{mR_e^2}\left(1-\sqrt{1-2R_e/a}\right)^2.
\end{equation}
The virtual state energy $E_V$ is given by the same expression for
$a<0$.  Equation~\eqref{eq:ED-reff} holds when the higher-order shape
terms in the effective range expansion are negligible.  Expanding it
to lowest order in $R_e/a$ gives
\begin{equation}
  E_D=-\frac{\hbar^2}{ma^2}\left(1+\frac{R_e}{a}+O(\frac{R_e^2}{a^2})\right),
\end{equation}
which is then a small correction to eq.~\eqref{eq:ED-scat}.

\subsection{Feshbach Resonances}

Feshbach resonances is one of the most important experimental tools
for ultra-cold atomic gases, see \cite{chin09} for a recent review.
These resonances allows for arbitrary tuning of scattering lengths, in
particular the $s$-wave scattering length which is relevant for our
purpose.  This has been a major key for controlling macroscopic
stability and decay of condensates \cite{dalfovo99}, and for creating
shallow molecular dimers \cite{chin09}.

In the zero-range approximation, the scattering length of a Feshbach
resonance can be described by the phenomenological magnetic field
dependent expression
\begin{equation}
  \label{eq:scat-fesh}
  a(B)=a_{bg}\left(1-\frac{\Delta B}{B-B_0}\right).
\end{equation}
Here $a_{bg}$ is the background scattering length away from resonance,
$B_0$ is the position of the Feshbach resonance and $\Delta B$ the
width. Experimentally, these resonance parameters are usually
determined indirectly via peaks in the loss rate of condensates.

\begin{figure}[htbp]
  \centering
  \includegraphics[scale=0.9]{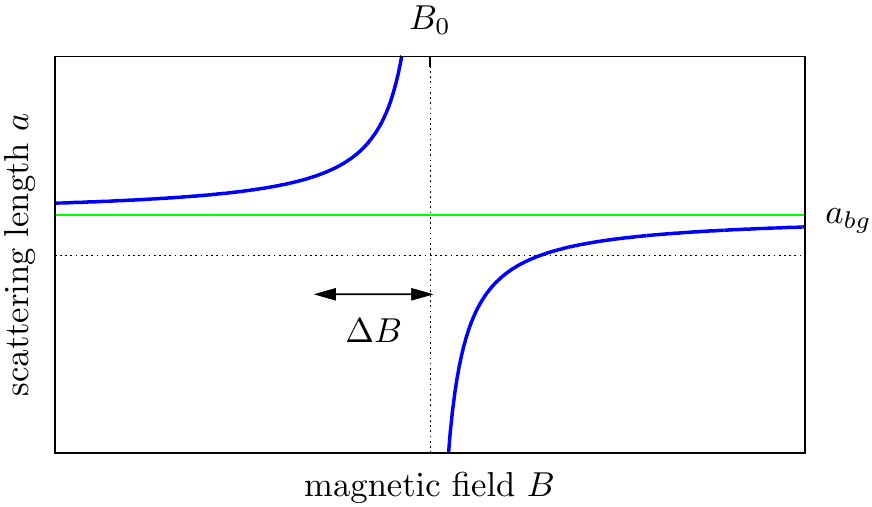}
  \caption{Scattering length $a(B)$ for the Feshbach resonance model
    eq.~\eqref{eq:scat-fesh} as function of magnetic field $B$. The
    resonance is centered around $B_0$ with width $\Delta B$ and has
    background scattering length $a_{bg}$. 
  }
  \label{fig:feshbach-res}
\end{figure}

\begin{figure}[htbp]
  \centering
  \begin{tabular}{cc}
    \includegraphics[scale=0.9]{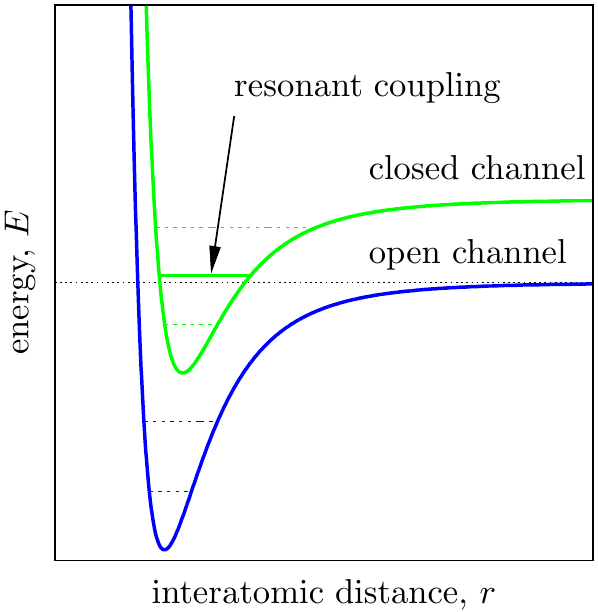}&
    \includegraphics[scale=0.9]{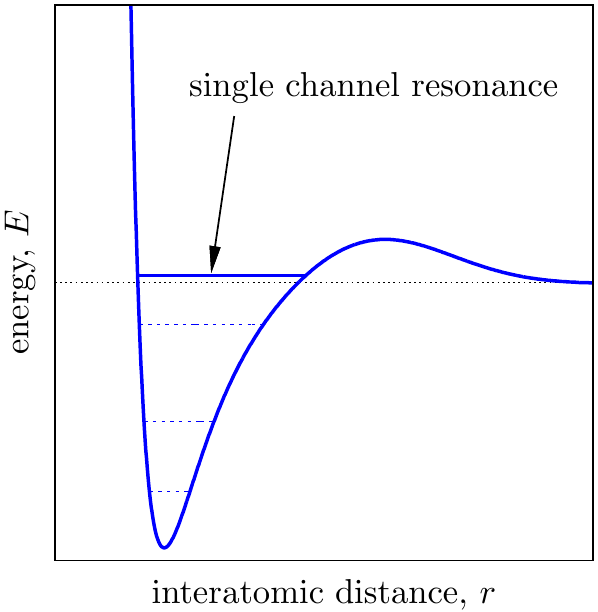}
  \end{tabular}
  \caption{Left: Schematic representation of a Feshbach resonance. A
    bound state of the closed channel has energy close to the (zero)
    incoming energy of the open channel. The relative position of the
    channels can be changed by an external magnetic field, thereby
    tuning the coupling. Right: A schematic (shape) resonance of a
    single-channel model. The Feshbach resonance can be modeled by a
    shape resonance if the elastic (open-open channel) scattering is
    the only contribution.}
  \label{fig:feshbach}
\end{figure}

The phenomenological behavior in eq.~\eqref{eq:scat-fesh} can be
reproduced in various models, a common example is the two-channel
model described in \cite{pethick02}. In this model the resonance width
$\Delta B$ turns out to be proportional to the matrix element
connecting the open and closed channels that overlap in energy and
cause the resonant behavior, see fig.~\ref{fig:feshbach}.

The scattering length, eq.~\eqref{eq:scat-fesh}, is obtained when a
two-channel model is reduced to an effective single-channel model. The
related effective range of the effective single-channel zero-range
model is determined from the resonance width as (see e.g.
\cite{bruun05}),
\begin{equation}
  \label{eq:reff-fesh-const}
  R_{e0}=-2 \frac{\hbar^2}{m\Delta\mu\Delta B a_{bg}},
\end{equation}
where $\Delta\mu$ is the difference between the magnetic moments in
the open and closed channel. This result holds on-resonance, i.e. near
$|a|=\infty$. In terms of common experimental units the effective
range reads
\begin{equation}
  R_{e0}= \frac{-5.16\cdot 10^{6}a_0}{
    \left(\tfrac{m}{m_u}\right)
    \left(\tfrac{a_{bg}}{a_0}\right)
    \left(\tfrac{\Delta \mu}{\mu_B}\right)
    \left(\tfrac{\Delta B}{G}\right)},
\end{equation}
where $a_0$ is the Bohr radius, $m_u$ is the unified atomic mass unit,
$\mu_B$ the Bohr magneton, and $\Delta B$ is measured in Gauss.  The
effective range is always negative but can be arbitrarily large for
narrow resonances. This is consistent with our observations
for the zero-range limit of finite-range potentials, see below.

The effective range, eq.~\eqref{eq:reff-fesh-const}, is constant
(independent of magnetic field strength, $B$) and only holds near
$|a|=\infty$. In chapter~\ref{chap:mean-field} we derive the more
general $B$-dependent effective range,
\begin{equation}
  \label{eq:reff-fesh-theory}
  R_e=R_{e0}\left(1-\frac{a_{bg}}{a(B)}\right)^2.
\end{equation}

\subsection{Atomic Interactions and Model Potentials}

\subsubsection{van der Waals Interaction}
The real atomic potential between neutral atoms can in the
Born-Op\-pen\-hei\-mer approximation mainly be described by a short-range
and a long-range part. The short-range repulsive core arises from the
overlapping electron clouds, while the attractive long-range van der
Waals interaction is due to the polarizability of the electron clouds.
Asymptotically this van der Waals tail goes as
\begin{equation}
  V(r) \simeq -\frac{C_6}{r^6}+ O\left(\frac{1}{r^{12}}\right).
\end{equation}
For ultra-cold gases, where the relative momentum between the atoms is
small, this is the dominating part of the interaction. The $C_6$
coefficient defines the van der Waals length
\begin{equation}
  \label{eq:vdw-length}
  l_{vdW}=\left(\frac{mC_6}{\hbar^2}\right)^{1/4},
\end{equation}
which is the typical interaction length scale between neutral atoms.
Examples of $l_{vdW}$ are $44.93a_0$ for $^{23}$Na and $82.58a_0$ for
$^{87}$Rb \cite{chin09}.  The effective range for the single-channel
van der Waals interaction can be estimated to $R_e\simeq 1.39 l_{vdW}$
for $a\gg l_{vdW}$ \cite{chin09}.

\subsubsection{Model Potentials}

Low energy scattering can be described by model potentials of finite
range. Typical choices are the finite square well or a van der Waals
tail with a hard core which can be solves analytically. For example, the
square well has scattering length and effective range
\begin{equation}
  \label{eq:squarewell-a-re}
  \frac{a}{r_0}=1-\frac{\tan s}{s},   \qquad 
  \frac{R_e}{r_0}=1-\frac{1}{3}\left(\frac{r_0}{a}\right)^2-\frac{1}{s^2}\frac{r_0}{a},
\end{equation}
in terms of the finite range $r_0$ and dimensionless depth $s=r_0\sqrt{mV_0}/\hbar>0$.
Each time a bound state is at the threshold, $a$ diverges
($s=\pi/2+n\pi$) and the effective range equals the finite range
$R_e=r_0$. When $a\to 0$ the effective range diverges as $-1/a^2$.
The scattering length and effective range for the van der Waals tail
($-C_6/r^6$ for $r>r_0$) with hard core ($\infty$ for $r<r_0$) is
found in \cite{braaten06}.

\begin{figure}[htbp]
  \centering
  \includegraphics[scale=0.9]{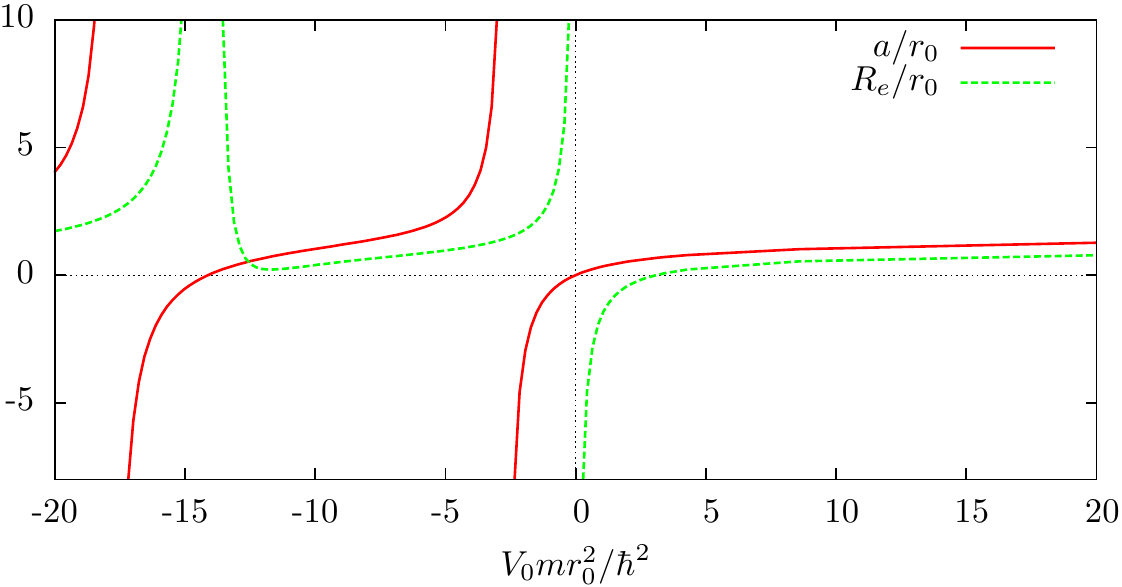}
  \caption{Scattering length and effective range for the Gaussian
    interaction, eq.~\eqref{eq:gauss-interaction}, in units of the range
    $r_0$. The first bound state occurs at $V_0=-2.68
    \hbar^2/(mr_0^2)$.}
  \label{fig:gauss-int}
\end{figure}

Another common choice is the Gaussian potential
\begin{equation}
  \label{eq:gauss-interaction}
  V(r)=V_0\exp(-r^2/r_0^2),
\end{equation}
which we will use later for numerical calculations. The scattering
length have features similar to the square well, but the effective
range is always positive for the attractive case, see fig.~\ref{fig:gauss-int}.

Feshbach resonances can also be modeled by single-channel potentials.
This is of course only possible if the energy is low and elastic
scattering is considered. The model potential should have the same
low-energy phase shift as the real Feshbach resonance. The diverging
scattering length can be be modeled by any potential, just by tuning a
two-body bound state to zero energy.  However, as we saw in
eq.~\eqref{eq:reff-fesh-theory} the effective range for a Feshbach
resonance can be very large and negative.  Only certain single-channel
model potentials can reproduce this.

In general the effective range of a finite-range potential is given by
\begin{equation}
  \label{eq:reff-integral}
  R_e=2\int_0^\infty \left[v_0^2(r)-u_0^2(r)\right]\ud r,
\end{equation}
where $u_0$ is the zero-energy wave-function normalized
asymptotically as $u_0\to 1-r/a$, and $v_0=1-r/a$ is the asymptotic
solution extended to all $r$. When considering a potential of finite range
$r_0$, we have $u_0=v_0$ in the outer region and the integral only
runs to $r_0$. Performing the integration over $v_0^2$ and noting that
the integral over $u_0^2$ is positive we get the bound \cite{phillips97}
\begin{equation}
  \label{eq:re-bound}
  R_e\le 2 r_0 \left(1-\frac{r_0}{a}+\frac{r_0^2}{3a^2}\right).
\end{equation}
In the resonant limit $a=\infty$, we find $R_e\le 2r_0$, in particular
the effective range must be negative if we also approach the
zero-range limit.\footnote{The ``Wigner bound''
  \cite{wigner55,phillips97} says that for zero-range interactions
  $k\cot\delta$ is monotonically decreasing with energy,
  i.e. $d(k\cot\delta(k))/d(k^2)\le 0$. Using
  eq.~\eqref{eq:eff-range-exp} this also leads to $R_e\le0$.}
We note that the large negative effective range occurs when the
amplitude of $u$ is considerably larger inside the potential
as compared to the asymptotic value ($v_0=1$).

\begin{figure}[htbp]
  \centering
  \includegraphics[scale=1.0]{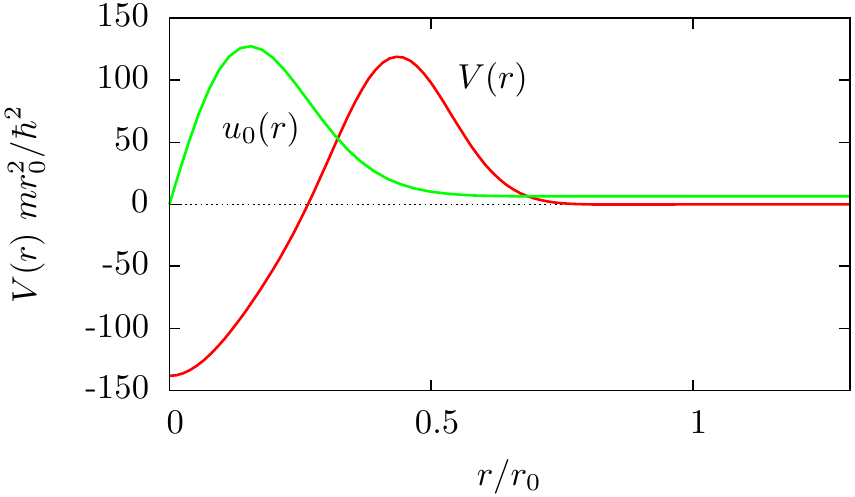}
  \caption{Soft two-body potential $V$ with barrier,
    eq.~\eqref{eq:yujun-pot}. The parameters $D=-138.27$,
    $B=128.49$ (in units $\hbar^2/(mr_0^2)$), and $\chi=4.6667$ are
    tuned to give large scattering length $a=556.88 $ and large
    negative effective range $R_e=-142.86$. The zero-energy
    wave-function $u_0(r)$ (arb. units) has large amplitude inside the
    barrier.}
  \label{fig:yujun-pot}
\end{figure}

This means that to reproduce a large negative effective range with a
single-channel finite-range potential we need to have a (shape)
resonance around zero energy, see e.g. fig.~\ref{fig:feshbach}. The
simplest way to do this is by having an attractive inner region and a
repulsive barrier.  This could for example be the square well with a
square barrier, which can be solved analytically \cite{lmjensen06}. We
instead consider the softer potential with barrier
\begin{equation}
  \label{eq:yujun-pot}
    V(r)=D\ \text{sech}^2\left(\chi\frac{r}{r_0}\right)
        +B \exp\left(-2(\chi\frac{r}{r_0}-2)^2\right),
\end{equation}
which is shown in fig.~\ref{fig:yujun-pot}. By tuning the depth of the
pocket and
height of the barrier we can make both $a$ and $|R_e|$ much larger
than $r_0$. This leads to a large amplitude of the zero-energy
wave-function $u_0$ inside the potential, as compared to the
asymptotic region. In the limit $|a|\to\infty, R_e\to-\infty$ all
probability is located inside the potential.
This means that universal features of the two-body problem can be lost
for $R_e\to-\infty$ even within the normal universal limit of large
scattering length. We investigate this point for the three-body Efimov
effect in chapter~\ref{chap:efimov2}.

In conclusion, when we talk about effective range it means
higher-order in the two-body scattering dynamics, i.e. the $k^2$ term
of the phase shift expansion eq.~\eqref{eq:eff-range-exp}. It can
either be related to the finite range of a single-channel model
potential (such as the van der Waals interaction) or the higher-order
term of a coupled-channel Feshbach resonance,
eq.~\eqref{eq:reff-fesh-theory}. However, large effective range
(compared to the interaction range $r_0$) can only occur near a
Feshbach or shape resonance, and must be negative in this case.

\section{Three-Body Systems}
We now describe the three-body problem with the hyper-spherical
adiabatic approximation. We consider the zero-range limit with a
Faddeev-type decomposition and appropriate boundary conditions to
derive the effective hyper-radial potential responsible for the Efimov
effect. Applications and experiments are outlined.

\subsection{Hamiltonian and Coordinates}
\begin{figure}[htbp]
  \centering
  \begin{minipage}[c]{0.47\textwidth}
    \centering \includegraphics[width=3cm, height=3cm]{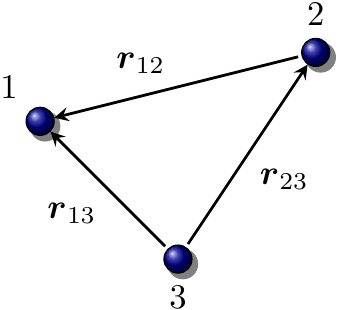}
  \end{minipage} \hfill
  \begin{minipage}[c]{0.47\textwidth}
    \centering \includegraphics[width=3cm, height=3cm]{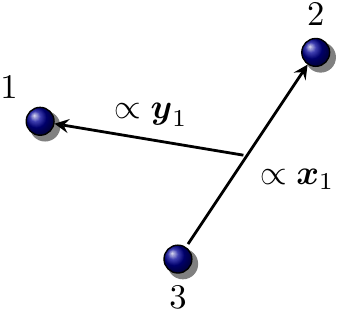}
  \end{minipage}
  \caption{Relative coordinates $\bm r_{ij}$ and one set
    ($i=1$) of Jacobi-coordinates $\bm x_1,\bm y_1$.}
  \label{fig:3B-coordinates}
\end{figure}

Let us first describe the hyper-spherical adiabatic approximations,
details can be found in \cite{nielsen01,fedorov01}.
We consider the three-body system with the center-of-mass (cm) Hamiltonian 
\begin{equation}
  \hat H=\hat T +\sum_{i<j} V(\bm r_{ij}),
\end{equation}
where $\hat T$ is the kinetic energy operator in the center-of-mass
frame and $V$ is the two-body interaction.  We denote the vector from
particle $j$ to $i$ by $\bm r_{ij}=\bm r_i-\bm r_j$, and the vector
from the center-of-mass of the pair $(j,k)$ to particle $i$ by $\bm
r_{i,(jk)}=\bm r_i-(\bm r_j+\bm r_k)/2$. In the equal mass system the
Jacobi coordinates $(\bm x_i,\bm y_i)$ for $i=1,2,3$ are defined
as\footnote{Different proportionality factors are used in literature,
  we use this one to be consistent with later definitions for $N>3$.}
\begin{equation}
  \label{eq:jacobi-coor}
  \bm x_i= \frac{1}{\sqrt{2}} \bm r_{jk},\quad
  \bm  y_i=\sqrt{\frac{2}{3}}\bm r_{i,(jk)},
\end{equation}
where $\{i,j,k\}$ is a cyclic permutation of $\{1,2,3\}$, see
fig.~\ref{fig:3B-coordinates}.

\begin{figure}[htbp]
  \centering
  \begin{minipage}[c]{0.47\textwidth}
    \centering \includegraphics[height=2.5cm]{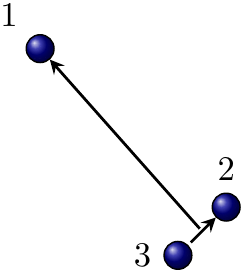}
  \end{minipage} \hfill
  \begin{minipage}[c]{0.47\textwidth}
    \centering \includegraphics[height=2.3cm]{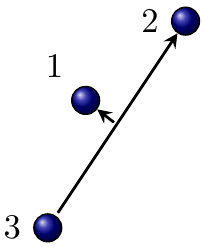}
  \end{minipage}
  \caption{Two configurations with the same hyper-radius $\rho$ but different
  hyper-angles $\alpha_1$. Left: $\alpha_1\sim 0$. Right:
  $\alpha_1\sim \pi/2$.}
  \label{fig:3B-coordinates-2}
\end{figure}

The six hyper-spherical coordinates
$(\rho,\alpha_i,\Omega_{xi},\Omega_{yi})$ are defined as
\begin{equation}
  \label{eq:hypersp-coor}
  x_i=\rho\sin\alpha_i,\quad y_i=\rho\cos\alpha_i,
\end{equation}
where $\rho\ge 0$ is the hyper-radius, $\alpha_i\in[0,\pi/2]$ is the hyper-angle,
$\Omega_{xi}=\{\vartheta_i,\varphi_i\}$ the angles describing the
direction of $\bm x_i$, and similarly $\Omega_{yi}$ for $\bm y_i$. The
hyper-radius
\begin{equation}
  \label{eq:rho-3}
  \rho^2=x_i^2+y_i^2 =\frac{1}{3}(r_{12}^2+r_{13}^2+r_{23}^2)
\end{equation}
is independent of the chosen Jacobi-set and describes the average size
of the three-body system. It is the only dimension-full coordinate of
the hyper-spherical coordinates. All the angles are denoted together by
$\Omega_i=(\alpha_i,\Omega_{xi},\Omega_{yi})$.
For fixed $\rho$, the hyper-angle
$\alpha_i$ describes (together with $(\Omega_{xi},\Omega_{yi})$) the
internal configuration. For $\alpha_i\sim0$ particles $j,k$ are
close to each other, and for $\alpha_i\sim\pi/2$ particle $i$ lies
between $j$ and $k$, see fig.~\ref{fig:3B-coordinates-2}.

The volume elements of the Jacobi and hyper-spherical coordinates are related by
\begin{equation}
  \label{eq:hypersp-vol-element}
  \begin{split}
    &\ud {\bm x_i} \ud {\bm y_i} = \rho^5 \ud \rho \ud \Omega, \\ 
    &\ud \Omega \equiv \sin^2 \alpha_i \cos^2 \alpha_i \ud \Omega_{xi}
    \ud \Omega_{yi}.
  \end{split}
\end{equation}
The kinetic energy operator in hyper-spherical coordinates is given by
\begin{equation}
  \label{eq:hypersp-kinetic-operator}
  T=\frac{\hbar^2}{2m}\left( 
    -\frac{\partial^2}{\partial\rho^2}
    -\frac{5}{\rho}\frac{\partial}{\partial\rho}  
    +\frac{\hat\Lambda^2}{\rho^2}
  \right),
\end{equation}
where the square of the grand angular momentum operator $\hat\Lambda$ is
\begin{equation}
  \label{eq:grand-angular-operator}
  \hat\Lambda^2 =-\frac{1}{\sin 2\alpha_i}
  \frac{\partial^2}{\partial\alpha_i^2}\sin(2\alpha_i)
  -4
  +\frac{\hat l_{xi}^2}{\sin^2\alpha_i}
  +\frac{\hat l_{yi}^2}{\cos^2\alpha_i}.
\end{equation}
Here $l_{xi}^2$ and $l_{yi}^2$ are the angular momentum operators
corresponding to $\bm x_i$ and $\bm y_i$. We only consider zero total
angular momentum, i.e. $l_{xi}=l_{yi}=0$.
Two different sets of Jacobi coordinates $i$ and $j$ are connected
through the so-called kinematic rotation \cite{nielsen01}
\begin{equation}
  \label{eq:kinematic-rotation}
  \begin{pmatrix}
    \bm x_j\\
    \bm y_j
  \end{pmatrix}
  =
  \begin{pmatrix}
    -\cos\gamma_{ij}& \sin\gamma_{ij}\\
    -\sin\gamma_{ij}&-\cos\gamma_{ij}
  \end{pmatrix}
  \begin{pmatrix}
    \bm x_i\\
    \bm y_i
  \end{pmatrix}.
\end{equation}
For three identical particles the ``rotation angle'' is $\gamma_{ij}=\sigma\{i,j,k\}\pi/3$, where
$\sigma\{i,j,k\}$ is the sign of the permutation $\{i,j,k\}$.
By calculating $\bm x_i^2$ from eq.~\eqref{eq:kinematic-rotation} and
using eq.~\eqref{eq:hypersp-coor}, the different hyper-angles are then
related by
\begin{equation}
  \label{eq:hyper-angles-relation}
  \begin{split}
    \sin^2\alpha_j
  &=\cos^2\gamma_{ij}\sin^2\alpha_i
  +\sin^2\gamma_{ij}\cos^2\alpha_i\\
  &-2 \sin\gamma_{ij} \cos\gamma_{ij} \sin\alpha_i \cos\alpha_i \cos \theta_i,
  \end{split}
\end{equation}
where $\theta_i\in[0,2\pi]$ is the angle between $\bm x_i$ and $\bm
y_i$. Using $|\cos \theta_i|\le1$ one can show that for a fixed
hyper-angle $\alpha_i$ in the coordinate set $i$, the hyper-angle in
set $j$ is restricted by
\begin{equation}
  \label{eq:hyper-angles-interval}
  \left|\frac{\pi}{3}-\alpha_i\right|\le\alpha_j\leq\frac{\pi}{2}-\left|\frac{\pi}{6}-\alpha_i\right|.
\end{equation}

\subsection{Hyper-Spherical Adiabatic Expansion}
In the hyper-spherical adiabatic expansion one first solve the angular
 part ($\Omega$) of the Schr\"odinger equation for fixed $\rho$,
\begin{equation}
  \label{eq:hyperang-eigen-eq}
  \left(\hat\Lambda^2+\frac{2m}{\hbar^2}\rho^2\sum_{i=1}^3
    V(\sqrt{2}\rho\sin\alpha_i)\right)\Phi_n(\rho,\Omega)
  =\lambda_n\Phi_n(\rho,\Omega).
\end{equation}
This gives a complete set of eigenfunctions $\Phi_n(\rho,\Omega)$ and
corresponding eigenvalues $\lambda_n(\rho)$ as functions of
$\rho$. We expand the total wave function as\footnote{The phase-factor
  $\rho^{-5/2}$ is simply conventional to simplify equations below.}
\begin{equation}
  \label{eq:total-wf-expansion}
  \Psi=\sum_n\rho^{-5/2}f_n(\rho)\Phi_n(\rho,\Omega).
\end{equation}
Inserting this into the full Schr\"odinger equation gives a coupled
set of hyper-radial equations for the coefficients $f_n(\rho)$,
\begin{equation}
  \label{eq:hyper-radial-equations}
  \left(-\frac{\partial^2}{\partial\rho^2}
    +\frac{2m}{\hbar^2}(V_{\textup{eff},n}(\rho)-E)\right)f_n(\rho)
  =\sum_{n'\ne n}(2P_{nn'}\frac{\partial}{\partial \rho}+Q_{nn'})f_{n'}(\rho),
\end{equation}
where the effective hyper-radial potentials are
\begin{equation}
  \label{eq:Veff-definition}
  V_{\textup{eff},n}(\rho)=\frac{\hbar^2}{2m}
  \frac{\lambda_n(\rho)+15/4}{\rho^2}-Q_{nn},
\end{equation}
and the non-adiabatic couplings are
\begin{equation}
  \label{eq:PQ-definition}
  \begin{split}
    P_{nn'}(\rho)&
    =\left\langle\Phi_n(\rho,\Omega)\left|\frac{\partial}{\partial\rho}
      \right|\Phi_{n'}(\rho,\Omega)\right\rangle_\Omega,\\
    Q_{nn'}(\rho)&
    =\left\langle\Phi_n(\rho,\Omega)\left|\frac{\partial^2}{\partial\rho^2}
      \right|\Phi_{n'}(\rho,\Omega)\right\rangle_\Omega.
  \end{split}
\end{equation}
The brackets denote integration over $\Omega$. The angular wave-functions
are normalized to unity for fixed $\rho$, i.e.
$\langle\Phi(\rho,\Omega)|\Phi(\rho,\Omega)\rangle_\Omega=1$. The
identity $P_{nn}=0$ also holds.  In the strict adiabatic limit where
all the off-diagonal coupling terms $P_{nn'}$ and $Q_{nn'}$ vanish, the
hyper-radial equations, eq.~\eqref{eq:hyper-radial-equations}, decouple.

\subsection{Hyper-Angular Faddeev Decomposition}
\begin{figure}[htbp]
  \centering
  \includegraphics[scale=0.8]{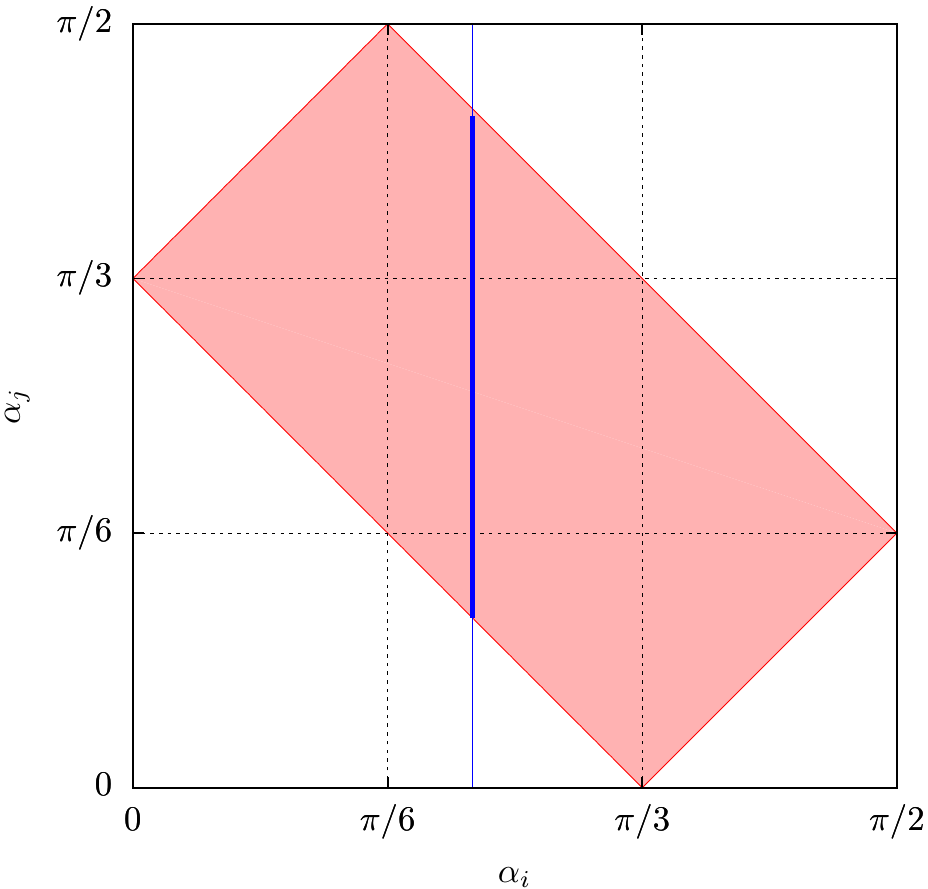}
  \caption{Integration limits for the rotation operator
    eq.~\eqref{eq:rot-operator}. For a given $\alpha_i$ the shaded
    region determines the interval
    of integration for $\alpha_j$. A specific value of $\alpha_i$ is
    indicated by the thick blue line.}
  \label{fig:rotate-limits}
\end{figure}

Let us first rewrite the angular equation
eq.~\eqref{eq:hyperang-eigen-eq} as
\begin{equation}
  \label{eq:angular-eq2}
  \left(
    -\frac{\partial^2}{\partial\alpha_i^2}
    -\nu^2
    +\sum_{i=1}^3 U(\alpha_i)
  \right)(\sin(2\alpha_i)\Phi)=0,
\end{equation}
where the reduced interaction is $U(\alpha_i)=\frac{2m}{\hbar^2}\rho^2
V(\sqrt{2}\rho\sin\alpha_i)$, the eigenvalue is $\nu^2=\lambda+4$,
and we have omitted the subscripts $n$.
We now split the hyper-angular wave-function into three identical
Faddeev components,
\begin{equation}
  \label{eq:faddeev-decom}
  \Phi(\alpha_i)=\frac{\psi_i+\psi_j+\psi_k}{\sin(2\alpha_i)},
\end{equation}
which fulfill the three Faddeev equations
\begin{equation}
  \label{eq:faddeev-eq}
  \left( -\frac{\partial^2}{\partial\alpha_i^2}
    -\nu^2 \right)\psi_i
    + U(\alpha_i)(\psi_i+\psi_j+\psi_k)=0.
\end{equation}
Adding these equations leads to eq.~\eqref{eq:angular-eq2}.
The components $\psi_i,\psi_j,\psi_k$ have the same functional form, but are expressed in
different Jacobi sets. If one function, say $\psi=\psi_i$, is given we
obtain the ``rotated'' component $\rot[\psi]=\psi_j$ by expressing
$\psi$ in coordinates $j$ and projecting onto $s$-waves. Since we have
identical particles the two rotated components are identical and we
may write
\begin{equation}
  \label{eq:faddeev-decom2}
  \Phi(\alpha_i)=\frac{\psi+2\rot[\psi]}{\sin(2\alpha_i)}.
\end{equation}
Specifically, the rotation operator is given by
\begin{equation}
  \begin{split}
      \frac{\rot[\psi](\alpha_i)}{\sin(2\alpha_i)}
      &=\iint\frac{1}{(4\pi)^2}\frac{\psi(\alpha_i)}{\sin(2\alpha_j)}\ud\Omega_{xi}\ud\Omega_{yi}\\
      &=\frac{1}{2}\int_0^{2\pi}\frac{\psi(\alpha_j)}{\sin(2\alpha_j)}\sin\theta_i\ud\theta_i,
  \end{split}
\end{equation}
where $\alpha_j$ depends on $\theta_i$ via
eq.~\eqref{eq:hyper-angles-relation} (with fixed $\alpha_i$). From
eq.~\eqref{eq:hyper-angles-relation} we also find
$\sin(2\alpha_i)\sin\theta_i\ud\theta_i=(4/\sqrt{3})\sin(2\alpha_j)\ud\alpha_j$. The
rotation operator can then be written as
\begin{equation}
  \label{eq:rot-operator}
  \rot[\psi](\alpha_i)=\frac{2}{\sqrt{3}}
  \int_{|\frac{\pi}{3}-\alpha_i|}^{\frac{\pi}{2}-|\frac{\pi}{6}-\alpha_i|}
  \psi(\alpha_j)\ud\alpha_j.
\end{equation}
The new integration limits were obtained from
eq.~\eqref{eq:hyper-angles-interval}. The limits are also shown in a
useful form in fig.~\ref{fig:rotate-limits}.

\subsection{Zero-Range Solution}

We now take the zero-range limit, i.e. $U=0$ in
eq.~\eqref{eq:faddeev-eq}. The solutions with boundary condition
$\psi(\pi/2)=0$ are
\begin{equation}
  \label{eq:zr-faddeev-comp}
  \psi(\alpha_i)=\mathcal N(\rho)\sin(\nu(\alpha_i-\frac{\pi}{2})),
\end{equation}
with normalization $\mathcal N(\rho)$ depending on the eigenvalue
$\nu(\rho)$. It is straightforward to rotate this solution with
eq.~\eqref{eq:rot-operator} and we obtain
\begin{equation}
  \label{eq:zr-rot-comp}
  \rot[\psi](\alpha_i) = \mathcal N\frac{4}{\sqrt{3}\nu} \left\{
 \begin{array}{ll}
   \sin(-\nu\alpha_i)\sin(\nu\frac{\pi}{6}) 
   &\text{,  } 0\le\alpha_i\le\frac{\pi}{3}\\
   \sin(\nu(\alpha_i-\frac{\pi}{2}))\sin(\nu\frac{\pi}{3}) 
   &\text{,  } \frac{\pi}{3}\le\alpha_i\le\frac{\pi}{2}\;.
 \end{array} \right.
\end{equation}
Note that this function is zero at $\alpha_i=0$ and $\pi/2$.  It
is continuous but not differentiable at $\alpha_i=\pi/3$, which comes
from the fact that the zero-range model has $\psi(0)\ne 0$ (in
contrast to any finite-range potential). However, the total
wave-function $\Phi$ is smooth everywhere.

We also need to impose a boundary condition at $\alpha_i=0$ to
determine the eigenvalue $\nu$. In the two-body case the asymptotic
radial wave-function is given by $u(r)=rR(r)\propto\sin(kr+\delta(k))$.  The
zero-range limit can be formulated as the free problem with boundary condition
\begin{equation}
  \label{eq:zr-boundary-2b}
  \left.\frac{\partial \ln (r R(r))}{\partial r}\right|_{r=0}
  =\left.k\cot\delta(k)\right|_{k=0}=-\frac{1}{a}\;.
\end{equation}
For the three-body problem we apply the same boundary condition in
each of the three Jacobi sets (for fixed $\rho$),
\begin{equation}
  \label{eq:zr-boundary-3b}
    \left.\frac{\partial \ln (r_{jk} \Phi)}{\partial r_{jk}}\right|_{r_{jk}=0}=-\frac{1}{a}\;.
\end{equation}
In the limit $r_{jk}\to0$ we have to lowest order
$r_{jk}=\sqrt{2}x_i\simeq\sqrt{2}\rho\alpha_i$. Using
eq.~\eqref{eq:faddeev-decom2} the logarithmic derivative then becomes
\begin{equation}
  \label{eq:zr-boundary-3b-2}
  \left.\frac{\partial \ln (r_{jk} \Phi)}{\partial r_{jk}}\right|_{r_{jk}=0}\simeq
  \left.\frac{1}{\sqrt{2}\rho}\frac{\partial \ln (\psi+2\rot[\psi])}{\partial \alpha_i}\right|_{\alpha_i=0}
\end{equation}
to lowest order in $\alpha_i$.  By inserting
eq.~\eqref{eq:zr-faddeev-comp} and eq.~\eqref{eq:zr-rot-comp} we
obtain
\begin{equation}
  \label{eq:zr-efimov-eq}
  \frac{1}{\sqrt{2}\rho} \ \frac{ -\nu\cos(\nu\frac{\pi}{2})+ \frac{8}{\sqrt{3}}\sin(\nu\frac{\pi}{6}) }
  {\sin(\nu\frac{\pi}{2})} 
  = -\frac{1}{a}\;,
\end{equation}
which determines the angular eigenvalues $\nu_n(\rho)$. This equation
was first obtained by V. Efimov \cite{efimov71}.

\subsection{The Efimov Effect, $|a|=\infty$}
\label{sec:efimoveffect-inf-a}
Let us consider the resonant limit $|a|=\infty$. Then
eq.~\eqref{eq:zr-efimov-eq} has a single imaginary
solution\footnote{The equation also have real (continuum) solutions
  $\nu_m=2m$, where $m$ is a positive integer, except $m=2$ which is
  spurious.}  $\nu_0\simeq 1.00624 i$ determined by
\begin{equation}
  -\nu\cos(\nu\frac{\pi}{2})+ \frac{8}{\sqrt{3}}\sin(\nu\frac{\pi}{6})=0.
\end{equation}
We will refer to $\nu_0$ or the related effective potential
$1/\rho^2$ as the Efimov solution.\footnote{Although it should also be
  attributed to G. S. Danilov \cite{danilov61}, see \cite{efimov71}.}
The related effective hyper-radial potential becomes
\begin{equation}
  \label{eq:efimov-potential}
  V_{\textup{eff},0}(\rho)=-\frac{\hbar^2}{2m}\frac{\xi^2+1/4}{\rho^2},
\end{equation}
where $\xi=|\nu_0|=1.60024$.  Since the angular solutions only depend
on $\rho$ via the constant eigenvalues $\nu_n$, all non-adiabatic
couplings $P_{n,n'}$, $Q_{n,n'}$ vanish and the adiabatic
approximation becomes exact in this limit.
The lowest hyper-radial equation becomes
\begin{equation}
  \label{eq:efimov-radial}
  \left(-\frac{\partial^2}{\partial\rho^2}
    -\frac{\xi^2+1/4}{\rho^2} \right)f_0(\rho)=0,
\end{equation}
where $\kappa=\sqrt{2m(-E)}/\hbar$.
Scaling the variables as $\tilde f_0=f_0/\sqrt{\rho}$, $\tilde\rho=\kappa\rho$
eq.~\eqref{eq:efimov-radial} turns into a Bessel equation. The
solution is the modified Bessel function of the second kind (of
imaginary order) $\tilde f_0(\tilde\rho)=K_{i\xi}(\tilde \rho)$
\cite[sec. 9.6.1]{abramowitz95}. Thus the hyper-radial solutions with
binding wave number $\kappa^{(n)}$ become
\begin{equation}
  \label{eq:radial-solutions}
  f_0^{(n)}(\rho)=\sqrt{\rho}K_{i\xi}(\kappa^{(n)}\rho)
  \simeq\sqrt{\rho}|\Gamma(i\xi)|\sin\left(\xi\ln(\kappa^{(n)}\rho)+\theta\right).
\end{equation}
The last approximation is for the low-distance limit $\kappa\rho\ll 1$
where we used several identities and expansions for the $\Gamma$ function. The
phase $\theta$ is given by
$\theta=\textrm{arg}\{\Gamma(-i\xi)2^{-i\xi}\}+\pi/2\simeq 0.874\pi$, and
$|\Gamma(i\xi)|\simeq 0.5148$.

We see that the solutions have infinitely many nodes at short
distances, meaning that there are infinitely many low-lying
states. This is the well known Thomas effect in the three-body
problem with zero-range interactions and fixed two-body binding energy
(or scattering length $a$) \cite{thomas35,nielsen01}.

\begin{figure}[htbp]
  \centering
  \includegraphics{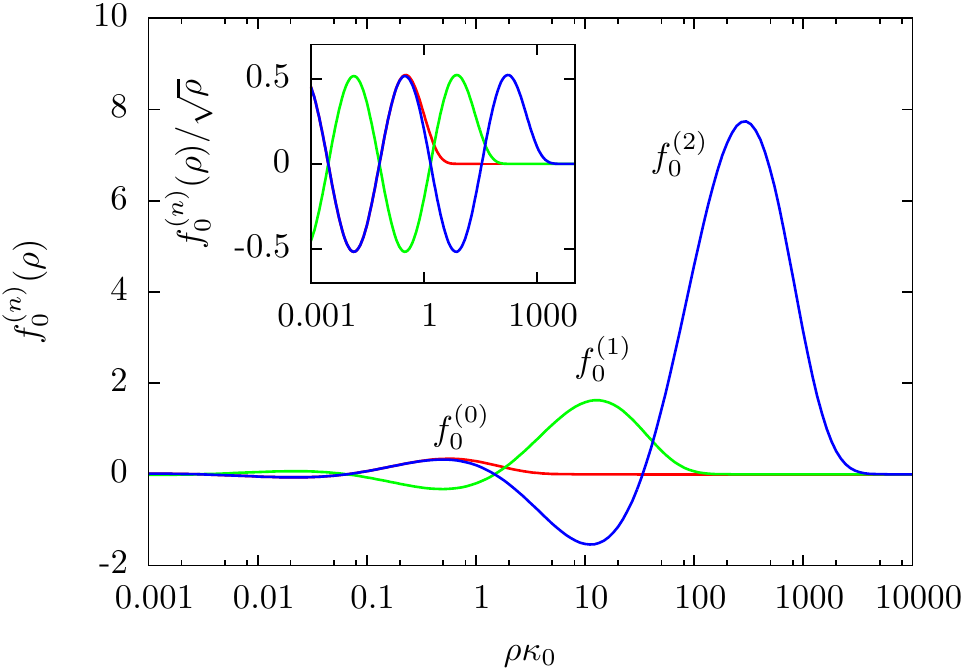}
  \caption{Hyper-radial wave-functions,
    eq.~\eqref{eq:radial-solutions}, for the Efimov states in the
    effective $1/\rho^2$ potential eq.~\eqref{eq:efimov-potential}.
    The scale $\kappa_0$ is the binding wave number for state $n=0$.
    The solutions have infinitely many nodes as $\rho\to0$, see inset. }
  \label{fig:besselKn}
\end{figure}

The Thomas collapse is unphysical since the real short-distance details
must be taken into account. Thus, we need to introduce a
regularization, e.g. by setting $f_0(\rho_0)=0$ at some short
arbitrary distance $\rho_0$. Using eq.~\eqref{eq:radial-solutions}
this leads to $\kappa^{(n)}=\exp(-n\pi-\theta)/\rho_0$, giving the
characteristic scaling between the energies,
\begin{equation}
  \label{eq:efimov-scaling-E}
  \frac{\kappa^{(n)}}{\kappa^{(0)}}=e^{-\pi n/\xi}\simeq 22.7^{-n},
  \quad\textrm{or}\quad
  \frac{E^{(n)}}{E^{(0)}}=e^{-2\pi n/\xi}\simeq 515.0^{-n}.
\end{equation}
This is the essence of the Efimov effect, where infinitely many
loosely bound states (Efimov trimers) accumulate a zero energy.  Three
hyper-radial wave-functions are shown in fig.~\ref{fig:besselKn}.  The
root-mean-square hyper-radii of the Efimov states scale accordingly as
\begin{equation}
  \label{eq:efimov-scaling-rho}
  \frac{\langle\rho^2\rangle_n^{1/2}}{\langle\rho^2\rangle_0^{1/2}}=e^{\pi n/\xi}\simeq 22.7^n.
\end{equation}

\subsection{Universal Scaling, $|a|<\infty$}
Let us finally discuss the case of finite but large scattering length.
The constant solution $\nu_0$ in eq.~\eqref{eq:zr-efimov-eq} still
holds, but now only for $\rho\lesssim |a|$, thus the effective
$1/\rho^2$ potential will be modified at $\rho\gtrsim |a|$. This means
that all the Efimov states with hyper-radii larger than $|a|$ (see e.g
fig.~\ref{fig:besselKn}) will be affected. However, Efimov states with
hyper-radii smaller than $|a|$ only have exponential tails in the
affected region and are therefore essentially unchanged. Note also,
that the regularization can be considered fixed: The infinite
scattering length is tuned by adjusting the two-body energy
infinitesimally around zero leaving the underlying finite-range
potentials fixed.

\begin{figure}[tbp]
  \centering
  \includegraphics{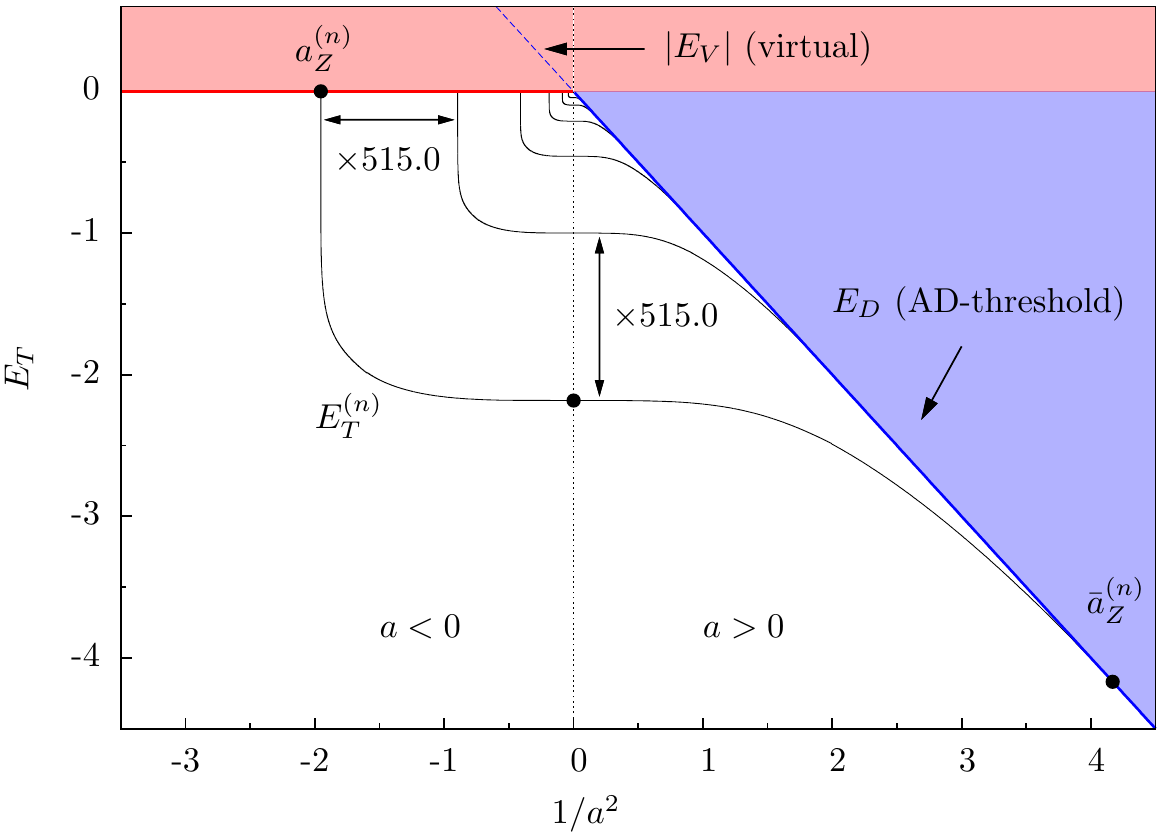}
  \caption{Discrete universal scaling of Efimov states. Trimer energy
    levels, $E^{(n)}_T$ are shown as functions of scattering length,
    $a$.  Points indicate the critical scattering lengths where the
    trimer crosses the trimer and atom-dimer thresholds.  Both axes
    were scaled by a power of ${1/8}$ to reduce the large scaling
    factor $515.0$ to $2.18$, i.e. the axes actually show
    $[\hbar^2/(ma^2E_\infty)]^{1/8}$ and $[E_T/E_\infty]^{1/8}$, with
    $E_\infty$ being an arbitrary regularization scale modulo 515.0
    (the energy of a trimer at $a=\infty$).}
  \label{fig:efimovplot}
\end{figure}

The resulting behavior of the Efimov trimer energies, $E_T^{(n)}$ ,
are shown in fig.~\ref{fig:efimovplot} as function of inverse squared
scattering length $1/a^2$. On the $a>0$ side the two-body system has a
shallow bound state (dimer) with energy $E_D=-\hbar^2/ma^2$,
eq.~\eqref{eq:ED-scat} (indicated by the blue diagonal line). For the
three-body case this defines the atom-dimer fragmentation threshold.
For $a<0$ the two-body system has a shallow virtual state $E_V$ and
the three-body system has a three-body fragmentation threshold at zero
energy (indicated by the red horizontal line). At $|a|=\infty$ the
infinite sequence of Efimov trimers is shown.

When $a$ is decreased from infinity to $a<0$ the effective hyper-radial
potential moves up at large distances $\rho\gtrsim|a|$, and the most
loosely bound Efimov trimers are pushed into the continuum. For
$a>0$ the effective potential decreases, now converging to the dimer
binding energy $V_\textup{eff}\to E_D<0$ for larger hyper-radii. Each Efimov
trimer moves down correspondingly, until it disappear into the
atom-dimer continuum.

Because of the scale-invariant $1/\rho^2$ potential all the Efimov
trimer energies scale the same way, and only a single universal
function needs to be calculated.  The specific form can be found by
solving the Schr\"odinger or Faddeev equations numerically for
short-ranged potentials (as done in chapter~\ref{chap:efimov}), or by
using effective field theory (EFT)
\cite{braaten03,braaten03a,braaten06}. In the exact universal limit
the energy $E_T^{(n)}$ of the $n$'th trimer state can be parametrized
as
\begin{equation}
  \label{eq:efimov_eq}
  E_T^{(n)}-\frac{\hbar^2}{ma^2}=e^{\Delta(\varphi)/\xi}e^{-2\pi n/\xi}E_\infty,
\end{equation}
where $\xi=1.00624$, $\Delta(\varphi)$ is a universal function
\cite{braaten06} of the angle $\varphi\in[-\pi;-\pi/4]$ given by
$\tan^2\varphi=-E_T^{(n)}m a^2/\hbar^2$ (i.e. $\varphi$ is the polar
angle in the coordinates of fig.~\ref{fig:efimovplot}).  $E_\infty$ is
the regularization scale (here the energy of state $n=0$ at
$a=\infty$) which cannot be determined within the zero-range theory.
It has to be supplied from phenomenological information.

The two critical scattering lengths $a^{(n)}_Z$ and $\bar{a}^{(n)}_Z$
where the Efimov trimer $E_T^{(n)}$ disappears into the three-body
and atom-dimer continuum, see fig.~\ref{fig:efimovplot}, are given
explicitly in the zero-range limit by \cite{braaten06,gogolin08}
\begin{equation}
  \label{eq:a_crit}
  a^{(n)}_{Z}=-1.50763/\kappa^{(n)},\quad  
  \bar{a}^{(n)}_{Z}=0.0707645/\kappa^{(n)},
\end{equation}
where $\kappa^{(n)}=\sqrt{-mE_T^{(n)}(a=\infty)}/\hbar$ is the wave
number for the $n$'th trimer energy at infinite scattering length.

\subsubsection{Three-Body Recombination in Experiments}

The critical scattering lengths $a^{(n)}_{Z}$ and $\bar{a}^{(n)}_{Z}$
can be observed via maxima and minima in the atom loss rate of cold
gases \cite{kraemer06}.  In the ultra-cold limit the loss due to
two-body collisions $B+B\to B_2$ are suppressed by energy-momentum
conservation.  Therefore three-body collisions $B+B+B\to B+B_2$ can be
considered as the dominant mechanism for the total loss rate. The
general behavior for the three-body recombination rate is $|a|^4$ for
large $a$ \cite{nielsen99b,esry99a,braaten06}. On top of this
background additional resonance or interference effects effects from
the Efimov trimers should be seen.

The first case where the Efimov trimer disappear into the three-body
continuum ($a<0$) leads to a resonant peak in the recombination
rate. In numerical hyper-spherical descriptions this is explained by a
tunneling into a small $\rho$-region and subsequent decay into
strongly bound molecular states \cite{esry06}.  The second case where
the trimer crosses the atom-dimer threshold ($a>0$) leads to a minima
in the recombination rate.\footnote{These minima are actually located
  between the positions where the Efimov trimer hit the threshold.}
This can be explained as two interfering pathways from the incoming
hyper-spherical channel to the weakly bound atom-dimer channel
\cite{esry06}.

Until now, only a few experiments measuring such effect have been
performed, namely in $^{133}$Cs gases \cite{kraemer06,knopp09} and
also in a three-component $^6$Li Fermi gas \cite{wenz09}. The
experiments will be conclusive when the scaling factor $22.7$ is
observed (or a correspondingly smaller factor for unequal mass
systems).

\section{Mean-Field Condensates}

The Gross-Pitaevskii (GP) equation is based on a mean-field ansatz with
the condensate wave-function (or order parameter) $\Psi(\bm r)$
together with the zero-range contact interaction
\begin{equation}
  \label{eq:zero-range-potential}
  V_{ZR}(r)=U_0\delta(r),\quad U_0=\frac{4\pi\hbar^2a}{m}.
\end{equation}
The coupling constant $U_0$ is chosen such that the correct energy
shift is reproduced, see chapter~\ref{chap:mean-field}.  This is
equivalent to a Born approximation where $a_{born}$ is replaced by the
physical scattering length $a$.  This leads to the GP energy
functional
\begin{equation}
  E(\Psi)=\int \ud{\bm r} 
  \left(\frac{\hbar^2}{2m}|\nabla \Psi|^2+V_{ext}(\bm r) |\Psi|^2
    +\frac{U_0}{2}|\Psi|^4 \right),
\end{equation}
which includes kinetic energy, the external trapping potential
$V_{ext}$, and the interaction term from
eq.~\eqref{eq:zero-range-potential}. A variation of this functional
with the constraint of fixed particle number $N$ leads to the GP
equation
\begin{equation}
  \left(-\frac{\hbar^2}{2m}\nabla^2+V_{ext}(\bm r)+U_0|\Psi(\bm
    r)|-\mu \right)\Psi(\bm r)=0,
\end{equation}
where $\mu$ is the chemical potential.

The GP equation (or variations and extensions of it) has been able to
describe a wide variety of BEC phenomena, e.g. the spatial condensate
profile in different traps, solitons in homogeneous condensates,
rotating BECs with vortex profiles, stability of low-energy modes etc.
\cite{pethick02,pitaevskii03}.  A specific useful approach is the
Thomas-Fermi (TF) approximation where the kinetic energy term is
neglected and analytical solutions are possible. The approximation
holds for repulsive gases ($a>0$) with many particles (e.g. $Na/b_t\gg
1$ in the harmonic trap of size $b_t=\sqrt{\hbar/(m\omega)}$). See
chapter~\ref{chap:tf} for elaborate discussions.

From the perspective that the contact interaction
eq.~\eqref{eq:zero-range-potential} should produce correct
energy shifts, it is only the lowest order approximation. In
chapter~\ref{chap:mean-field} we derive the higher-order interaction
term proportional to $\delta(\bm r)\nabla^2+\nabla^2\delta(\bm r)$,
and the related modified GP equation with an extra term. The purpose
of chapter~\ref{chap:mean-field} is to investigate possible effects of
such higher-order terms via variational and numerical solutions.
Chapter~\ref{chap:mean-field} approaches the same question
analytically in the TF limit.

\section{Stochastic Variational Method}

The stochastic variational method (SVM) \cite{suzuki98,sorensen05} is
a numerical minimization technique to find approximate solutions of
few- and many-body problems.  The method is based on two basic
ingredients, i) stochastic random sampling and ii) deterministic
linear variation.

One of the major advantages of the SVM is that one can calculate a
large number of bound states, obtaining both the spectrum and
wave-functions directly. Another feature is that one can control the
amount of inter-particle correlations allowed in the calculated states
-- depending on the application one may e.g. look for highly
correlated few-body states or weakly correlated many-body states. The
method is also numerically efficient and it can be parallelized with
close to linear scalability.

\subsection{The Linear Variational Principle}
\label{sec:linear-variation}

Let us consider the stationary many-body Schr{\"o}dinger equation
$\hat H \Psi = E\Psi$ and denote the eigenfunctions and eigenvalues by
$\Psi_n$ and $E_n$, respectively.
The well-known Rayleigh-Ritz variational principle states that the
variational energy $\mathcal E$ evaluated with an arbitrary
trial-function $\Psi$ gives an upper bound to the exact ground state
energy of the Hamiltonian $\hat H$, i.e.
\begin{equation}
  \label{eq:ray-ritz}
  \mathcal E= 
  \frac{\langle \Psi|\hat H|\Psi\rangle}{\langle \Psi|\Psi\rangle} \ge
  E_1.
\end{equation}
The approach below extends this principle to any number of excited
states.

To find approximate variational solutions to the full Schr{\"o}dinger
equation we first restrict the problem to a smaller space spanned by a
(possibly over-complete) set of $K$ basis functions $\psi_k$. Let us
first express the wave-function in this space, i.e.
\begin{equation}
  \label{eq:wf-expansion}
  \Psi=\sum_{k=1}^K c_k \psi_k,
\end{equation}
where $c_k$ are expansion coefficients. These coefficients determine
$\Psi$ completely within the space $\{\psi_k\}$, although the
expansion may not be unique. By inserting eq.~\eqref{eq:wf-expansion}
in the Schr{\"o}dinger equation and projecting onto the $j$'th basis
state $\psi_j$ we obtain
\begin{equation}
  \label{eq:generalized-eval-eq}
  \sum_{k=1}^K H_{jk}c_k =
  \mathcal E \sum_{k=1}^K S_{jk} c_k.
\end{equation}
Here the Hamiltonian matrix elements and the overlap matrix in the
basis $\{\psi_k\}$ are given by
\begin{equation}
  \label{eq:H-S-matrixrep}
  H_{jk}=\langle \psi_j|\hat H|\psi_k\rangle,\quad
  S_{jk}=\langle \psi_j|\psi_k\rangle,
\end{equation}
where the brackets mean integration over all coordinates in $\psi_k$.
In terms of matrix multiplication, eq.~\eqref{eq:H-S-matrixrep} reads
$\bm H \bm c =\mathcal E \bm S \bm c$, where ${\bm
  c}=(c_1,\dots,c_K)^{T}$ and $\bm H,\bm S$ have the matrix elements
$H_{jk},S_{jk}$.  This is a generalized eigenvalue equation of size $K$
with $K$ real eigenvalues $\mathcal E_1 \le \dots \le \mathcal E_K$
and corresponding eigenfunctions $\bm c_1,\dots,\bm c_K$. 

The above procedure shows how to find approximate solutions to the
full Schr{\"o}dinger equation within a smaller subspace. Let us now
specify what we mean by the term ``approximate''.
We take the space $\{\psi_k\}$ to be fixed, and treat the linear
coefficients $c_k$ in eq.~\eqref{eq:wf-expansion} as variational
parameters. The variational energy, eq.~\eqref{eq:ray-ritz}, expressed
in the basis $\{\psi_k\}$ then becomes
\begin{equation}
  \label{eq:ray-ritz-matrixrep}
  \mathcal E= 
  \frac{\bm c^\dagger \bm H \bm c}{\bm c^\dagger \bm S \bm c},
\end{equation}
where $\bm c^\dagger=(c_1^*,\dots,c_K^*)$. The stationary solutions
within our subspace correspond to $\partial\mathcal E/\partial
c_i=\partial\mathcal E/\partial c_i^*=0$ for all $i$. We multiply
eq.~\eqref{eq:ray-ritz-matrixrep} by $\bm c^\dagger \bm S \bm c$ and
take the partial derivative $\partial/\partial c_i^*$ (with fixed
$c_i$ and $c_j,c_j^*$, $j\ne i$). This leads directly to the
generalized eigenvalue equation
\begin{equation}
  \label{eq:generalized-eval-eq2}
  \bm H \bm c =\mathcal E \bm S \bm c.
\end{equation}
Thus, the energies $\mathcal E$ found by projecting the
Schr{\"o}dinger equation onto the fixed subspace $\{\psi_k\}$ are
actually the best possible energies as seen from the
variational perspective.

Furthermore, it can be shown rigorously \cite{suzuki98} that the
generalized (ordered) eigenvalues $\mathcal E_1 \le \dots \le \mathcal
E_K$ are strict upper bounds to the exact (ordered) eigenvalues
$E_1\le\dots\le E_K\le\dots$, namely
\begin{equation}
  \label{eq:energy-bounds}
  E_1\le \mathcal E_1,\ \cdots,\  E_K\le \mathcal E_K.
\end{equation}
It can also be shown \cite{suzuki98} that by adding another basis
state $\psi_{K+1}$, while keeping the first $\psi_1,\dots,\psi_K$
fixed, the upper bounds in eq.~\eqref{eq:energy-bounds} can only get
better.
This linear variational principle extends the Rayleigh-Ritz
variational principle to an arbitrary number of bound states.

\subsection{Basis States and Minimization Procedure}
\label{sec:svm-strategy}

To employ the variational principle above we must generate a set of
basis states $\psi_k$. These states should form a complete set,
or at least be able to describe the physical degrees of freedoms of
interest. For the approach to be numerically tractable one should also
be able to evaluate the matrix elements $H_{ij}$ and $S_{ij}$
analytically. Common choices \cite{suzuki98,sorensen05} are
exponential functions or Gaussians (see below). Each basis state
$\psi_k$ is then parametrized by one or more nonlinear parameters,
which we denote by the single symbol $\alpha^{(k)}$. 

There are many different minimization strategies for the SVM. The
simplest one is to randomly pick a large number of functions $\phi_k$
(i.e. nonlinear parameters $\{\alpha_k\}$) and solve
\eqref{eq:generalized-eval-eq2}. This approach has two minor
disadvantages, i) many of the random basis states $\phi_k$ may not be
necessary to describe the real eigenstates, and ii) solving the full
generalized eigenvalue equation of large dimension is numerically very
inefficient. Luckily both issues can be solved by adding or modifying
only a single state at a time.

Let us assume that the generalized eigenvalue equation has been solved
in the $K$-dimensional space spanned by $\psi_1,\dots,\psi_K$ with
resulting eigenvalues $\mathcal E_1,\dots,\mathcal E_K$ and
eigenstates $\phi_1,\dots,\phi_K$. We now expand the space by a single
basis state $\psi_{K+1}$. Using the Gram-Schmidt process we
construct $\phi_{K+1}$ from $\psi_{K+1}$ which is orthogonal to all
other $\phi$'s. In the basis $\phi_1,\dots,\phi_{K+1}$ the generalized
eigenvalue problem \eqref{eq:generalized-eval-eq2} is almost diagonal
and reduces to finding the roots of the simple function
\begin{equation}
  \label{eq:root-finding}
  D(\mathcal E')=\sum_{k=1}^K\frac{|h_k|^2}
  {\mathcal E_k-\mathcal E'}-\mathcal E'-h_{K+1},
\end{equation}
where $h_k=\langle\phi_k|\hat H|\phi_{K+1}\rangle$.  $D(\mathcal E')$
has exactly $K+1$ roots $\mathcal E_1,\dots,\mathcal E_{K+1}$ which
are the new eigenvalues. Thus, by adding only a single basis state at
a time, the generalized eigenvalue problem reduces to one-dimensional
root-finding.

\begin{figure}[htb]
  \centering
  \includegraphics[scale=0.9]{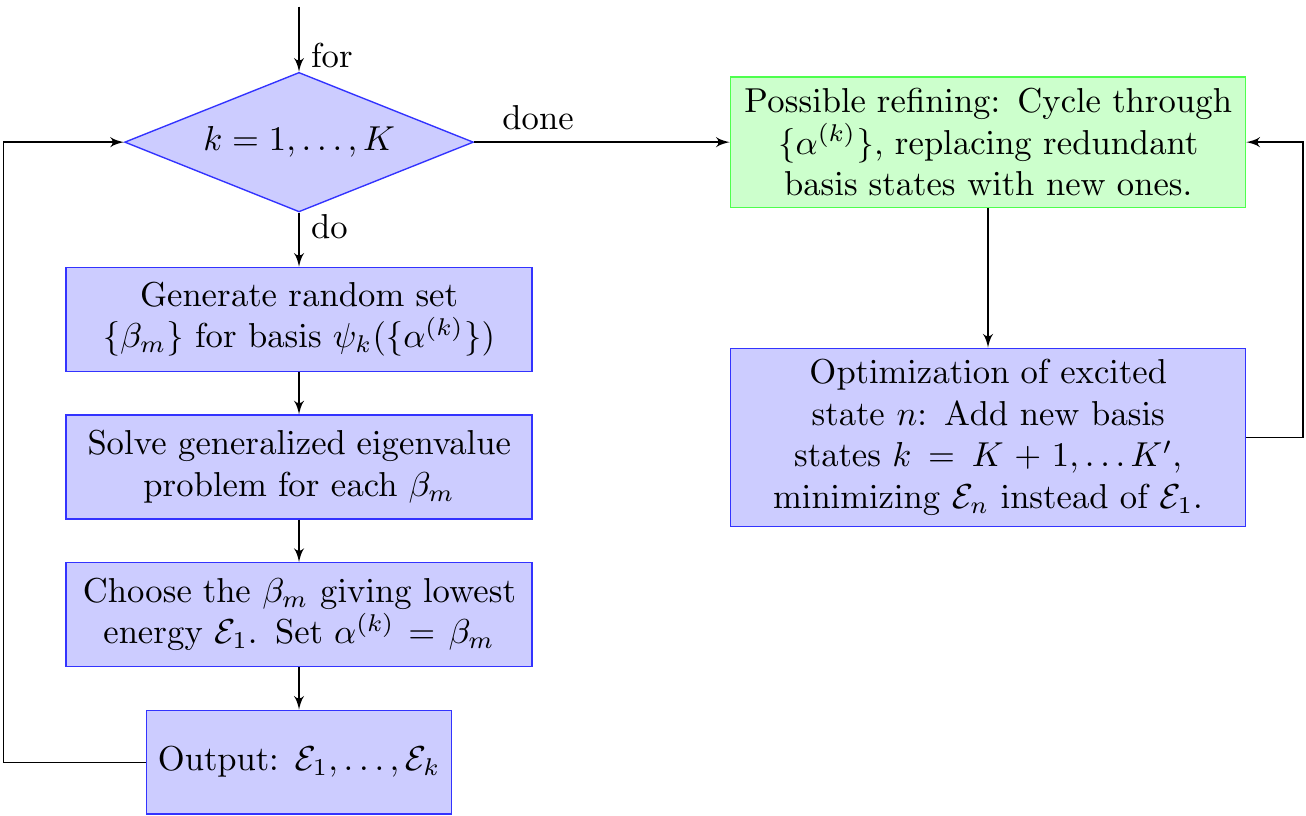}
  \caption{A simple example of the SVM minimization procedure. The
    ground state is found first, followed by a possible refining and
    minimization of excited states. Since this strategy only changes
    one basis state at a time, the generalized eigenvalue problem
    eq.~\eqref{eq:generalized-eval-eq2} reduces to a faster root-finding
    problem.}
  \label{fig:svm-flow}
\end{figure}

Let us give a short example on how the minimization procedure can be
done within the SVM, as shown schematically in
fig.~\ref{fig:svm-flow}. We start by adding one basis state at a time,
minimizing the ground state. For the $k$'th basis state $\psi_k$ we
stochastically generate a set $\{\beta_m\}$ of candidates for the
parameters $\{\alpha^{(k)}\}$. For each candidate we solve the
generalized eigenvalue equation eq.~\eqref{eq:generalized-eval-eq2}
from eq.~\eqref{eq:root-finding}.  We choose the $\beta_m$ giving the
lowest ground state $\mathcal E_1$. All other eigenvalues are also
guaranteed to be improved. After all $K$ eigenstates are found one may
refine the basis: Redundant basis states can be replaced by new ones
still optimizing $\mathcal E_1$.\footnote{This is at the expense of
  the excited states which almost certainly go up.} The entire
procedure can then be repeated for excited states.

This example is quite simple compared to the actual SVM implementation
we use \cite{sorensen05}, but it contains the essential details.  The
method can be tuned in many ways, e.g. continuous optimization of the
random sampling algorithm and selection of appropriate basis states
\cite{sorensen05}.

\subsection{Center-of-Mass and Symmetrization}
The specific many-body Hamiltonian used with the SVM is given by
\begin{equation} 
  \label{eq:hamiltonian}
  H=-\frac{\hbar^{2}}{2m}
  \sum_{i=1}^{N}\frac{\partial^2}{\partial\mathbf{r}_{i}^{2}}
  +\frac{m\omega^{2}}{2}\sum_{i=1}^{N}r_{i}^{2}\
  +\sum_{i<j}V(\left|\mathbf{r}_{i}-\mathbf{r}_{j}\right|),
\end{equation}
with a harmonic oscillator trap with angular frequency $\omega$ and
trap length $b_t=\sqrt{\hbar/(m\omega)}$, and two-body interactions
$V(r_{ij})$ chosen either as the zero-range or Gaussian model
potentials, eqs.~\eqref{eq:zero-range-potential}
and~\eqref{eq:gauss-interaction}.  Since the confining trap is
harmonic, the center-of-mass (cm) $\bm R=\sum_{i=1}^N\bm r_i/N$ can be separated out. We
will only consider the center-of-mass ground state motion, i.e.
\begin{equation}
  \label{eq:psi-cm}
  \Psi_\text{cm}(\bm R)=\exp(-\frac{NR^2}{2b_t^2}),
\end{equation}
with corresponding energy $E_{cm}=\frac{3}{2}\hbar\omega$.

Since we consider identical bosons the total wave-function must be
completely symmetric with respect to interchange of every pair. This
is accomplished with the symmetrization operator
\begin{equation}
  \label{eq:symmetrizer}
  \hat S=\frac{1}{\sqrt{N!}}\sum_\sigma \hat P_\sigma.
\end{equation}
Here $\hat P_\sigma$ is the permutation operator corresponding to the
permutation $\sigma$, and the sum is over all possible
permutations. The symmetrization does not affect the conclusions of
subsection~\ref{sec:linear-variation} and \ref{sec:svm-strategy} since
it is a linear operation.
The specific wave-function is then represented as
\begin{equation}
  \label{eq:psi-general}
  \Psi(\bm r_1,\dots,\bm r_N)=\Psi_\text{cm}(\bm R)\,\hat S
  \sum_{k=1}^K c_k\, \psi_k(\{\alpha_{ij}^{(k)}\},\{\bm r_{ij}\}).
\end{equation}
The basis functions, $\psi_k$, depend on all the internal coordinates
$\{r_{ij}\}$, and are parametrized by a, possibly large, number of
variables $\{\alpha_{ij}^{(k)}\}$. 

\subsection{Correlated Gaussians}
As the specific basis functions $\psi_k$ we use the so-called
explicitly correlated Gaussians,
\begin{equation}
  \label{eq:psi-full}
  \Psi_\text{full}=\Psi_\text{cm}\,\hat S\,\sum_{k=1}^K c_k
  \exp\left(-\frac{1}{2}\sum_{i<j}^N\alpha_{ij}^{(k)}r_{ij}^2\right).
\end{equation}
This will also be referred to as the full correlated basis.  The basis
is complete for zero total angular momentum states and it allows all
types of clustering in the system. This basis choice allows
analytical evaluation of matrix elements for the Gaussian interaction,
see \cite{suzuki98,sorensen05}.  However, because of the
symmetrization the computational complexity is of order
$\mathcal{O}(N!)$, and the method is only possible for relatively
small number of particles, say $N\lesssim 5$.

The zero-range potential, eq.~(\ref{eq:zero-range-potential}), requires
an uncorrelated wave-func\-tion which we choose in the form of the
linear combination of the hyper-radial basis-functions
\begin{equation}
  \label{eq:psi-rho}
  \Psi_\rho=\Psi_\text{cm}\,\sum_{k=1}^K c_k
  \exp\left(-\frac{1}{2} N\alpha^{(k)}\rho^2\right),
\end{equation}
where $\rho$ is the hyper-radius for $N$ particles,
\begin{equation}
  \label{eq:rho-N}
    \rho^{2} 
    = \frac{1}{N} \sum^N_{i<j} r_{ij}^2
    = \sum^{N}_{i=1}(\bm r_i-\bm R)^2
    = \sum^{N}_{i=1} r^2_i - N\bm R^2,
\end{equation}
generalizing the $N=3$ definition, eq.~\eqref{eq:rho-3}. This function
is totally symmetric and thus does not require the symmetrization
operator ${\hat S}$.
It is a specific choice for the parameters of eq.~\eqref{eq:psi-full}
where the different pairs of particles have the same parameters
$\alpha^{(k)}=\alpha_{ij}^{(k)}$, see eq.~\eqref{eq:rho-N}.  This is
reminiscent of a mean-field approximation, since all particles are
treated identically. Note that only one basis function $K=1$ with
$\alpha^{(1)}=1/Nb_t^2$ corresponds to the single-particle product of
the non-interacting case. The zero-range potential with the
hyper-radial variational wave-function eq.~\eqref{eq:psi-rho} provides
results similar to the Gross-Pitaevskii equation \cite{sorensen05a}.

For a typical system of trapped atoms even when the scattering length
is large the density of the system remains small, $n r_0^3 \ll 1$, and
one can assume that only binary collisions play a significant role in
the system dynamics.  In this approximation the variational
wave-function can be simplified by allowing only two-body correlations
in the basis-functions,
\begin{equation}
  \label{eq:psi-2B}
  \Psi_\text{2B}=\Psi_\text{cm}\,\hat S\,\sum_{k=1}^K c_k
  \exp\left(-\frac{1}{2} N\alpha^{(k)}\rho^2-\frac{1}{2}\beta^{(k)}r_{12}^2\right),
\end{equation}
where
$\alpha^{(k)}$ and $\beta^{(k)}$ are the nonlinear parameters.
This form is equivalent to the Faddeev-Yakubovski expansion
$\Psi=F(\rho)\sum_{i<j}\phi(\rho,r_{ij})$ used in
\cite{sogo05a,sogo05b} where each of the two-body amplitudes only
depend on the distance between two-particles.
The symmetrization of this function can be done analytically
\cite{sorensen05,sorensen05a} by collecting similar terms which
greatly simplifies the numerical calculations. The two-body correlated
approach is therefore very useful in dilute many-body systems.

In the trap of size $b_t$ with interactions of range $r_0$, the
non-linear parameters $\alpha_{ij}^{(k)}$ are typically optimized
stochastically in an interval from $1/b_t^2$ to $1/r_0^2$, or even
larger. This allows both short and long-range correlations in the
system.

%% file: efimov.tex
\section{Introduction}

Much effort has been devoted to extract universal features in many
branches of physics, because the applications become more transparent.
Several topics concerned with Efimov physics are of special interest
here: i) the Efimov effect \cite{efimov70,efimov90} where anomalous
three-body properties arise at the threshold of binding of two
particles, ii) halos and Borromean systems \cite{jensen04} where
required scaling properties are equivalent to large probabilities in
non-classical regions, iii) universality for few-body systems derived
from a zero-range interaction \cite{amorim97,frederico99,braaten06},
iv) general properties of Bose-Einstein condensates \cite{pethick02},
and v) their instability due to three-body recombination
\cite{kraemer06}.

In all these five cases the two-body $s$-wave scattering length $a$ is
the crucial and only characterizing parameter which is independent of
the details of the responsible potentials.  In fact the same
scattering length can be achieved by disparate potentials.  It is
highly desirable to assess the uncertainties in the results from the
leading order terms and extend to include correction terms.  An
extension inevitably needs more details which again should be
expressed in terms of model-independent parameters.  The obvious
choice is then to exploit the effective range expansion of the
two-body phase shifts where the leading term containing $a$ is given
by the zero-energy limit, and the second term is proportional to the
energy and the effective range $R_e$.  Inclusion of even higher-order
terms is usually not productive because it is either inefficient or
difficult, conceptually and practically.

The purpose of this chapter is to go beyond the scattering length
approximation for Efimov physics.  We express the corrections in terms
of the model-independent, low-energy scattering parameters, scattering
length and effective range.  For negative scattering lengths an
attempt in this direction was made by varying the form of the
potentials and computing the critical strengths for few-body binding
\cite{richard94,moszkowski00}. This was aimed at finding the Borromean
window where three particles are bound even though all the two-body
systems are unbound.  The results in \cite{moszkowski00} are however
not expressed in terms of model-independent parameters.  Other
attempts to include effective range terms within effective field
theory are reviewed in \cite{braaten06}.

Specifically, we calculate the low-lying energy spectrum for three
trapped identical bosons interacting via finite-range two-body
potentials.  The spectrum is compared to results from the zero-range
model. The thresholds for trimer binding and atom-dimer binding are
extracted and related to effective range corrections. Effective range
corrections to Efimov physics and Borromean binding are two aspects of
the same effect, and we connect these two regions qualitatively.  The
Borromean window becomes slightly narrower for substantial effective
ranges. The structure at the atom-dimer threshold is an atom far away
from the dimer and the major energy correction is due to the change of
dimer energy with effective range. This structure becomes less
pronounced when the effective range increases. Comparisons with recent
results from effective field theory are carried out.

\section{Procedure}

We consider $N=3$ identical bosons with mass $m$ and coordinates $\bm
r_i$ in a spherical harmonic trap with frequency $\omega$ and
corresponding trap length $b_t=\sqrt{\hbar/m\omega}$.
As the two-body interaction we mainly use an attractive Gaussian,
eq.~\eqref{eq:gauss-interaction}, of fixed range $r_0$, see
fig.~\ref{fig:gauss-int}.
We also include results for many other potential shapes based on input
from \cite{moszkowski00}. The strength $V_0$ is varied within the
interval where the potential either cannot support bound states
(scattering length $a<0$) or have only one bound state ($a>0$). The
positive effective range $R_e$ is then also a given function of
$V_0$. However, near the resonance, $|a|=\infty$, $R_e$ varies slowly
as function of $V_0$. We choose the trap to be much larger than the
potential range, $b_t/r_0=3965$.

The wave functions and energies are found with the stochastic
variational method described in chapter~\ref{chap:theory} using the
fully correlated Gaussians, eq.~\eqref{eq:psi-full}. The non-linear
parameters $[\alpha_{ij}^{(k)}]^{-1/2}$ in eq.~\eqref{eq:psi-full} are
optimized stochastically in a large interval covering values of order
$r_0$ up to values of order $b_t$. This allows both short and
long-range correlations. The separated center-of-mass motion is given
by the lowest oscillator wave function with the energy
$E_{cm}=\frac{3}{2}\hbar\omega$.

\section{Zero-Range Approximation} 
Let us briefly refresh the results of the zero-range model in
chapter~\ref{chap:theory}.  In the zero-range limit of the two-body
potential, the scattering length $a$ is the only remaining interaction
parameter.  For $a>0$ the weakly bound dimer has an energy
$E_D=-\hbar^2/(ma^2)$, eq.~\eqref{eq:ED-scat}, while the virtual
state $E_V$ is given by the same expression for $a<0$.  The correction
to $E_D$ and $E_V$ in the effective range expansion is given by
eq.~\eqref{eq:ED-reff}.  In all numerical applications in this
chapter we have found agreement with eq.~\eqref{eq:ED-reff} and
higher-order terms on the two-body level are not needed.

For three particles many bound Efimov trimers may exist even when the
two-body system cannot support any bound states or is weakly bound.
The energies $E_T^{(n)}$ of these states are given in the zero-range
limit by eq.~\eqref{eq:efimov_eq}, see also fig.~\ref{fig:efimovplot}.
The energies scale geometrically with a factor of 515.0, while the
hyper-radii scale with 22.7.  The energies are only determined up to a
three-body regularization scale, $E_{\infty} \equiv
E_T^{(0)}(a=\infty)$, which cannot be determined within the zero-range
theory. It must be fixed by real finite-range calculations or
experimental data.

For $a<0$ the Efimov trimers cross the threshold for three-body
binding, while in the opposite direction of $a>0$ the thresholds are
crossed for binding of the atom-dimer system.  The two corresponding
critical scattering lengths $a^{(n)}_Z$ and $\bar{a}^{(n)}_Z$,
corresponding to $E_T^{(n)}=0$ and $E_T^{(n)}=E_D$, are given in the
lowest-order zero-range limit by eq.~\eqref{eq:a_crit},
\begin{equation}
  \label{eq:a_crit_repeat}
  a^{(n)}_{Z}=-1.50763/\kappa^{(n)},\quad  
  \bar{a}^{(n)}_{Z}=0.0707645/\kappa^{(n)}.
\end{equation}
Here $\kappa^{(n)}=(-mE_T^{(n)}(a=\infty))^{1/2}/\hbar$ is the wave
number for the $n$'th trimer energy at infinite scattering length.
The critical scattering lengths $a^{(n)}_{Z}$ ($\bar{a}^{(n)}_{Z}$)
can be observed via maxima (minima) in the three-body recombination
rate of cold atomic gases \cite{kraemer06} as described in
chapter~\ref{chap:intro} and \ref{chap:theory}.
Since, in this lowest order model, the wave number is given by
$\kappa=1/a$, eq.~\eqref{eq:a_crit_repeat} can instead be written as
\begin{equation}
  \label{eq:kappa_crit}
  1/\kappa^{(n)}_{Z}=-1.50763/\kappa^{(n)},\quad  
  1/\bar{\kappa}^{(n)}_{Z}=0.0707645/\kappa^{(n)},
\end{equation}
where $\kappa^{(n)}_{Z}$ and $\bar{\kappa}^{(n)}_{Z}$ are the
respective critical wave numbers.

\section{Results}

\subsection{Three-Body Energies: Overview}
\begin{figure}[htbp]
  \centering
  \includegraphics[scale=1.05]{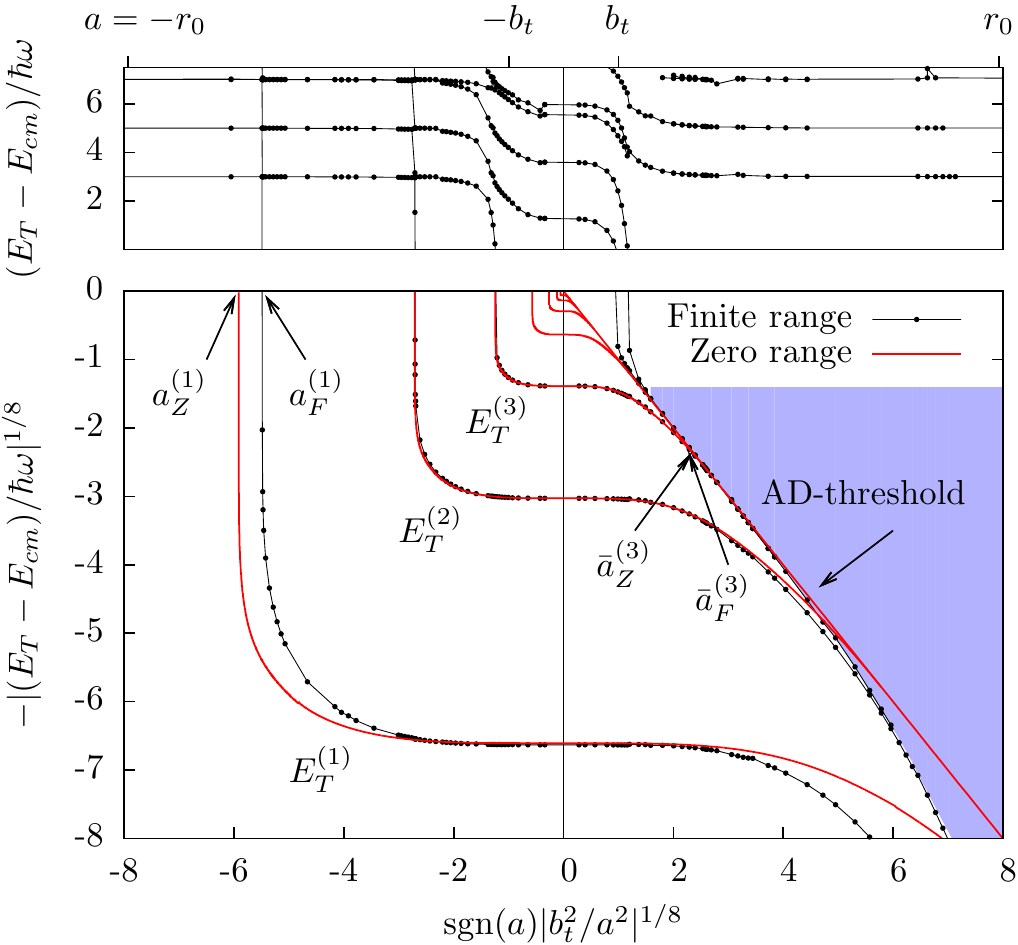}
  \caption{Three-body energy levels, $E^{(n)}_T$ measured relative to
    the center-of-mass energy $E_{cm}$, as function of scattering
    length, $a$. Finite- and zero-range results are shown as points
    and curves, respectively. As finite-range interaction we use a
    Gaussian with adjustable depth. The upper and lower parts,
    corresponding to positive and negative energies, are shown
    separately.  The central value of $|a|=\infty$ corresponds to the
    threshold for binding two particles. The smallest $a$ (far left
    and far right) corresponds to the interaction range, $\lvert
    a\rvert=r_0$.  There is a quasi-continuum for $a>0$ (shaded) which
    shows the threshold for atom-dimer binding (individual states are
    not shown). }
  \label{fig:univ-1}
\end{figure}

We compute the energies and wave-functions corresponding to the
Hamiltonian in eq.~\eqref{eq:hamiltonian} with a Gaussian interaction
by variation of the form in eq.~\eqref{eq:psi-full}.  The resulting
three-body energies are shown in fig.~\ref{fig:univ-1} as function of
scattering length.  The center-of-mass energy is subtracted and the
axes have been scaled by a power of $1/8$ in order to obtain a well
proportioned figure.  The critical scattering lengths and effective
ranges for our finite-range calculations are denoted, in analogy to
\eqref{eq:a_crit_repeat}, with a subscript $F$, i.e. $a^{(n)}_{F}$,
$\bar{a}^{(n)}_{F}$, $R^{(n)}_{F}$ and $\bar{R}^{(n)}_{F}$.

The three-body energies obtained in the zero-range model are
characterized by the arbitrary regularization scale $E_{\infty}$.  We
choose it to equal our second trimer energy, i.e.  $E_{\infty} =
E_T^{(2)}(a=\infty)$.  The lowest energy, $E_T^{(1)}(a=\infty)$, is
avoided for regularization because the ground state energy could be
influenced by the finite-range properties of the potential whereas the
excited spectrum must have the features of the Efimov sequence.
However, in fig.~\ref{fig:univ-1} also the lowest energy coincides
with the zero-range result for $|a|=\infty$.

The finite-range calculation only shows three bound trimer states of
negative energy in contrast to the infinite series of Efimov states in
the free zero-range model. This reflects that the higher lying states
are pushed up by the walls of the external field.  They now appear in
the region of positive energies, still corresponding to bound states,
but now determined by the properties of the trap and not the two-body
potential.

For $a<0$ the trimer states become less bound and cross the trimer
threshold. Deviations from the zero-range model is largest for the
lowest state, $n=1$, which is moved to the right in the plot, i.e.
towards larger $|a|$. The states $n=2,3$ move in the same direction,
but the corrections are much smaller.

For $a>0$ the straight diagonal line shows the atom-dimer zero-range
threshold for binding.  The corresponding finite-range result is lower
and quantitatively in agreement with the dimer energy in
eq.~\eqref{eq:ED-reff}. Only effective range corrections are needed
for an accurate description of the dimer energy here.  Above the
atom-dimer threshold a quasi-continuum is present arising from the
dense atom-dimer spectrum confined by the harmonic oscillator
potential. The small spacing between these levels is then of order
$\hbar\omega$.  The three Efimov trimers near the atom-dimer threshold
lie below the zero-range result.  This is mainly explained by
corrections on the dimer energy, as we discuss later.

\subsection{The $|a|=\infty$ Spectrum}
We already commented on the energies at $|a|=\infty$ above, let us now
consider the quantitative features. The energies
$E_T^{(n)}/\hbar\omega$ for $n=1,\dots,7$ are given in the first row
of tab.~\ref{tab:efimov-E}.  The normal Efimov scaling sequence
without a trap, is also shown in tab.~\ref{tab:efimov-E}. It can of
course only describe the lowest bound Efimov states, $n=1,2,3$, and not
the trap-like state for $n\ge 4$. Our finite-range results agree well
with these zero-range values. The ground state energy, $n=1$, is
remarkably close to the zero-range result. The purely attractive
Gaussian interaction gives a slightly lower energy, but one could have
expected more model-dependence for the ground state. The $n=3$ energy
is slightly higher, which is due to the trap, see below.  The energy
ratios for the finite-range calculations are
$(E_T^{(n)}-E_{cm})/(E_T^{(n+1)}-E_{cm}) = 530, 522$ for $n=1,2$, also
in agreement with the zero-range result, $e^{2\pi/s_0}=515.0$.

In \cite{jonsell02a} the energy of three identical bosons in a
harmonic trap was considered. Only zero-range interactions and
$s$-waves were included. In the limit $|a|/b_t=\infty$ the three-body
energies, $E_T^{(n)}$, are given semi-analytically from the equation
\begin{equation}
  \label{eq:jonsell02a-energy}
  \theta=-\textrm{arg}\left\{\frac{\Gamma(\frac{1}{2})
      -\frac{E_T^{(n)}}{2\hbar\omega}-\frac{s_0}{2}}
    {\Gamma(1-s_0)}\right\},
\end{equation}
where $s_0=1.00624 i$ and $\theta$ is a free regularization parameter
or three-body phase shift (equivalent to $E_\infty$).  We fix
$\theta=0.404\pi$ to match $E_T^{(2)}$ of our finite-range
calculations. The resulting spectrum is given in
tab.~\ref{tab:efimov-E}. All energies agree perfectly with our
finite-range results, the deviations being due to numerical errors
only (except for the ground state).

{\renewcommand{\arraystretch}{1.5} \renewcommand{\tabcolsep}{0.2cm}
\begin{table}[tbhp]
  \centering
  \begin{threeparttable}
    \begin{tabular}{c|ccc|cccc}
      \hline 
      \hline
      & \multicolumn{3}{c|}{Efimov states} & \multicolumn{4}{c}{Trap-like states}\\
      $n$ & 1 & 2 & 3 & 4 & 5 & 6 & 7\\
      \hline
      FR (trap) & $-3.744\cdot10^{6}$ & $-7067$ & $-12.04$ & $2.78$ & $5.10$ & $7.05$ & $7.47$\\
      ZR (free)\tnote{$\dagger$} & $-3.64\cdot10^{6}$ & $-7067$\tnote{$\star$} & $-13.72$ & -- & -- & -- & --\\
      ZR (trap)\tnote{$\ddagger$} & $-3.64\cdot10^{6}$ & $-7067$\tnote{$\star$} & $-12.20$ & $2.76$ & $5.05$ & $6.97$ & $7.20$\\
      \hline
      \hline
    \end{tabular}
    \begin{tablenotes}
      \scriptsize
    \item[$\dagger$] Zero-range model in free space, i.e. universal
      $515.0$ scaling.
    \item[$\ddagger$] Zero-range model in trap, \cite{jonsell02a},
      with $\theta=0.404\pi$ equivalent to $E_\infty$.
    \item[$\star$] This value was used to fix the three-body scale,
      $E_\infty$.
    \end{tablenotes}
  \end{threeparttable}
  \caption{Efimov trimer energies $E_T^{(n)}$ in a trap in units of
    $\hbar\omega$ for $a=\infty$. The finite-range (FR) calculations
    are compared with the zero-range (ZR) theory.}
  \label{tab:efimov-E}
\end{table}
}

\subsection{Trap-Like States}
We already commented on the trap-like states for $|a|=\infty$. Let us
now consider the positive energy spectrum for $a<0$ in the upper part
of fig.~\ref{fig:univ-1}.  This is a ``quasi-continuum'' with the
spacing of order $\hbar\omega$.  The asymptotic limit at small $|a|$
is simply the oscillator spectrum for three particles which remains when
the two-body interaction is negligibly small. The lowest positive
energy level for $a=0$ is $3\times 3\hbar\omega/2$ and the excitation
spectrum is obtained by adding $2 \hbar\omega$. The almost vertical
lines for the Efimov states, continuing into the positive energy
region, cross the oscillator states.  For small $|a|$ the coupling is
also small and the avoided crossing appears like true crossings.  As
$a$ increases the couplings increase and smooth avoided crossings
appear.  For $a>0$ a similar spectrum exists, but the huge amount of
avoided crossings is not computed.

More generally, the avoided crossings in the upper part of
fig.~\ref{fig:univ-1} resemble the Zel'dovich level rearrangement for
two particles, where an attractive long-range potential is perturbed
by a strong attractive short-range potential \cite{combescure07}. In
our case the long-range interaction is the harmonic trap, and we have
three particles instead of two. 
Also, in \cite{pub-fedorov09-1} we showed how to calculate narrow resonances
of three-body systems. The approach was to discretize the continuum by
introducing a large artificial oscillator trap. When varying the trap
size avoided crossings occurred, and resonance energies and widths
could be extracted. Such avoided crossing are closely related to the
crossings in fig.~\ref{fig:univ-1}. The only difference is whether the
trap or interaction is varied, while the other is fixed.

Recently, Efimov physics in a finite square box with large positive
scattering length has been investigated within effective field theory.
In the preprint \cite{kreuzer09a} it is concluded that
``By decreasing the box size, the binding energy decreases and
eventually the state is shifted into the positive energy regime. The
finite volume corrections are most important for the shallowest states
which are largest in size and feel the finite volume first''.
These two conclusions agree with our results above where all but the
three lowest Efimov states were pushed up by the harmonic trap. It is
also consistent with \cite{jonsell02a}. The behavior is easily
understood from the fact that a confining volume induces extra kinetic
(zero-point) energy.
However, in the published version \cite{kreuzer09b} the figures were
changed and the conclusion was replaced by ``If the box size is
decreased, the binding \textit{increases}[ed.]''. This conclusion
contradicts all physical intuition and is not elaborated in
\cite{kreuzer09b}.

\subsection{Finite-Range Borromean Window Corrections}
\label{sec:borromean-window}

The differences between finite- and zero-range results for $a<0$ can
be extracted from fig.~\ref{fig:univ-1}.  We focus on the measurable
quantities expressing that an energy threshold has been crossed.

\begin{figure}[htbp]
  \centering
  \includegraphics{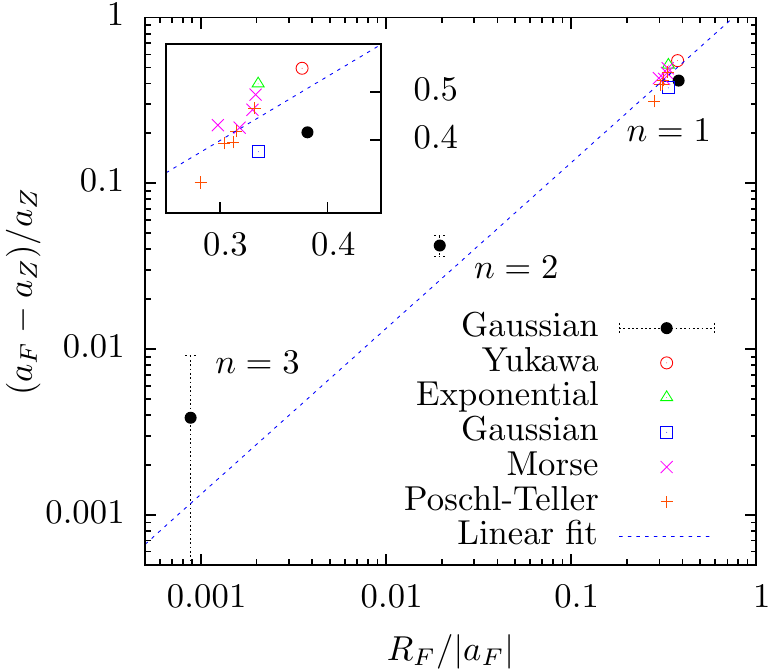}
  \caption{The relative shift between finite- and zero-range critical
    scattering lengths, $a^{(n)}_{F}$ and $a^{(n)}_{Z}$ as function of
    the corresponding critical effective range $R^{(n)}_{F}$.  The
    Gaussian data from fig.~\ref{fig:univ-1} are shown with numerical
    error bars.  The other data are for the ground states of different
    finite-range potentials from \cite{moszkowski00}. The zero-range
    energy scale $E_T^{(n)}(a=\infty)$ was chosen equal to the
    individual finite-range energies at $|a|=\infty$. The fit is given
    in eq.~\eqref{eq:fig2-fit}.}
  \label{fig:univ-2}
\end{figure}

The finite-range results for the negative critical scattering lengths
$a^{(n)}_{F}$ are shown in tab.~\ref{tab:efimov-aR} for $n=1,2,3$
together with the corresponding effective ranges $R^{(n)}_{F}$. The
related zero-range values $a^{(n)}_{F}$ were calculated from
eq.~\eqref{eq:a_crit_repeat} with the regularization fixed by the
finite-range energies $E_T^{(n)}-E_{cm}$ from tab.~\ref{tab:efimov-E}.
The values are shown in fig.~\ref{fig:univ-2}, where we plot the
relative shift in critical scattering lengths as functions of
effective range. The systematics is discussed below.
{ \renewcommand{\arraystretch}{1.5} \renewcommand{\tabcolsep}{0.2cm}
\begin{table}[tbhp]
  \centering
  \begin{threeparttable}
    \begin{tabular}{c|ccc}
      \hline \hline 
      & \multicolumn{3}{c}{Efimov states}\\
      $n$ & 1 & 2 & 3\\
      \hline 
      $a_{F}^{(n)}$ & $-4.376$ & $-74.09$ & $-1631$\\
      $a_{Z}^{(n)}$ & $-3.089$ & $-71.10$ & $-1625$\\
      $R_{F}^{(n)}$ & $1.670$ & $1.449$ & $1.436$\\
      \hline \hline 
    \end{tabular}
  \end{threeparttable}
  \caption{Critical negative scattering lengths, $a_F^{(n)}$, for binding of
    the $n$'th Efimov trimer state. The finite-range interaction (F) is a
    Gaussian with range $r_0$. The zero-range (Z) results are fixed by
    the trimers energies at $|a|=\infty$. The effective range at the
    position for binding is also shown. All values are in units of
    $r_0$.}
    \label{tab:efimov-aR}
\end{table}
}

We note that the first threshold, $a_F^{(1)}$, is also significant by
marking the interval of scattering lengths where two particles cannot
bind while three particles can.  This Borromean window between the
threshold and $|a|=\infty$, has been calculated in \cite{moszkowski00}
for a number of different radial shapes of the two-body potentials.
Their results are expressed as ratios, $S=g_3/g_2$, of critical
strengths required precisely to bind the two and three body systems,
respectively.  This ratio is limited to $2/3\le S\le 1$, but for most
ordinary potentials $S\simeq 0.8$.  We find $S=0.793$ for the the
Gaussian potential in agreement with $0.79$ in \cite{moszkowski00}.
With the critical potential strengths given in \cite{moszkowski00} we
calculate $a^{(1)}_{F}$ and $R^{(1)}_{F}$. The results are compared to
the Gaussian values $a^{(n)}_{F}$ and the zero-range results in
fig.~\ref{fig:univ-2}. As in fig.~\ref{fig:univ-1}, the effect of the
positive effective range is a systematic shift of the critical
scattering length towards larger absolute values.  The difference
decreases with increasing $n$ which implies that most significant
effects are related to small $|a|$-values. The available systematics
in fig.~\ref{fig:univ-2} can be described by a straight line, i.e.
\begin{equation}
  \label{eq:fig2-fit}
  \frac{a^{(n)}_{F}-a^{(n)}_{Z}}{a^{(n)}_{Z}}= 1.3
 \frac{R^{(n)}_{F}}{|a^{(n)}_{F}|} \; .
\end{equation}
The wide range of potential shapes used, indicate that this result
most probable is model-independent.
Although the conclusions are based on calculations with $R_e>0$ only,
we speculate that eq.~\eqref{eq:fig2-fit} has the same form for
$R_e<0$. This implies that the shift of critical scattering length
changes sign, i.e. the energies move to the left in
fig.~\ref{fig:univ-1}.

The shift in eq.~\eqref{eq:fig2-fit} should be measurable for cold
atomic gases as positions of extrema of the three-body recombination
rate near a Feshbach resonance \cite{kraemer06}.
For broad resonances the background channel (the van der Waals
interaction) will determine the effective range. Thus $R_e$ will be
positive and of the order $l_{vdW}$, and the shift will be toward
larger $|a|$. On the other hand, for narrow Feshbach resonances $R_e$
is negative and given by eq.~\eqref{eq:reff-fesh-const}, so the shift
will be towards smaller $|a|$.

\subsection{Finite-Range Atom-Dimer Corrections} 

\begin{figure}[htbp]
  \centering
  \includegraphics[scale=1.1]{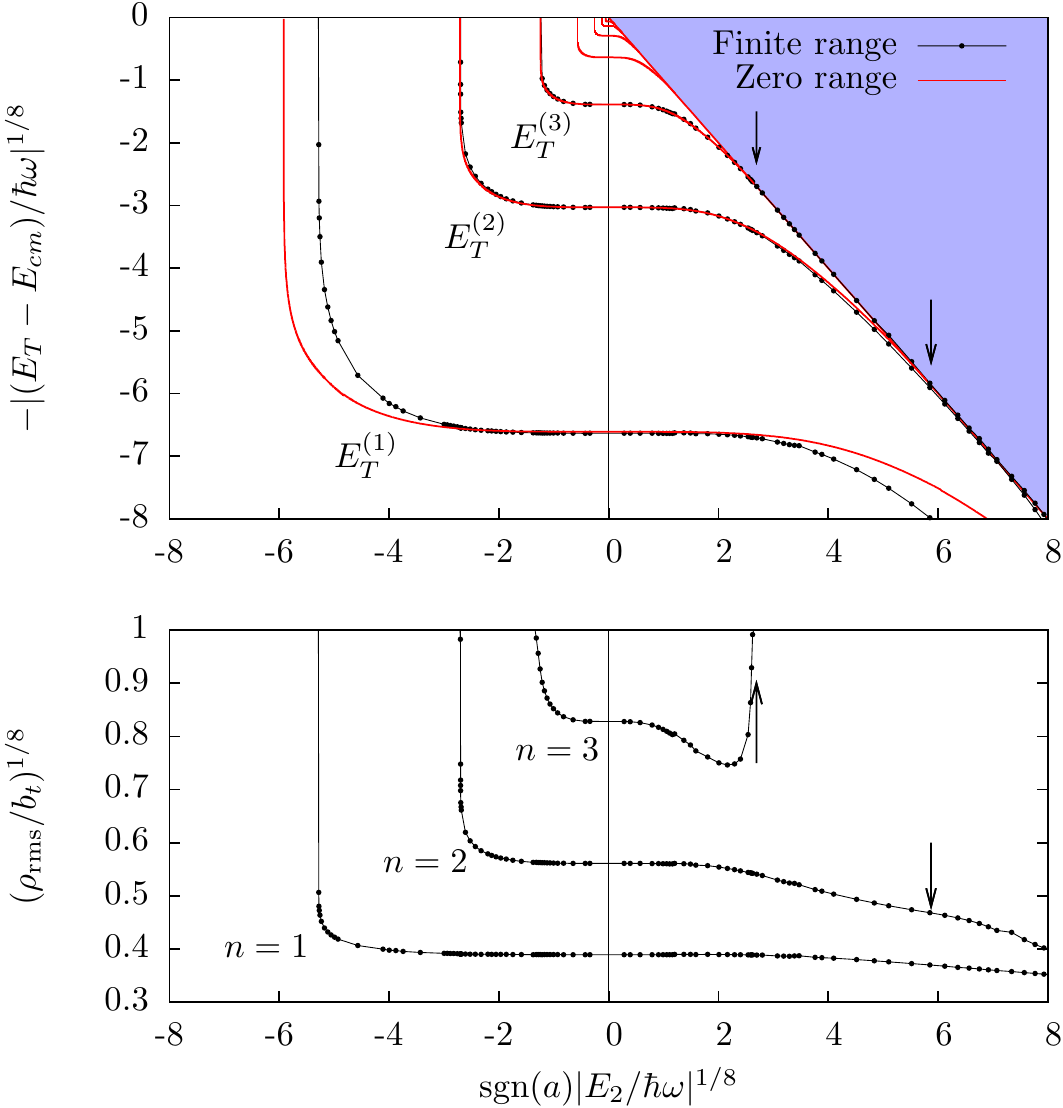}
  \caption{Top: Three-body energies, $E^{(n)}_T$, as in
    fig.~\ref{fig:univ-1}, but plotted as function of two-body energy
    instead of scattering length. On the right part ($a>0$) $E_2=E_D$
    is the weakly bound dimer, eq.~\eqref{eq:ED-reff}, while on the
    left part ($a<0$) $E_2=E_V$ is the shallow virtual dimer (also
    eq.~\eqref{eq:ED-reff}). Bottom: The root-mean-square hyper-radii,
    $\rho_\text{rms}$, as function of two-body energy. The zero-range
    predictions of the crossing of the Efimov states with the
    atom-dimer threshold are indicated by arrows. }
  \label{fig:univ-virtual}
\end{figure}

The effective range corrections are more complicated for $a>0$ because
the atom-dimer threshold is also shifted, see fig.~\ref{fig:univ-1}.
This effect is investigated in fig.~\ref{fig:univ-virtual} (upper
part) where the trimer energy is now shown as function of the
finite-range dimer energy, $E_D$.  Then the thresholds coincide for
finite- and zero-range models.

The ground state size is comparable to the range of the attractive
finite-range potential, and hence decreases more strongly with
increasing $E_D$.  As expected this state remains in the discrete
region below the atom-dimer threshold.  The energies of the excited
trimer states $n=2,3$ are in perfect agreement with the zero-range
prediction. Only $E_T^{(2)}$ is slightly below the zero-range model
near the atom-dimer threshold. We conclude that the major shift in
trimer energies and threshold scattering lengths are due to effective
range corrections to the dimer energy alone.

This is consistent with the explanation in \cite{efimov70,efimov90}
that these Efimov states disappear into the atom-dimer continuum.  As
also described in \cite{braaten06} this implies that their structure,
as the threshold is approached, converges to the dimer-state with a
loosely bound atom at large distance.

This understanding is tested by computing the root-mean-square
hyper-radii $\rho_\text{rms}=\langle\rho^2\rangle^{1/2}$,
eq.~\eqref{eq:rho-3}, shown in the lower part of
fig.~\ref{fig:univ-virtual} as function of $E_D$.  We first consider
the $n=3$ state. When $E_D$ increases, both the energy and the size of
the Efimov state decreases.  However, when the state approaches the
atom-dimer threshold, the structure changes rather abruptly towards a
dimer and a free atom, and hence the radius increases correspondingly
fast towards the upper limit, $b_t$, defined by the trap. From Fig
\ref{fig:univ-virtual}, this occurs at
($\bar{a}^{(3)}_{F},\bar{R}^{(3)}_{F}) = (84.59, 1.423)r_0$.  This is
consistent with the zero-range result $\bar{a}^{(3)}_{Z}=80.9 r_0$
(indicated by the arrow), since the effective range is small compared
to the scattering length. The energy of the $n=2$ Efimov state in
fig.~\ref{fig:univ-virtual} is also in good agreement with the
zero-range result.  However, when the state approaches the atom-dimer
threshold no drastic increase in size is observed, as for ordinary
Efimov states. This can be ascribed to the effective range which is
comparable to the scattering length in this regime, namely
$(\bar{a}^{(2)}_{F}, \bar{R}^{(2)}_{F})=(4.1, 1.18)r_0$. This effect
needs a more careful analysis in the future.

\subsection{Effective Range Corrections in EFT}
After the publication of the above results in
\cite{thogersen08-3}, many of these results have been confirmed
qualitatively by effective field theory \cite{platter09,platter09-2}.
In this subsection we sum up the common effects and also give
quantitative comparisons. In \cite{platter09} the linear effective
range corrections to the three-body energies were calculated for
$R_e>0$. The main features are presented in their fig.~1, which
essentially is identical to fig.~\ref{fig:univ-virtual} in this
chapter. The following conclusions are given in
\cite{platter09,platter09-2}:
\begin{enumerate}
\item[i)] The three-body spectrum at $|a|=\infty$ is unperturbed when
  introducing a linear effective range correction.
\item[ii)] For $a<0$ the critical scattering lengths at the trimer
  threshold move towards higher $|a|$. The effect is largest for the
  lowest states.
\item[iii)] For $a>0$ the effects are in general small (after the
  dimer energy correction is taken into account). The trimer only
  becomes slightly more bound.
\end{enumerate}
All these qualitative effects are in perfect agreement with the
previous conclusions of this chapter.

\begin{figure}[tbhp]
  \centering
  \includegraphics{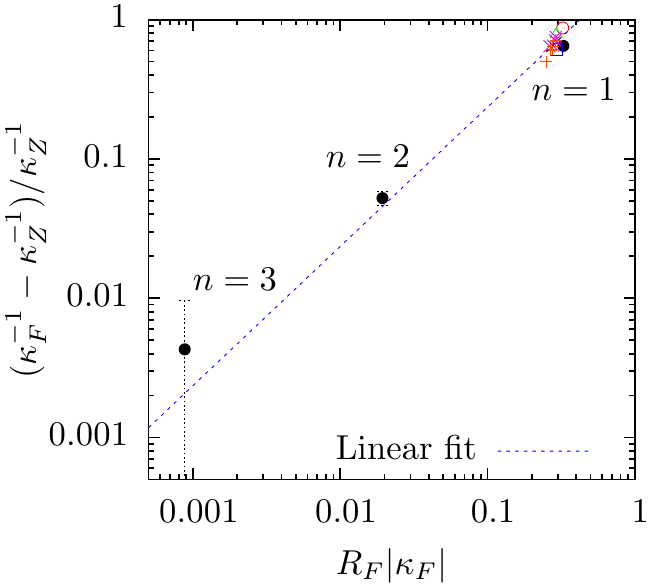}
  \caption{Same data as fig.~\ref{fig:univ-2}, but with scattering
    length $a$ replaced by inverse binding wave number, $\kappa^{-1}$.
    Legends are the same as in fig.~\ref{fig:univ-2} and the fit is
    given in eq.~\eqref{eq:fig2-virtual-fit}.  }
  \label{fig:univ-virtual2}
\end{figure}

Let us now turn to a quantitative comparison of the shift in critical
scattering length for $a<0$.  Since \cite{platter09} use the two-body
bound state pole (i.e. energy) instead of scattering length, we must
repeat the analysis of subsection~\ref{sec:borromean-window} using
$\kappa^{-1}$ instead of $a$. We include the full effective range
corrections in $\kappa$ as given as in
eq.~\eqref{eq:kappa-wavenumber}. The result is shown in
fig.~\ref{fig:univ-virtual2}.
The use of $\kappa^{-1}$, as compared to fig.~\ref{fig:univ-2}, mostly
affects the $n=1$ data. The linear relationship becomes marginally
better using $\kappa^{-1}$ instead of $a$. The linear fit gives
\begin{equation}
  \label{eq:fig2-virtual-fit}
  \frac{1/\kappa^{(n)}_{F}-1/\kappa^{(n)}_{Z}}{1/\kappa^{(n)}_{Z}}= 2.4
  R^{(n)}_{F} |\kappa^{(n)}_{F}|.
\end{equation}
We extract values for the same shift in fig.~1 of \cite{platter09}.
By reading off the critical values for the lowest curve%
\footnote{Note that they use the notation $\gamma$ instead of
  $\kappa$, and use a scaled plot.} %
we estimate their lowest order result to be
$\kappa_{LO}/\kappa_0=-5.68$ and the next-to-lowest order result
(including linear $R_e$) to be $\kappa_{NLO}/\kappa_0=-5.19$. Here
$\kappa_0$ is an arbitrary momentum scale.  The effective range was
chosen to be constant, $R_e\kappa_0=0.01$.  This gives
\begin{equation}
  \label{eq:platter-shift}
  \frac{\kappa^{-1}_{NLO}-\kappa^{-1}_{LO}}{\kappa^{-1}_{LO}}
  / R_e|\kappa_{NLO}|
  = 1.8\pm 0.5.
\end{equation}
This value agrees with the coefficient in
eq.~\eqref{eq:fig2-virtual-fit}.  The difference can be attributed to
numerical and read-off errors.  Also, the models are based on very
different backgrounds and formalisms: In \cite{platter09} only linear
effects in $R_e$ are included in an EFT approach, while our
finite-range calculations in principle can have minor effects from
higher-order shape parameters (i.e. potential model-dependence) and
the trap. Taking all these remarks in consideration, the agreement is
actually extraordinarily good.

The discussion above is only based on a single value from
\cite{platter09}, i.e. the predictions of the linear relationship in
eq.~\eqref{eq:fig2-virtual-fit} is not confirmed by EFT yet. Such work
is currently in progress, and preliminary results confirm the linear
prediction with the correct coefficient%
\footnote{Private communications, David L. Canham, Universit\"at Bonn,
  Germany.}.

\section{Conclusions and Outlook}

We have used a basis of correlated Gaussians to calculate the spectrum
for three identical bosons for a finite-range potential and an
external trap. The scattering length was tuned to arbitrary values,
and the many excited states were computed accurately as function of
scattering length.
The universal spectrum at infinite scattering length was in perfect
agreement with analytical models.

We then focused on two thresholds at negative and positive scattering
lengths, i.e. the Borromean window where two particles cannot bind
while three can form many bound states, and the atom-dimer threshold
where dimers can bind and an atom may be bound to the dimer.

We have extracted the universal behavior of both these thresholds,
including corrections expressed in terms of effective range divided by
scattering length.  We also conclude that effective range corrections
to Efimov physics and Borromean binding are two aspects of the same
effect and these two regions can be connected quantitatively. The
linear shift in critical scattering lengths at the trimer threshold
agrees quantitatively with effective field theory.  The main effect is
that the universal scaling factor becomes smaller for the lowest
Efimov states (smallest scattering length) on the negative scattering
length side.

We also show that the structure as the atom-dimer threshold is
approached is a weakly bound atom moving away from the dimer. The main
effective range correction to the trimer energies near the atom-dimer
threshold then comes from the dimer only. This seems to induce a
larger scaling factor for the lowest Efimov states.

The results have observable consequences for three-body recombination
rates in atomic Bose-Einstein condensates and should be taken into
account when the scattering length is comparable to the effective
range \cite{kraemer06,knopp09}.  
The effects are also applicable for halo nuclei and Borromean systems.

Future investigations could repeat a similar analysis with potentials
of negative effective range, which could be useful for modeling narrow
Feshbach resonances.  We speculate that the linear relationship
eq.~\eqref{eq:fig2-fit} also holds for negative effective range.  This
conjecture is strongly supported in the next chapter, where we
analytically derive bounds for the Efimov effect in the case of large
negative effective range.

%% file: efimov2.tex
\section{Introduction}
We have seen in previous chapters how universal scaling properties in
three-body systems arise when the scattering length $a$ is much larger
than the range $r_0$ of the underlying two-body potential. In this
regime several three-body observables are universal.
The universal scaling of Efimov trimers is usually said to exist for
rms-sizes between $r_0$ and $a$
\cite{efimov73,efimov91,braaten06,kraemer06}. The effective range
$R_e$ from a low-energy phase shift expansion is sometimes used
instead of $r_0$ in this statement
\cite{fedorov01,thogersen08-3,platter09}.  This ambiguity occurs
because $r_0$ and $R_e$ are often of the same order. However, for
narrow Feshbach resonances in atomic gases $R_e$ can be much larger
than $r_0$ \cite{bruun05}, and the implications for such systems need
to be explored.

Zero-range models, in particular in combination with the
hyper-spherical approximation \cite{fedorov01,jonsell04,braaten06},
have been successful in semi-quan\-ti\-ta\-tive descriptions of
three-body systems in the universal regime. Semi-rigorous finite-range
corrections have been attempted by including higher-order terms in
the effective range expansion \cite{fedorov01,platter09} as a step
towards the full finite-range calculations as in
\cite{suno02,thogersen08-3} while maintaining the conceptual and
technical simplicity of the zero-range approximation.

The obvious generalization of the zero-range model,
eqs.~\eqref{eq:zr-boundary-3b}--\eqref{eq:zr-efimov-eq}, is to
substitute $-1/a$ with $-1/a+(R_e/2)k^2$, where $k$ is the two-body
wave number, in the relevant expressions for the logarithmic
derivative of the total wave-function at small separation of the
particles. However, in three-body systems neither the two-body wave
number nor the small separation are uniquely defined, and rigorous
inclusion of all terms of the given order is non-trivial.  The lack of
rigor in previous works could have serious implications for
applications where finite-range effects are important, such as the
stability conditions for condensates in traps, and in particular for
Efimov physics. Experimental progress \cite{kraemer06,knopp09} will
soon require this increased accuracy near the boundaries of the
universal regime.

In this chapter we consider a system of three identical bosons near a
Feshbach resonance in the universal regime with large scattering
length usually described by model-independent zero-range potentials.
We employ the adiabatic hyper-spherical approximation described in
chapter~\ref{chap:theory} and derive the rigorous
large-distance equation for the adiabatic potential for finite-range
interactions. The equation is suitable for the analytic studies of the
finite-range corrections in the three-boson problem. 
We investigate the finite-range corrections to the adiabatic potential
and the non-adiabatic term and compare with the zero-range
approximation. The effective range correction to the zero-range
approximation must be supplemented by a new term of the same order.
The non-adiabatic term can be decisive.
Efimov physics is always confined to the range between effective range
and scattering length. Our analytical results agree with numerical
calculations for realistic potentials.

\section{Adiabatic Eigenvalue Equation}
We first generalize the adiabatic eigenvalue equation,
eq.~\eqref{eq:zr-efimov-eq}, to arbitrary finite-range potentials. The
derivation is similar to that of chapter~\ref{chap:theory}, but with
some modifications.

We consider three identical bosons of mass $m$ and coordinates $\bm
r_i$ interacting via a finite-range two-body potential $V$, where we
assume $V(r_{jk})=0$ for $r_{jk} = \abs{\bm r_j - \bm r_k} >r_0$.  Only
relative $s$-waves are included.  
We use the hyper-radius $\rho^2=(r_{12}^2+r_{13}^2+r_{23}^2)/3$,
eq.~\eqref{eq:rho-3} and hyper-angles
$\tan\alpha_i=(r_{jk}/r_{i,(jk)})\sqrt{3}/2$, where
$r_{i,(jk)}=\abs{\bm r_i-(\bm r_j+\bm r_k)/2}$. In the following we
shall use one set of coordinates and omit the index.

The adiabatic hyper-spherical approximation treats the hyper-radius
$\rho$ as a slow adiabatic variable and the hyper-angle $\alpha$ as the
fast variable. The eigenvalue $\lambda(\rho)\equiv\nu^{2}(\rho)-4$ of
the fast hyper-angular motion for a fixed $\rho$ serves as the
adiabatic potential for the slow hyper-radial motion. The eigenvalue is
found by solving the Faddeev equations, eq.~\eqref{eq:faddeev-eq}, for fixed
$\rho$,
\begin{equation} 
  \label{eq:faddeev}
  \Big[-\frac{\partial^2}{\partial\alpha^2}-\nu^2 +U\Big]\psi=-2U\rot[\psi].
\end{equation}
Here $\psi(\rho,\alpha)$ is the Faddeev hyper-angular component,
\begin{equation}
  \label{eq:U}
  U(\rho,\alpha)=V(\sqrt{2}\rho\sin\alpha) \frac{2m\rho^2}{\hbar^2}
\end{equation}
is the rescaled potential, and
\begin{equation}
  \label{eq:rotation}
  \rot[\psi](\rho,\alpha)
  \equiv\frac{2}{\sqrt{3}}\int_{\abs{\frac{\pi}{3}-\alpha}}^{\frac{\pi}{2}-\abs{\frac{\pi}{6}-\alpha}}\psi(\rho,\alpha^{\prime})\ud\alpha^{\prime}
\end{equation}
is the operator that rotates a Faddeev component into another Jacobi
system and projects it onto $s$-waves,
eq.~\eqref{eq:rot-operator}. The hyper-angular wave-function of the
three-body system is
\begin{equation}
  \label{eq:Phi}
  \Phi(\rho,\alpha)=\frac{\psi(\rho,\alpha)+2\rot[\psi](\rho,\alpha)}{\sin2\alpha}.
\end{equation}
In the adiabatic approximation the hyper-radial function $f(\rho)$
satisfies the ordinary hyper-radial equation,
eq.~\eqref{eq:hyper-radial-equations}, with the effective potential
\begin{equation}
  \label{eq:V_eff}
  V_\textup{eff}(\rho)=\frac{\hbar^{2}}{2m}\left(\frac{\nu^{2}-1/4}{\rho^{2}}-Q\right), \quad\,
  Q=\left\langle\Phi\right|\frac{\partial^{2}}{\partial\rho^{2}}\left|\Phi\right\rangle,
\end{equation}
where $Q$ is the non-adiabatic term and $\Phi$ is normalized to unity
for fixed $\rho$.

We first divide the $\alpha$-interval $[0;\pi/2]$ into two regions:
(I) where $U \neq 0$, and (II) where $U = 0$.  The regions are
separated at $\alpha = \alpha_0$ where
$\sin\alpha_0\equiv r_0/(\sqrt{2}\rho)$. 

\noindent\emph{Region (II)}: Here $U=0$ and we have the
free solution to eq.~\eqref{eq:faddeev},
\begin{equation}
  \label{eq:phiII}
  \psi^{II}(\alpha)=\mathcal N(\rho)\sin(\nu(\alpha-\pitwo)),
\end{equation}
with the boundary condition $\psi^{II}(\pi/2)=0$ and normalization
$\mathcal N(\rho)$. 

\begin{figure}[htbp]
  \centering
  \includegraphics[scale=0.8]{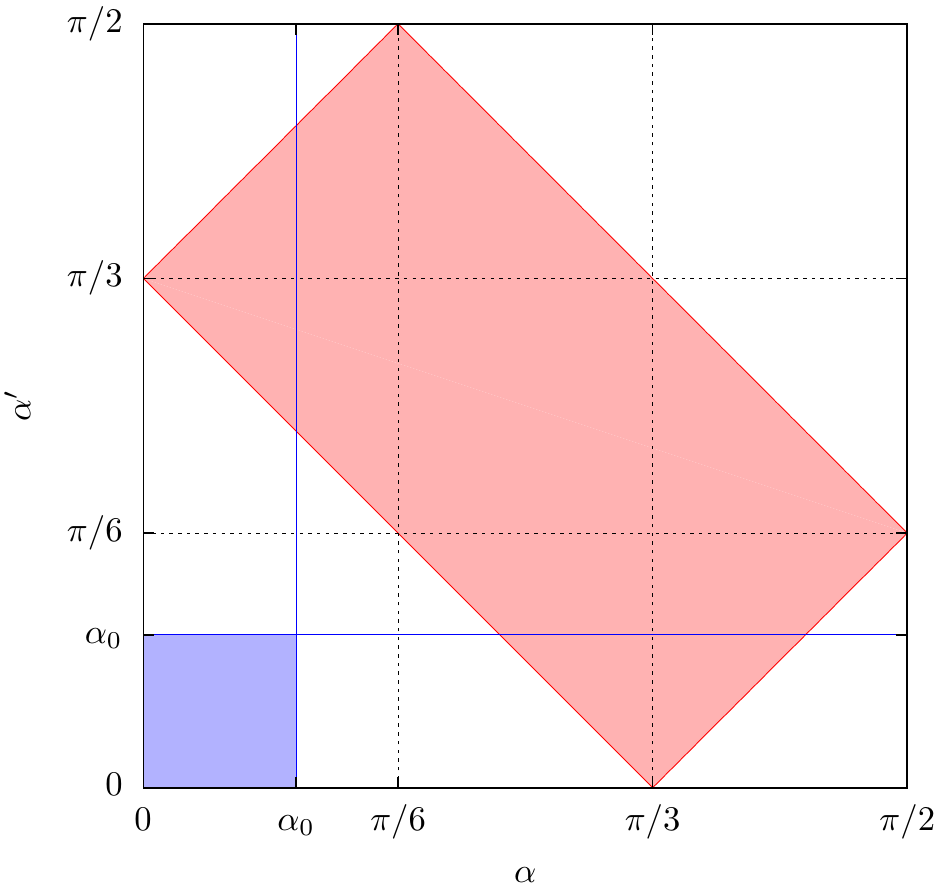}
  \caption{Integration limits for the rotation operator
    eq.~\eqref{eq:rotation}.  For $\rho>\rho_c$ the shaded regions
    does not overlap and the three-body problem reduces to an
    effective two-body problem.  When $\alpha\le\alpha_0\le \pi/6$ the
    integration limit for $\alpha'$ is strictly outside the potential
    range $[0:\alpha_0]$ giving $\rot[\psi]=\rot[\psi^{II}]$.}
  \label{fig:rotate-limits2}
\end{figure}

\noindent\emph{Region (I)}: We now restrict the problem to be outside the
``collapsed'' region defined by $\rho \ge \rho_c \equiv \sqrt{2}r_0$.
This implies that $\alpha_0\le \pi/6$. Since $\alpha\le \alpha_0$ the
rotation operator, eq.~\eqref{eq:rotation}, only integrates over the
free solution $\psi^{II}$, see fig.~\ref{fig:rotate-limits2}.
Equation~\eqref{eq:faddeev} then simplifies to
\begin{equation}
  \label{eq:faddeev-like_eq}
  \Big[-\frac{\partial^2}{\partial \alpha^2} - \nu^2 + U \Big]\psi^{I}
  = -2 U \rot[\psi^{II}].
\end{equation}
The solution is $\psi^{I} = \psi^{Ih} - 2 \rot[\psi^{II}]$, where
$\psi^{Ih}$ and $-2\rot[\psi^{II}]$ are homogeneous and inhomogeneous
solutions, respectively. $\psi^{Ih}$ is the regular solution to
\begin{equation}
  \label{eq:modif_angular_eq}
  \Big[
    -\frac{\hbar^2}{m}\frac{\partial^2}{\partial r^2}
    -\frac{\hbar^2 k_\rho^2}{m} 
    +V_\rho(r)
  \Big]\psi^{Ih}=0,
\end{equation}
\begin{equation} 
  \label{eq:potrho}
  V_\rho(r) \equiv V(\sqrt{2}\rho\sin(\frac{r}{\sqrt{2}\rho})),
\end{equation}
where $k_\rho=\nu/(\sqrt{2}\rho)$ and $r=\sqrt{2}\alpha\rho$. When $\alpha\to\alpha_0$,
\begin{equation}
  \label{eq:phiI}
  \psi^{Ih}\propto \sin(k_\rho r+\delta_\rho),
\end{equation}
where the modified phase shift $\delta_\rho(k_\rho)$
arises from the modified two-body potential, $V_\rho$.

The solutions $\Phi$ in region (I) and (II) are now matched smoothly, giving
\begin{equation}
  \label{eq:logderiv_match}
  \left. \frac{\partial}{\partial\alpha}\ln\psi^{Ih}\right|_{\alpha_0}=
  \left.\frac{\partial}{\partial\alpha}\ln\left(\psi^{II}+2\rot[\psi^{II}]\right)\right|_{\alpha_0}.
\end{equation}
After inserting eqs.~\eqref{eq:phiII} and \eqref{eq:phiI}, this
equation becomes
\begin{equation}
  \label{eq:generalized_efimov_eq}
  \frac{1}{\sqrt{2}\rho} \ \frac{ -\nu\cos(\nu\pitwo)+ \frac{8}{\sqrt{3}}\sin(\nu\pisix) }
  {\sin(\nu\pitwo)} 
  = k_\rho\cot\delta_\rho(k_\rho),
\end{equation}
which defines $\nu$ as function of $\rho$. 
This generalizes the $-1/a$ expression for the lowest-order zero-range
model, eq.~\eqref{eq:zr-efimov-eq}.

The right-hand-side deviates from the zero-range approximations
\cite{jonsell04,platter09} which use the normal two-body phase shift,
i.e. the right-hand-side is $k_\rho\cot\delta(k_\rho)$. We use the rigorously defined phase
shift $\delta_\rho$ for $V_\rho$ instead of $\delta$. Our result is
exact for $\rho\ge \rho_c$.

\section{Effective Range Expansion}

\begin{figure}[tb]
  \centering
  \includegraphics[scale=1.1]{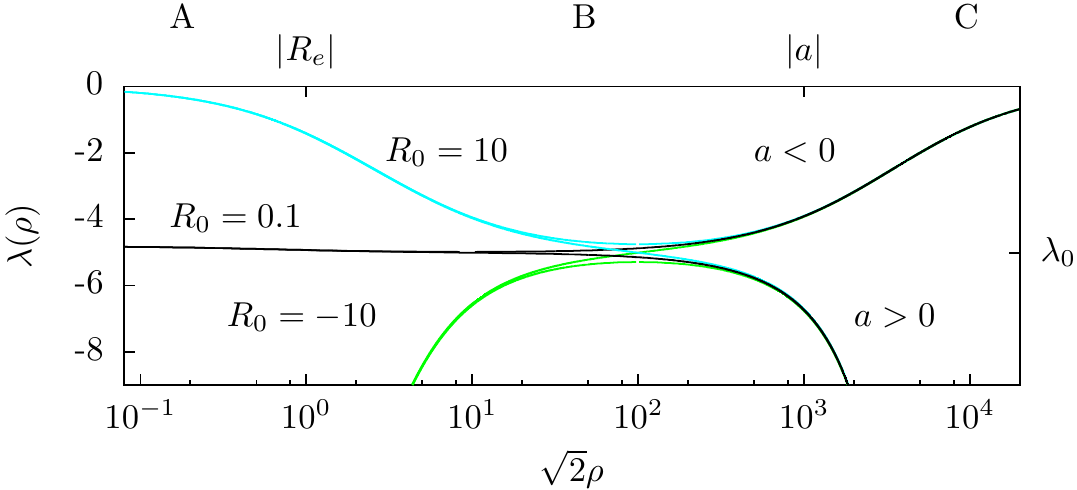}
  \caption{Adiabatic eigenvalues $\lambda(\rho)$ from
    eq.~\eqref{eq:generalized_efimov_eq2} as function of hyper-radius
    $\rho$, for large scattering length $a$ and negative $R_e$.
    Different values of the model-dependent length $R_0$ are used,
    showing that the universal solution $\lambda_0$ exists in the
    region $\abs{R_0}\ll\sqrt{2}\rho\ll\abs{a}$. Lengths are in
    units of $\abs{R_e}$.}
  \label{fig:lambda-analytic}
\end{figure}

In the limit $\rho\gg \rho_c$, the $\rho$-dependent potential, $V_\rho(r)$,
approaches $V(r)$, and consequently
$\delta_\rho$ approaches $\delta$.  
The $\rho$-dependent low-energy effective range expansion corresponding to $V_\rho$ is
then to second order
\begin{equation} 
  \label{eq:effexp}
  k_\rho\cot\delta_\rho(k_\rho)\Big|_{k_\rho\to0}
  \approx
-\frac{1}{a(\rho)}+\frac{R_e(\rho)}{2}k_\rho^2,
\end{equation}
where $a(\rho)$ and $R_e(\rho)$ are functions of $1/\rho^2$ that
converge to $a$ and $R_e$ for $\rho\to\infty$.
Up to $1/\rho^2$ in eq.~\eqref{eq:effexp} we get
\begin{equation}  \label{eq:scatexp}
      \frac{1}{a(\rho)} \approx\frac{1}{a} +
      R_V \frac{1}{2\rho^2},\quad
      R_e(\rho)\approx R_e.
\end{equation}
The model-dependent expansion parameter $R_V$, which we shall call the
``scattering length correction'', is found to be\footnote{The integral
  representation of the effective range, eq.~\eqref{eq:reff-integral},
  is usually proved by integrating and subtracting the radial equation
  for two different energies, comparing to the free solutions, and
  then letting the energies go to
  zero. Equation~\eqref{eq:c02_integral_rep2} can be proved in a
  similar manner by using different potentials ($V$ and $V_\rho$) at the
  same energy and letting $\rho\to\infty$.}
\begin{equation}
  \label{eq:c02_integral_rep2}
  R_V 
  =\frac{m}{6\hbar^2}\langle V'r^3 \rangle_u
  =\frac{m}{6\hbar^2}\int_0^{r_0}V'(r)r^3 u(r)^2\ud r,
\end{equation}
where $u$ is the zero-energy two-body radial wave-function,
asymptotically equal to $1-r/a$. Equation~\eqref{eq:generalized_efimov_eq}
then becomes
\begin{equation}
  \label{eq:generalized_efimov_eq2}
  \frac{1}{\sqrt{2}\rho} \ \frac{ -\nu\cos(\nu\pitwo)+ \frac{8}{\sqrt{3}}\sin(\nu\pisix) }
  {\sin(\nu\pitwo)} 
  = -\frac{1}{a}+\frac{R_e}{2}\ \frac{\nu^2}{2\rho^2}-\frac{R_V}{2\rho^2}.
\end{equation}
This equation without the last two finite-range terms has the
well-known purely imaginary solution $\nu_0=1.00624 i$, or
$\lambda_0=-5.0125$, for $\sqrt{2}\rho\ll\abs{a}$. This solution
gives $V_\textup{eff}\propto-1/\rho^2$ which is the basis of Efimov
physics. The $R_e$-term was included in \cite{fedorov01,platter09},
but not the model-dependent $R_V$-term. The latter term makes the finite-range
corrections to the zero-range adiabatic eigenvalues explicitly
non-universal. The last two terms in
eq.~\eqref{eq:generalized_efimov_eq2} restrict the solution
$\lambda_0$ to the region $\abs{R_0} \ll \sqrt{2}\rho \ll \abs{a}$,
where
\begin{equation}
  \label{eq:RE}
  R_0 \equiv \frac{R_e}{2}\nu_0^2-R_V,
\end{equation}
as seen in fig.~\ref{fig:lambda-analytic} where the lowest solution to
eq.~\eqref{eq:generalized_efimov_eq2} is shown for different parameter
choices. Thus, naively one would think that the lower limit for
Efimov physics is determined by the model-dependent length
$\abs{R_0}$. However, we will show later that $Q$ restores
universality and recovers the model-independent effective range,
$R_e$.

\begin{figure}[tb]
  \centering
  \includegraphics[scale=1.1]{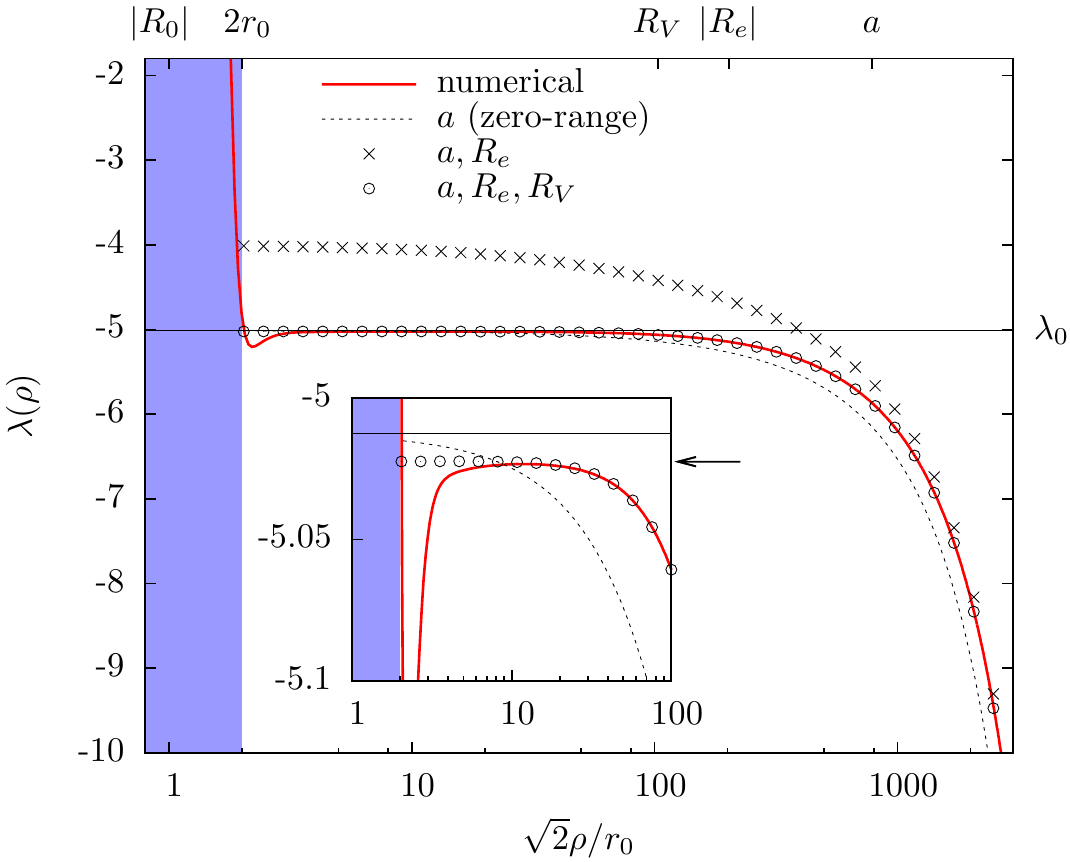}
  \caption{Exact numerical adiabatic eigenvalues $\lambda(\rho)$ for a
    potential with barrier, eq.~\eqref{eq:potential-with-barrier}
    (solid red line), compared to solutions of the eigenvalue
    equation, eq.~\eqref{eq:generalized_efimov_eq2}.  The zero-range
    model (dotted line) includes only $a$.  Crosses include
    $a,R_e$-terms and circles include $a,R_e,R_V $-terms. The inset
    shows details around $\lambda_0$. The arrow indicates the effect
    of the correction $\lambda_0-2R_0 /R_e$. The shaded region is
    $\rho<\rho_c$.}
  \label{fig:lambda-numerics}
\end{figure}

First, to illustrate the necessity of both $1/\rho^2$-terms in
eq.~\eqref{eq:generalized_efimov_eq2} we consider a large negative
effective range corresponding to a narrow Feshbach resonance
\cite{bruun05}. To model the large $\abs{R_e}$ we pick the attractive
potential with barrier, eq.~\eqref{eq:yujun-pot},
\begin{equation}
  \label{eq:potential-with-barrier}
    V(r)=D\ \text{sech}^2\left(\chi\frac{r}{r_0}\right)
        +B \exp\left(-2(\chi\frac{r}{r_0}-2)^2\right),
\end{equation}
where $D=-138.27$, $B=128.49$ in units of $\hbar^2/(mr_0^2)$, and
$\chi=4.6667$. The potential is negligible outside the range $r_0$.
The low-energy parameters are $a=556.88 $, $R_e=-142.86 $, $R_V
=73.031 $, and $R_0 =-0.71 $ in units of $r_0$.  In
fig.~\ref{fig:lambda-numerics} we compare $\lambda(\rho)$ obtained by
exact numerical solution of the Schr{\"o}dinger equation \cite{suno02}
containing the interaction eq.~\eqref{eq:potential-with-barrier} with
the solution of eq.~\eqref{eq:generalized_efimov_eq2}.
In the zero-range model (including only $1/a$), the $-\rho^2$
divergence for large $\rho$ is below the numerical solution. At small
distances, $\lambda$ approaches $\lambda_0$, above the numerical
solution.  Inclusion of the $R_e$-term, as in
\cite{fedorov01,platter09}, provides a better large-distance behavior
(since the dimer binding energy is corrected), but overshoots
dramatically for $\sqrt{2}\rho \lesssim a$ by approaching
$\lambda=-4$.  Including consistently both $R_e$- and $R_V$-terms
leads to complete numerical agreement with the exact numerical
solution except for very small $\rho$-values near $\rho_c$ where
higher-order terms are needed in
eq.~\eqref{eq:generalized_efimov_eq2}.

\section{Non-Adiabatic Corrections}

We now show that the non-adiabatic term restores model-independence
and recovers $\abs{R_e}$ as the limit for the region of Efimov
physics. For simplicity we only consider the limit $\abs{a}=\infty$
and assume $\abs{R_0}\ll\abs{R_e}$.\footnote{This assumption holds for
  the potential eq.~\eqref{eq:yujun-pot} as well as many other
  potentials with large effective range.}  We first consider
$\sqrt{2}\rho\ll\abs{R_e}$ (region A in
fig.~\ref{fig:lambda-analytic}).  Expansion of
eq.~\eqref{eq:generalized_efimov_eq2} to first order in $(\nu-\nu_0)$
gives a small constant correction
\begin{equation}
  \label{eq:nu_efimov_large_Re}
  \nu=\nu_0-\frac{R_0}{\nu_0 R_e}\left(1+O(\frac{\rho}{R_e})\right).
\end{equation}
This correction is marked by the arrow in
fig.~\ref{fig:lambda-numerics} (it is out of the range of
fig.~\ref{fig:lambda-analytic}).  This gives
\begin{equation}
  \label{eq:V_eff_efi_large_Re}
  V_\textup{eff}(\rho)=\frac{\hbar^2}{2m} 
  \left(\frac{\nu_0^2-1/4-2R_0 /R_e}{\rho^2} - Q\right).
\end{equation}
To evaluate $Q$ we note that a large negative effective range (for
$\abs{a}=\infty$) implies that the two-body wave-function $u$ is
localized mainly inside the potential range, as discussed in
chapter~\ref{chap:theory}. Then the angular three-body wave-function
$\Phi$ can be approximated by $u/\sin(2\alpha)$, i.e. $\psi\simeq u$
and neglecting rotated terms $\rot[\psi]$. The result is
$Q=c/\rho^2$, where $c\simeq -5/4$ as confirmed numerically.  This
term cancels the main $1/\rho^2$-part in
eq.~\eqref{eq:V_eff_efi_large_Re} and hence prohibits Efimov physics
for $\sqrt{2}\rho\ll\abs{R_e}$.  The intuitive reason is that the
two-body wave-function is essentially zero outside the potential,
despite the large scattering length, and hence three particles can not
interact at large distances.

When $\sqrt{2}\rho\gg\abs{R_e}$ (region B in
fig.~\ref{fig:lambda-analytic}) we find
\begin{equation}
  \label{eq:nu_efimov}
  \nu=\nu_0+\nu_0 c_0 \frac{R_0}{\sqrt{2}\rho}\left(1+O(\frac{R_e}{\rho})\right),
\end{equation}
where
\begin{equation}
  \label{eq:kE}
    c_0=\frac{\sin(\nu_0\pitwo)/\nu_0}{\frac{4\pi}{3\sqrt{3}}\cos(\nu_0\pisix)
      -\cos(\nu_0\pitwo)+\nu_0\pitwo\sin(\nu_0\pitwo)} \simeq -0.671.
\end{equation}
This gives the effective hyper-radial potential
\begin{equation}
  \label{eq:V_eff_efi}
    V_\textup{eff}(\rho)=\frac{\hbar^2}{2m} \left(\frac{\nu_0^2-1/4}{\rho^2}
    +\frac{c_0\nu_0^2}{\sqrt{2}\rho^3}
    \left( R_e\nu_0^2-2R_V \right) - Q\right).
\end{equation}
The $1/\rho^3$ dependence of the correction to the Efimov potential
$1/\rho^2$ was expected \cite{efimov91}. The model-independent term
proportional to $R_e/\rho^3$ was recently calculated in
\cite{platter09}. However, we also get a model-dependent term
$R_V/\rho^3$ which is of the same order.
$Q$ generally receives contributions both from distances inside and
outside the finite-range potential. Zero-range models only have the
external part of the wave-function, which depends on $\rho$ only though
the eigenvalue $\nu(\rho)$. The zero-range result for $Q$ is then
\begin{equation}
  Q_\textup{ZR} =M_0(\frac{\partial\nu}{\partial\rho})^2 
  =M_0 c_0^2 \nu_0^2 \frac{R_0^2}{2\rho^4},
\end{equation}
where $M_0=\langle
\Phi|\partial^2\Phi/\partial\nu^2\rangle|_{\nu=\nu_0}$. This fourth
order correction can be neglected in eq.~\eqref{eq:V_eff_efi}, as was
done in \cite{platter09}. However, the internal part of the
wave-function contributes to order $1/\rho^3$.
To estimate this $1/\rho^3$-term we take the analytically solvable
finite square well potential of range $r_0$ and $\abs{a}=\infty$. This
fixes $R_e=r_0$ and $R_V=n^2\pi^2r_0/24$ where $n$ is the number of
bound states (including the zero-energy state).  We find
\begin{equation}
  \label{eq:Qbox}
  Q_\textup{box}=c_0\nu_0^2(\frac{R_e}{2}-2R_V)\frac{1}{\sqrt{2}\rho^3},
\end{equation}
neglecting $1/\rho^4$-terms. The model-dependent $R_V$-terms in
eqs.~\eqref{eq:V_eff_efi} and \eqref{eq:Qbox} cancel exactly, giving
\begin{equation}
  \label{eq:V_eff_efi_box}
    V_\textup{eff}^\textup{box}(\rho)=\frac{\hbar^2}{2m} \left(
    \frac{\nu_0^2-1/4}{\rho^2}
    +c_0\nu_0^2(\nu_0^2-\frac{1}{2})\frac{R_e}{\sqrt{2}\rho^3}
    \right).
\end{equation}
So the effective potential receives a $R_e/\rho^3$ correction where
the model-dependent coefficient is different from zero-range models
\cite{platter09} because of the inclusion of $Q$.  We also expect the
$R_V$-terms to cancel for general potentials. In conclusion, the
Efimov effect persists for $\sqrt{2}\rho\gg\abs{R_e}$.

\section{Atom-Dimer Potential}

We have seen that model-dependent corrections to $\lambda_0$ are
cancelled by equivalent terms in $Q$. A similar effect occurs 
for the atom-dimer channel potential.
Suppose the binding energy
is $B_D=\hbar^2k_D^2/m$ with corresponding wave number $k_D>0$. Then
$\nu = i k_D \sqrt{2}\rho$ is an asymptotic solution to
eq.~\eqref{eq:generalized_efimov_eq} and $\lambda$ diverges as
$-\rho^2$ corresponding to a bound dimer and a free particle. For this
solution, the effective range expansion eq.~\eqref{eq:effexp} does not
hold, since asymptotically $k_\rho\to ik_D$ is finite. Instead
eq.~\eqref{eq:modif_angular_eq} reduces to the radial two-body
equation, with a normalized bound state $s$-wave function $u_D(r)$.
Treating $V_\rho-V\propto 1/\rho^2$ as a perturbation gives the
correction
\begin{equation}
  \label{eq:lambda_AD-corrections}
  \frac{\lambda+4}{2\rho^2}=-k_D^2
  -\frac{1}{2}\int_0^\infty \!\!\! r^3 u_D^2\frac{mV'(r)}{6\hbar^2} \ud r \frac{1}{\rho^2} 
  + O(\frac{1}{\rho^4}).
\end{equation}
Since $\rot[\psi]$ is exponentially small for the atom-dimer solution,
$Q$ can be computed using the unperturbed wave-function
$\psi=\sqrt{\rho}u_D(\sqrt{2}\alpha\rho)$, giving
\begin{equation}
  \label{eq:Q00_AD}
  Q=-\frac{1}{4\rho^2}
  +\int_0^\infty \!\! u_D(ru_D'+r^2u_D'')\ud r \frac{1}{\rho^2}  + O(\frac{1}{\rho^4}).
\end{equation}
The term $-1/(4\rho^2)$ for the atom-dimer solution is well-known and
sometimes referred to as the Langer correction \cite{nielsen99b}.

By using the two-body radial equation and partial
integration the two integrals in eqs.~\eqref{eq:lambda_AD-corrections}
and~\eqref{eq:Q00_AD} cancel. Thus the $1/\rho^2$-terms in the
effective potential eq.~\eqref{eq:V_eff} cancel exactly,
giving $V_\textup{eff}(\rho)=-B_D$ up to order $1/\rho^4$. Thus
$V_\textup{eff}$ only depends on $R_e$ through $B_D$.

\section{Applications near Feshbach Resonances}
The effective range near a narrow Feshbach resonance can be estimated
as in eq.~\eqref{eq:reff-fesh-const}.
As an example with a noticeable effect we take the alkali atoms
$^{39}$K with the very narrow Feshbach resonance at $B=825$G having
parameters $\Delta B=-32$mG, $\Delta\mu=-3.92\mu_B$, and
$a_{bg}=-36a_0$ \cite{derrico07}. This gives the large effective range
$R_e=-2.93\times 10^4 a_0$.  For $^{39}$K, $r_0$ is of the order of
the van der Waals length $l_\textup{vdW}=1.29\times 10^2a_0$
\cite{braaten06}.  Since $\abs{R_e}\gg r_0$, $\abs{R_e}$ determines
the lower limit for Efimov physics and corrections to the universal
regime are of order $R_e/a$ (not $l_\textup{vdW}/a$). Thus, the window
for universal physics is reduced.

Another example is the Feshbach resonance at $B_0=-11.7$G ($\Delta
B=28.7$G, $a_{bg}=2.30a_0$, $\Delta\mu=2.3\mu_B$ \cite{chin09}) in
$^{133}$Cs where \cite{kraemer06} apparently have observed a single
Efimov state.  In this case the effective range estimated from the
Feshbach model, eq.~\eqref{eq:reff-fesh-const}, is $R_e=-0.34a_0$.
This is much smaller than $l_{vdW}=202.0 a_0$ \cite{chin09}, thus
$l_{vdW}$ is the dominating scale for effective range corrections.
The linear effective range corrections of chapter~\ref{chap:efimov}
could be applied here.

\section{Conclusions and Outlook}

We have considered a three-body system of identical bosons with large
scattering length modeling a Feshbach resonance. The Efimov physics
occurring in this universal regime is customarily accounted for by
zero-range models. We used the adiabatic hyper-spherical approximation
and derived rigorously a transcendental equation to determine the
adiabatic potential for a general finite-range potential.  We have
solved this equation for large scattering length, investigated
finite-range effects, and compared with exact numerical results.

Inclusion of the effective range correction to the adiabatic
potential is insufficient in general.  Crucial corrections of the same
order must also be included from both a ``scattering length
correction'' ($R_V$) and the
non-adiabatic term.  These two contributions may separately be large
but they tend to cancel each other.
Accurate results in zero-range models must account for these new
corrections.
In conclusion, for large negative effective range the window for
Efimov physics is precisely open between the effective range (not the
potential range) and the scattering length.

%% file: N-efimov.tex
\section{Introduction}

The Efimov effect appears in quantum three-body systems when
attractive interactions between at least two pairs of particles are
such that the scattering length is much larger than the range of the
interaction; in other words two of the three two-body subsystems are
close to the threshold of binding. Under these conditions a
characteristic series of weakly bound and spatially extended Efimov
states appears in the system.  These states appear due to specific
long-range two-body correlations between particles caused by the large
scattering length.
The effect is easiest to see in the hyper-spherical adiabatic
approximation, chapter~\ref{chap:theory}, where the slow adiabatic
variable is the hyper-radius $\rho$.  It has been
shown~\cite{fedorov93} that close to the two-body threshold the
effective adiabatic potential $V_\textup{eff}(\rho)$ is attractive and
asymptotically proportional to the inverse square of the hyper-radius,
\begin{equation}
  \label{eq:Nefi-Veff}
  V_\textup{eff}(\rho)=-\frac{\hbar^2}{2m} \frac{\xi^2+1/4}{\rho^2},
\end{equation}
where $m$ is the atomic mass, and $\xi$ is a constant depending on masses
of the particles~\cite{nielsen01,nielsen98}. For three identical
bosons this constant is given by $\xi_3=1.00624$ leading to the
scaling factors
\begin{equation}
  \label{eq:scalefac}
  S_3=\exp(\pi/\xi_3)\simeq 22.7\quad\text{and}\quad S_3^2\simeq515.0,
\end{equation}
for radii and energies, respectively. This is equivalent to a
geometric series of bound states with exceedingly small energies,
$E_n\propto e^{-2\pi n/\xi_3}$, and exceedingly large root-mean-square
hyper-radii, $\langle\rho^2\rangle_n^{1/2}\propto e^{\pi n/\xi_3}$, where
$n$ is the state number.

It was shown by Amado and Greenwood that in an $N$-body system with
$N>3$ the Efimov effect does not exist at the $N-1$
threshold~\cite{amado73}.  In particular, for $N=4$ there is no
infinite sequence of four-body states at the trimer threshold. This
does not rule out an $N$-body Efimov effect at the two-body threshold
where the scattering length $a$ is infinite. Of course, weakly bound
$N$-body states at $a=\infty$ can not be true bound states, since at
this point the clusters with three and higher number of particles are
generally deeply bound. The states would then be unstable due to lower
lying thresholds and would decay into deeply bound cluster states.

However it has been suggested in~\cite{sorensen02a} that a sequence
of meta-stable $N$-body states with the characteristic exponential
energy dependence can yet show up at the two-body threshold.  Using the
$N$-body hyper-spherical method it has been shown that an $N$-body system
at the two-body threshold has a hyper-spherical adiabatic potential with
inverse-square dependence. This peculiar adiabatic potential appears
due to the same mechanism as for three particles and thus gives rise to
$N$-body Efimov states with a structure similar to that of three-body
Efimov states: An (otherwise) uncorrelated system with very specific
two-body correlations caused by the large scattering lengths.

This specific hyper-spherical adiabatic potential is not the lowest
one as different bound clusters with lower thresholds create lower
lying adiabatic potentials.  However, although not truly bound, these
$N$-body Efimov states might still exist as meta-stable states slowly
decaying into clusters, much like the Bose-Einstein condensate states.
The structure of the Efimov states is determined by the long-range
two-body correlations and is thus quite dissimilar to the structure of
clusterized states with short-range many-body correlations. Therefore
the overlap between Efimov states and clusterized states should be small
and consequently the life-time should be large.

The conclusions about meta-stable $N$-body Efimov states were obtained
in~\cite{sorensen02a} in an hyper-spherical adiabatic approximation
where the couplings to all other channels were neglected.  In this
chapter we report on a more realistic calculation of $N$-body Efimov
states with a different method, namely the two-body correlated
Gaussian stochastic variational method where no adiabatic
approximation is assumed.  We calculate spectra of trapped systems of
$N=3,4,5,6$, and 7 bosons: For each system the calculations reveal a
series of Efimov states where the energies and the radii exhibit the
characteristic exponential dependence upon the state number.  We
obtain the new scaling factors for $N>3$ and compare with analytical
results from the hyper-spherical adiabatic approach. Finally, we
discuss observable consequences.

\section{Numerical Results}

\subsection{The System and Procedure}

We consider a system of $N$ identical bosons with mass $m$ and
coordinates $\mathbf{r}_{i}$ in a spherical harmonic trap with
frequency $\omega$ and trap length $b_t=\sqrt{\hbar/(m\omega)}$.  The
two-body potential is an attractive Gaussian,
eq.~\eqref{eq:gauss-interaction}, with range $r_0$ and depth $V_0$.
As in chapter~\ref{chap:efimov} we choose the trap to be much larger
than the potential range, $b_t/r_0=3965$. The depth is tuned around
$V_0\simeq-2.68\hbar^2/(mr_0^2)$ where the first bound state is at the
threshold, giving $|a|\gg b_t$.

The Schr\"odinger equation for the Hamiltonian in
eq.~\eqref{eq:hamiltonian} is solved numerically with the stochastic
variational method of chapter~\ref{chap:theory}, using the two-body
correlated basis, eq.~\eqref{eq:psi-2B}.  The non-linear parameters in
the wave-function are optimized
stochastically by random sampling in a
region that covers the spatial distances from $r_0$ to $b_t$, which
allows us to describe states over many length scales and energies.

We look for $N$-body Efimov states with the structure analogous to that
of the three-body Efimov states.  The spatial extension of such state
should be much larger than the range of the potential $r_0$ and
smaller than the trap length $b_t$.

\subsection{Universal Energies and Radii}

\begin{figure}[htbp]
  \centering
  \includegraphics{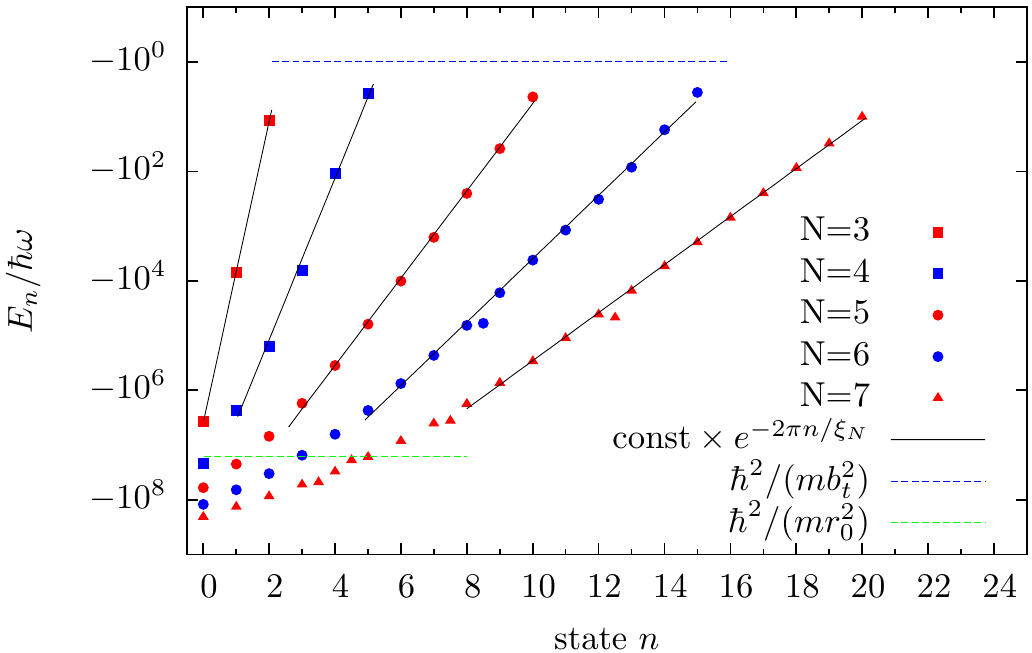}
  \caption{Energy $E_n$ as function of the state number $n$ for
    negative energy states of a system of $N$ bosons in a harmonic
    trap with length $b_t$ interacting via an attractive potential of
    range $r_0$ with scattering length much larger than $b_t$. The
    horizontal lines demarcate the region where the Efimov states can
    exist in trapped gases. The lines show the exponential fits of the
    form $E_n\propto e^{-2\pi n/\xi_N}$ drawn through the points in
    the indicated region.}
  \label{fig:Nefi-energy}
\end{figure}

The energies of Efimov states in a trap should be on one hand much
larger than the typical energy scale of a cluster state,
${\hbar^2/(mr_0^2)}$, and on the other hand smaller than the oscillator
energy $\hbar\omega={\hbar^2/(mb_t^2)}$.  We have calculated the
spectrum for $N=3,4,5,6$, and 7 bosons. The calculated energies are
shown in fig.~\ref{fig:Nefi-energy}. Indeed, in the indicated energy
region each of the $N$-body systems have a series of states with
exponential dependence upon the state number, $E_n\propto e^{-2\pi
  n/\xi_N}$.  The exponential fits give the scaling constants
$S_N=\exp{\pi/\xi_N}$, namely $S_3=23.6(4)$, $S_4=5.47(40)$,
$S_5=2.45(4)$, $S_6=1.93(2)$, and $S_7=1.66(1)$.  The indicated
statistical errors are from the fits, while the uncertainty related to
numerical convergence is not included. This is expected to be largest
for $N=3,4$ which use only a few states for the fit.  The $N=3$ result
agrees with the known analytical result of 22.694 \cite{braaten06}.

\begin{figure}[htbp]
  \centering
  \includegraphics{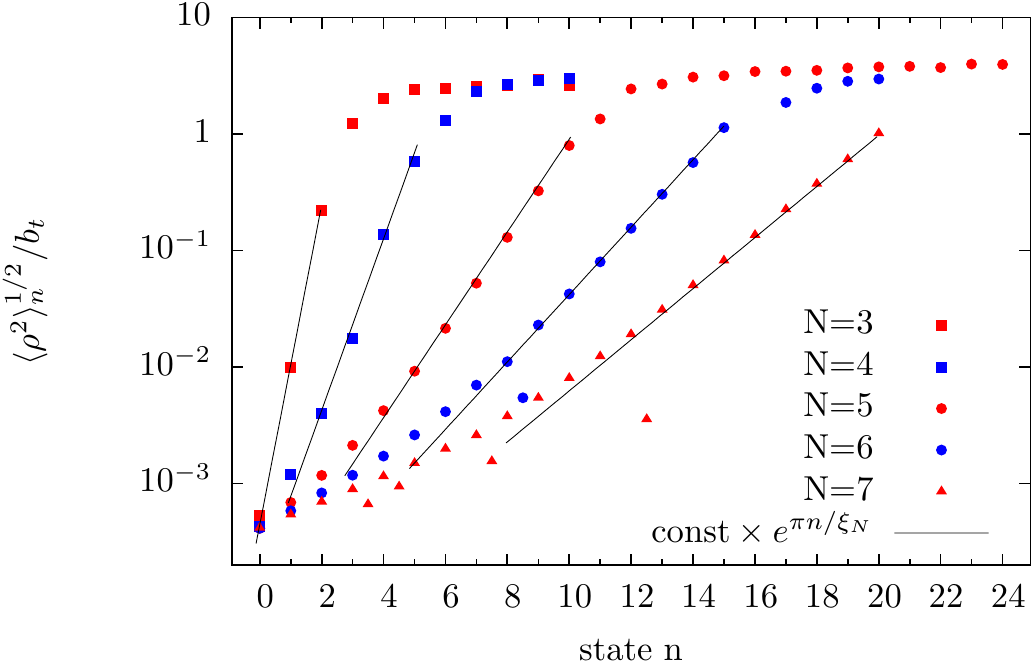}
  \caption{Root-mean-square hyper-radius
    $\langle\rho^2\rangle_n^{1/2}$ as function of the state number $n$
    of a system of $N$ bosons from fig.~\ref{fig:Nefi-energy}.  The
    lines show the fitting curves of the form const$\times e^{\pi
      n/\xi_N}$ drawn through the same points and with the same
    parameters $\xi_N$ as on fig.~\ref{fig:Nefi-energy}.  }
  \label{fig:Nefi-rms}
\end{figure}

The spatial extension of Efimov states in trapped systems must also be
much larger than the interaction range $r_0$ and much smaller than
the trap length $b_t$. The average size of the state can be described
with the ($N$-body) hyper-radius $\rho$, 
\begin{equation}
  \label{eq:Nefi-rho-N}
    \rho^{2} 
    = \frac{1}{N} \sum^N_{i<j} r_{ij}^2
    = \sum^{N}_{i=1}(\bm r_i-\bm R)^2
    = \sum^{N}_{i=1} r^2_i - N\bm R^2,
\end{equation}
where $r_{ij}$ is the pair distance and $\bm R$ the center-of-mass.
On fig.~\ref{fig:Nefi-rms} are shown the calculated r.m.s. hyper-radii
$\langle\rho^2\rangle_n^{1/2}$ as function of state number $n$.  The
radii of the Efimov states identified on fig.~\ref{fig:Nefi-energy}
are reproduced well with the exponentials
$\langle\rho^2\rangle_n^{1/2}\propto e^{\pi n/\xi_N}$ where the
parameters $\xi_N$ are taken from the fits on
fig.~\ref{fig:Nefi-energy}.  Apparently all these states fall within
the correct boundaries and the values of the radii follow the correct
exponential trend.

There are a few states in the $N=6$ and 7 systems with radii much
smaller than those of the typical states in the series with similar
energies, see fig.~\ref{fig:Nefi-rms}. Clearly these states are not
Efimov states but rather relatively compact states with a different
structure. We believe that these states are caused by missing
convergence or numerical inaccuracies.

\section{Comparison with Analytic Results}
The geometric scaling found in the previous subsection indicates that
an underlying $1/\rho^2$ effective potential exists.  In this section
we briefly outline the $N$-body results of \cite{sogo05a,sogo05b} and
find analytical scaling factors $S_N=\exp(\pi n/\xi_N)$ for arbitrary
$N$.

\subsection{$N$-Body Hyper-Radial Potential}

In \cite{sogo05a,sogo05b} the $N$-body zero-range problem was
investigated within the hyper-spherical adiabatic approximation. The
method is equivalent to the $N=3$ discussion in
chapter~\ref{chap:theory}, still using the hyper-radius as the
adiabatic variable and denoting all other coordinates by $\Omega$.
The basic approach is now to use a Faddeev-Yakubovski expansion
$\Phi(\rho,\Omega)=\sum_{i<j}\phi(\rho,r_{ij})$ for the angular
wave-function where each of the identical two-body amplitudes only
depend on the distance between two-particles.
This approximation is equivalent to our two-body correlated basis,
eq.~\eqref{eq:psi-2B}, used in the previous section.

The adiabatic eigenvalue is denoted $\tilde\nu(\rho,N)$ which is similar
to the eigenvalue $\nu(\rho)$ for the three-body case.  
If the effective hyper-radial potential is written in the form
eq.~\eqref{eq:Nefi-Veff}, the relation between $\xi(\rho)$ and
$\tilde\nu(\rho)$ is $4\tilde\nu=5-3N+2\sqrt{-\xi^2}$ (note that
$\xi^2$ is real). By imposing the zero-range boundary condition
similar to eq.~\eqref{eq:zr-boundary-3b}
and collecting similar terms from the two-body amplitudes $\phi$, the
adiabatic eigenvalue equation becomes
\begin{equation}
  \label{eq:Nefi-eq}
  \frac{1}{\sqrt{2}\rho}\frac{B(\tilde\nu)+R(\tilde\nu)}{A(\tilde\nu)}=-\frac{1}{a}.
\end{equation}
This form is similar to eq.~\eqref{eq:zr-efimov-eq} in the three-body case.
The terms $A$ and $B$ are found to be \cite{sogo05a,sogo05b}
\begin{equation}
  \label{eq:Nefi-AB}
  A=-\frac{\sin(\pi\tilde\nu)}{\sqrt{\pi}}
  \frac{\Gamma(\tilde\nu+\frac{3N-6}{2})}
  {\Gamma(\tilde\nu+\frac{3N-5}{2})} ,\qquad
  B=\frac{2\cos(\pi\tilde\nu)}{\sqrt{\pi}}
  \frac{\Gamma(\tilde\nu+\frac{3}{2})}{\Gamma(\tilde\nu+1)},
\end{equation}
while the ``rotated'' terms
$R=2(N-2)R_{34}+\frac{1}{2}(N-2)(N-3)R_{34}$ are
\begin{equation}
  \label{eq:Nefi-R13}
  R_{13}=\frac{2}{\sqrt{\pi}}
  \frac{\Gamma(\frac{3N-6}{2})}{\Gamma(\frac{3N-9}{2})}
  \left(\frac{2}{3}\right)^{(3N-8)/2} I_{13},
\end{equation}
\begin{equation}
  \label{eq:Nefi-R34}
  R_{34}=\frac{4}{\sqrt{\pi}} 
  \frac{\Gamma(\frac{3N-6}{2})}{\Gamma(\frac{3N-9}{2})}
  \left(\frac{1}{2}\right)^{(3N-6)/2} I_{34},
\end{equation}
with the integrals $I_{13}$ and $I_{34}$ given by
\begin{equation}
  \label{eq:Nefi-I13-I34}
  \begin{split}
    I_{13}&=\int_{-1}^{1/2}\!\!\!\ud x\ \sqrt{1+x}\ 
    (\frac{1}{2}-x)^{(3N-11)/2}P_{\tilde\nu}^{(3N/2-4,1/2)}(x),\\
    I_{34}&=\int_{-1}^{1}\ud x\ \sqrt{1+x}\ 
    (1-x)^{(3N-11)/2}P_{\tilde\nu}^{(3N/2-4,1/2)}(x).
  \end{split}
\end{equation}
Here $P_{\tilde\nu}^{(a,b)}$ are the Jacobi functions, see e.g.
\cite[App.~A]{nielsen01}. Equation~\eqref{eq:Nefi-eq} defines
$\xi(\rho)$ and hence the effective hyper-radial potential.

\subsection{Scaling Factors}

\begin{figure}[htbp]
  \centering
  \includegraphics{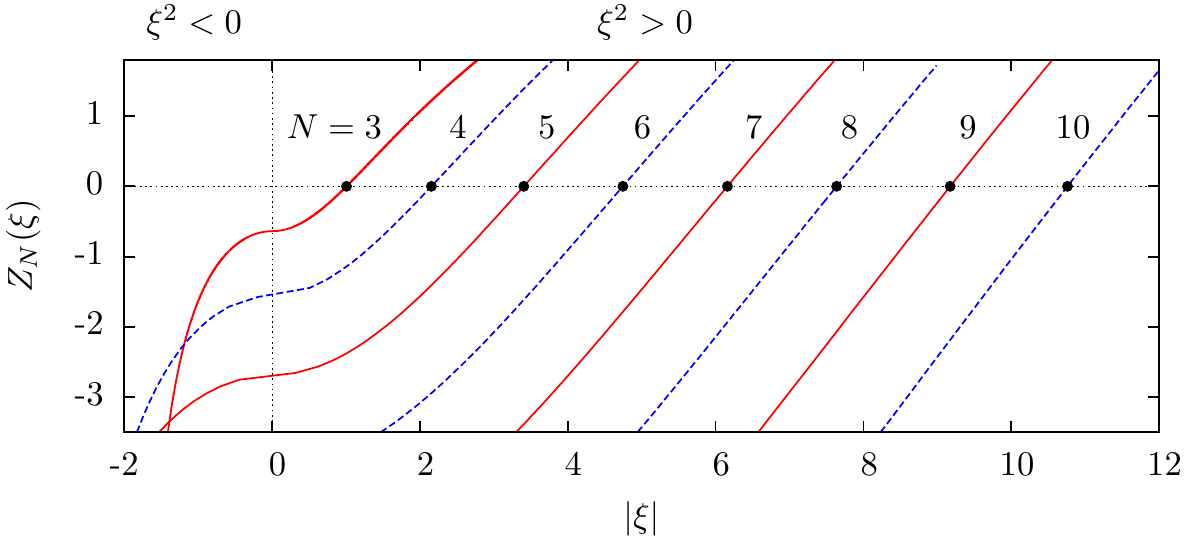}
  \caption{Roots $\xi_N$ of the function $Z_N(\xi)$ for boson number
    $N=3,\dots,10$. Only the lowest adiabatic potential (largest
    $\xi$) is shown.}
  \label{fig:Nefi-zeros}
\end{figure}

The solutions of eq.~\eqref{eq:Nefi-eq} was discussed in
\cite{sogo05a,sogo05b}, but only in the asymptotic limit
$N\to\infty$. We now solve the equation numerically for finite $N$ and
infinite scattering length. The solutions are defined by the roots of
the real function
\begin{equation}
  \label{eq:Nefi-Z}
  Z_N(\xi^2)=-\frac{B(\xi^2)+R(\xi^2)}{A(\xi^2)},
\end{equation}
which is calculated numerically and shown in fig.~\ref{fig:Nefi-zeros}
for $N=3,\dots,10$. Each root defines a constant solution $\xi_N$,
leading to the corresponding scale factor $S_N=\exp(\pi/\xi_N)$
(see section~\ref{sec:efimoveffect-inf-a}). 

\begin{figure}[htbp]
  \centering
  \includegraphics[scale=0.9]{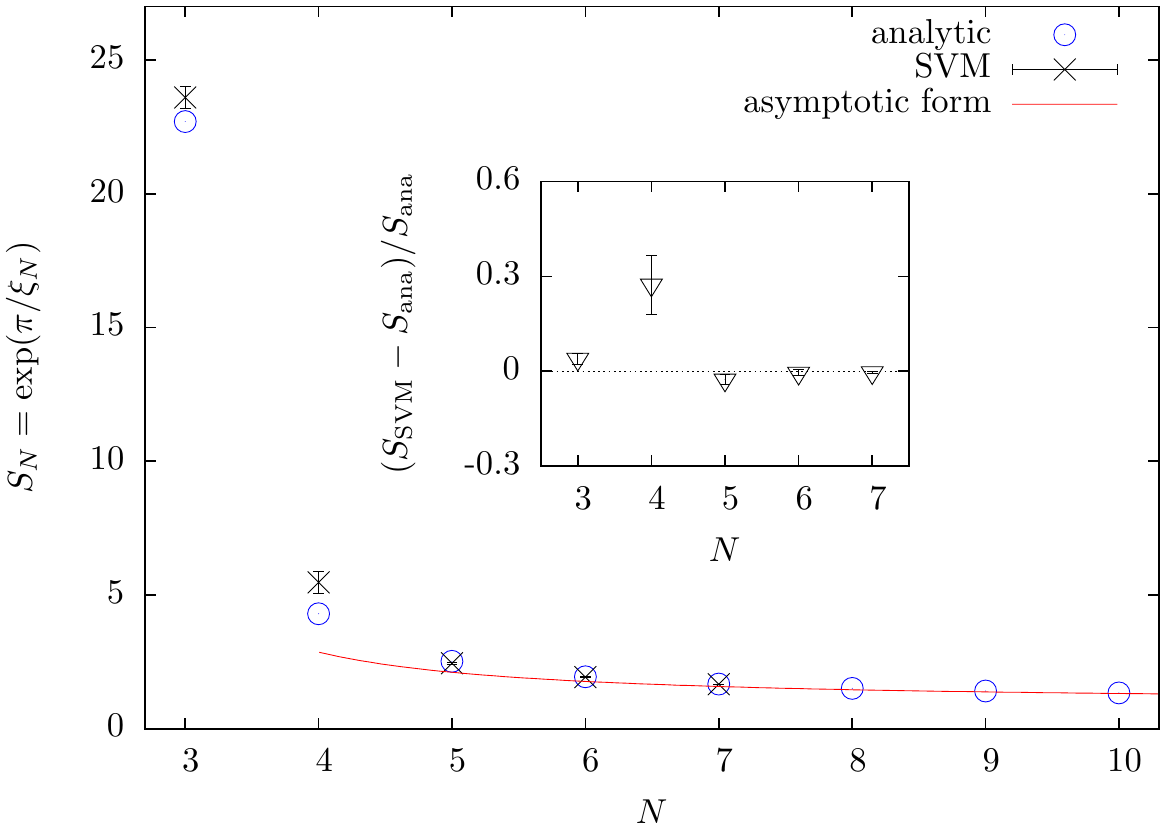}
  \caption{The scaling factors $S_N$ as function of the boson number
    $N$.  Both analytical roots and numerical (SVM) results are shown.
    The $N=3$ value is the well-known factor $22.7$. The asymptotic
    form is given by eq.~(\ref{eq:zN}).  }
  \label{fig:Nefi-scalefac}
\end{figure}

The values are summarized in tab.~\ref{tab:scaling} and in
fig.~\ref{fig:Nefi-scalefac}, where we also compare with the numerical
results. There is generally very good agreement between analytic and
numerical results. We note that the scaling factors decrease with
increasing $N$. Already for four particles it has been reduced to 4.29
as compared to 22.7.
We also show the asymptotic expression for large $N$,
\begin{equation}\label{eq:zN}
  \xi_N =\sqrt{
    \frac{5}{3}N^\frac{7}{3}\left(1-\frac{2}{N}\right)
    -\frac{(3N-4)(3N-6)}{4}-\frac{1}{4}
  }\; ,
\end{equation}
which was established in~\cite{sogo05a,sorensen03}. Our results are
consistent with the asymptotic estimate.

\begin{table}[htbp]
  \centering
  \begin{tabular}{c|ccc|c}
    \hline\hline 
    N & $\xi_N$ & $S_{N}$ & $S_{N}^{2}$ & $S_{N}$(SVM)\\
    \hline \hline 
    3 & 1.00624 & 22.6942 & 515.028 & $23.6(4)\ $\\
    4 & 2.15584 & 4.29413 & 18.4396 & $5.47(40)$\\
    5 & 3.40602 & 2.51523 & 6.32638 & $2.45(4)\ $\\
    6 & 4.74528 & 1.93875 & 3.75877 & $1.93(2)\ $\\
    7 & 6.15939 & 1.66537 & 2.77347 & $1.66(1)\ $\\
    8 & 7.63821 & 1.50878 & 2.27641 & --\\
    9 & 9.17426 & 1.40837 & 1.98352 & --\\
    10& 10.7618 & 1.33900 & 1.79291 & --\\
    \hline\hline
  \end{tabular}
  \caption{Scaling factors $S_N=\exp(\pi/\xi_N)$ for the $N$-body
    Efimov effect. Both analytical and numerical (SVM) results are
    shown. Indicated errors are from the fits in
    fig.~\ref{fig:Nefi-energy}. The deviation from the analytical results
    is largest for $N=3,4$ which use only a few states for the fit.}
  \label{tab:scaling}
\end{table}

\section{$N$-Body Recombination}

The $N$-body Efimov states could in principle be identified by their
contribution to the recombination rates in a cold gas as function of
scattering length. As the scattering length is increased (using the
Feshbach technique), the $N$-body Efimov states should produce peaks
in the recombination rate when crossing the threshold similar to the
three-body case. Such features would occur at
critical scattering lengths spaced by the scaling factors $S_N$.

The non-resonant background rate for $N$-body recombination can be
estimated as follows.  Let us first consider a three-body reaction.
The rate of the loss of particles from a cold gas due to the
three-body recombination reaction into a shallow dimer with the energy
$\hbar^{2}/(ma^{2})$ is given by Fermi's golden rule,
\begin{equation}
  \label{eq:dndt}
  -\frac{dN_{0}}{dt}=
  3\frac{N_{0}^{3}}{6}\frac{2\pi}{\hbar}\left|T_{fi}\right|^{2}
  \frac{d\nu_{f}}{dE_{f}} ,
\end{equation}
where the factor 3 is there since each recombination reaction removes
three particles from the cold gas, $N_{0}^{3}/6$ is the number of
triples in the gas of $N_{0}$ particles, $d\nu_{f}$ is the number of
final states (dimer plus atom) with relative momentum
$q_{f}=2/(\sqrt{3}a)$ and the relative kinetic energy
\begin{equation}
  E_{f}\equiv\frac{\hbar^{2}q_{f}^{2}}{2(\frac{2}{3}m)}=
  \frac{\hbar^{2}}{ma^{2}},
\end{equation}
and $T_{fi}$ is the transition matrix element from the initial
three-body state to the final atom-dimer state. The number of final
states is given by
\begin{equation}
  d\nu_{f}=\frac{V d^{3}q_{f}}{(2\pi)^{3}}
  =\frac{2}{3\sqrt{3}\pi{}^{2}}\frac{V m}{\hbar^{2}a}dE_{f}\;.
\end{equation}
In the typical experimental regime, where the scattering length is
still much smaller than the size of the trap, $a\ll b_{t}$, the
transition matrix element for the non-resonant three-body
recombination rate from a cold gas state into a shallow dimer state
can be estimated perturbatively, substituting the asymptotic
expressions for the initial and final wave-functions,
\begin{equation}
  \label{eq:tfi1}
  \begin{split}
    T_{fi}&=\int \ud \bm r \ud \bm R  \left[
      \psi_{\mathrm{d}}(r)\frac{e^{i\mathbf{q}_{f}\mathbf{R}}}{\sqrt{V}}
    \right]\\
    &\times\left(  U(\mathbf{R}-\frac{\mathbf{r}}{2})
      +U(\mathbf{R}+\frac{\mathbf{r}}{2}) \right)
    \left[ \frac{e^{i\mathbf{k}\mathbf{r}}}{\sqrt{V}}
       \frac{e^{i\mathbf{q}\mathbf{R}}}{\sqrt{V}} \right] ,
  \end{split}
\end{equation}
where $\mathbf{r}$ is the distance between two particles, $\mathbf{R}$
is the distance between their center-of-mass and the third particle,
$V\propto b_{t}^{3}$ is the normalization volume, $k$ and $q$ ($k\sim
q\propto b_{t}^{-1}\ll a^{-1}$) are the initial momenta of the cold
gas particles, and $\psi_{\mathrm{d}}(r)$ is the $s$-wave function of
the shallow dimer with binding energy $\hbar^{2}/(ma^{2})$.  Using the
zero-range approximation for the transition interaction,
$U(\mathbf{r})=U_0\delta(\mathbf{r})$, $U_0=4\pi\hbar^2 a/m$ and for
the dimer wave-function, $\psi_{\mathrm{d}}(r)\propto \exp(-r/a)/(ra^{1/2})$, the matrix element
eq.~\eqref{eq:tfi1} in the limit $k\sim q\ll a^{-1}$ is estimated as
(cf. \cite{fedichev96})
\begin{equation}
  T_{fi}\propto\frac{\hbar^{2}a^{5/2}}{V^{3/2}m}.
\end{equation}
Finally, the non-resonant factor of the three-body recombination
rate becomes (cf. \cite{nielsen01,fedichev96})
\begin{equation}
  \left.-\frac{dn}{dt}\right|_{\mathrm{3-body}}\propto n^{3}\frac{\hbar
    a^{4}}{m}\;,
\end{equation}
where $n=N_{0}/V$ is the density.

For an $N$-body recombination reaction into a shallow $N-1$ body
Efimov state with the binding energy of the order of
$\hbar^{2}/(ma^{2})$ the modifications to the expression for the rate
in eq.~\eqref{eq:dndt} include the extra $3(N-3)$ spatial dimensions
in the integral in eq.~\eqref{eq:tfi1}, which gives an extra factor
$(a^{3/2}V^{-1/2})^{2(N-3)}$, and also the factor $N_{0}^{3}/6$ is
substituted by $N_{0}^{N}/N!$. Thus for the $N$-body recombination
rate ($N\ge 3$) we have
\begin{equation} 
  \label{eq:N-body-rate}
  \left.-\frac{dn}{dt}\right|_{\mathrm{N-body}}
  \propto  n^{3}\frac{\hbar a^{4}}{m}(na^{3})^{N-3}\;.  
\end{equation}
We emphasize that these simple estimates only refer to the
non-resonant contributions and they also do not include other types of
decays where the final state structure might be substantially
different from the shallow $N-1$ cluster plus one particle.  The
result eq.~\eqref{eq:N-body-rate} was recently confirmed in \cite{metha09}.

Although in the regime $na^{3}\ll1$ the $N$-body recombination has an
additional small factor $(na^{3})^{N-3}$ it might still be possible to
observe the $N$-body Efimov states as a sequence of resonant peaks in
the recombination rate as function of scattering length, scaling as
$S_N$.

\section{Connection to Universal Four-Body States}

Recent theoretical work \cite{vonstecher09,dincao09,metha09} has
predicted a universal four-body effect where a single atom is loosely
bound to an Efimov trimer. For each trimer energy two four-body states
should exist below. This leads to two extra recombination peaks at the
critical scattering lengths \cite{vonstecher09}
\begin{equation}
  \label{eq:efimov-atom}
  a_{4B,1}\simeq0.43a_{3B},\qquad a_{4B,2}\simeq 0.90a_{3B},
\end{equation}
relative to the Efimov trimer resonance $a_{3B}$ (denoted $a_Z^{(n)}$
in chapter~\ref{chap:theory}). The second state $a_{4B,2}$ is very
close to the Efimov trimer. Two such features were reported
experimentally in \cite{ferlaino09}.

The universal scaling factor for this atom-Efimov effect is identical
to the three-body case, i.e. $a_{4B,j}^{(n+1)}/a_{4B,j}^{(n)}=22.7$
for $j=1,2$, since these positions simply scale with the three-body
states. This is different from our four-body Efimov effect which have
the scaling factor $S_4=4.29413$.  However, we note that twice the
ratio of the two special four body states is $2 a_{4B,2} /
a_{4B,1}\simeq 4.2$. This is very close $S_4$ and probably within the
numerical accuracy of \cite{vonstecher09}. It could indicate an
underlying fundamental relationship or be a mere coincidence.

We point out that the two four-body effects are distinct, since
different degrees of freedom are considered.  For both of the effects
the four-body states are embedded in the complete Efimov spectrum and
in this respect all four-body states are meta-stable states. To treat
these states as bound states one must pick out the relevant degrees of
freedom, and this is where the models differ.  The degrees of freedom
in the atom-Efimov model is obvious, while the four-body Efimov effect
presented in this chapter includes full two-body correlations.

Both effects could be relevant for Borromean four-body systems, as also
pointed out in \cite{metha09}. The connections and applications need
to be explored.

\section{Conclusions and Outlook}
We have calculated the spectrum of trapped $N$-boson systems with
$N=3,4,5,6,7$ using the stochastic variation method with a correlated
Gaussian basis. Only two-body correlations were allowed in the
variational space.  Thus the cluster states were a priori mostly
excluded from the variation space, which made the calculations
technically possible.
Inclusion of the cluster states would turn the Efimov states into
meta-stable states. However the life-time of these states should be
comparable with that of Bose-Einstein condensates which have 
similar structure and decay modes.

For each system a series of states is found with specific exponential
dependence of the energies and r.m.s hyper-radii on the state number,
which is a characteristic feature of Efimov states. For the $N=3$
system the obtained scaling factor agrees well with the known
analytical value.
We also calculated the scaling factors analytically for finite $N$ by
employing a specific Faddeev-Yakubovski decomposition in two-body
amplitudes. The results are in perfect agreement with the numerical
results.

It might be possible to observe the four-body Efimov states as peaks
in the recombination rate of a condensate as function of the
scattering length, similar to the three-body case. For a dilute gas
the four-body recombination rate is a factor $na^3$ smaller than the
three-body rate, therefore the accuracy constraints on the experiment
would be higher.  On the other hand, the large scaling factor 22.7 for
three-body recombination is reduced to 4.29 for four-body
recombination.  This would make the universal scaling easier to
observe. The four-body case may be related to other recently
investigated atom-Efimov states and could be applicable to Borromean
systems.

We finally note that if the $N$-body Efimov states are found, the
higher-order corrections can be derived by combining the
Faddeev-Yakubovski decomposition of this chapter with the analysis of
chapter~\ref{chap:efimov2}. This would lead to shifts in the critical
scattering lengths as discussed in chapter~\ref{chap:efimov}.

In conclusion we have lent theoretical support to the possibility of
existence of long lived meta-stable $N$-body Efimov states in trapped
Bose gases.

%% file: correlations.tex
\section{Introduction}

The density of trapped cold gases is generally low under typical
experimental conditions \cite{pethick02}, such that the parameter $n
r_0^3$ is small, where $n$ is the particle density and $r_0$ is the
range of the inter-particle potential. In other words the typical
distance between particles is much larger than the range of the
potential, and the typical relative momentum between particles is much
smaller than the inverse range of the potential.

In this regime the system of particles exhibits universality (also
called model-independence or shape independence) \cite{braaten06}. The
system is not sensitive to the details of the potential and the
properties of the system are essentially determined by only a few
low-energy parameters of the potential. Very different interaction
models then provide quantitatively similar results as soon as the
low-energy parameters are the same.

In a two-body system the universality is manifested in the well known
effective range expansion, where the low-energy $s$-wave phase shift
$\delta$ is determined by only two parameters, the scattering length $a$
and the effective range $R_e$, see eq.~\eqref{eq:eff-range-exp}.
The effective range is typically of the order of the range of the
potential while the scattering length can vary greatly.
In three-body systems the universality manifests itself in the
Thomas collapse \cite{thomas35}, the Efimov effect \cite{efimov70}, the
Phillips line \cite{chen91,bedaque00,fedorov02}, and other low-energy
phenomena \cite{nielsen99b}, also in two dimensions \cite{nielsen97}.

Cold gases also exhibit universality: In the dilute limit their
properties, in particular the energy per particle, are independent of
the shape of the inter-particle potential and are determined by the
scattering length alone. This universality is customarily employed by
using the zero-range (pseudo) potential for theoretical descriptions
of Bose-Einstein condensates. The zero-range potential has only
one parameter, the scattering length. Combined with the Hartree-Fock
product wave-function the zero-range potential leads to the
Gross-Pitaevskii equation \cite{dalfovo99, pitaevskii03}.

The limits of the zero-range model have been tested by numerical
calculations with finite-range potentials.  In particular, repulsive
potentials have been employed with Monte-Carlo methods
\cite{blume01,giorgini99,dubois01} in the case of large positive
scattering length.  These investigations showed that as the scattering
length is increased the energy of the Bose gas with a repulsive
potential exceeds the zero-range predictions and the condensate
fraction becomes considerably depleted.

However, repulsive potentials have a problem when modeling an
increasingly large positive scattering length: The potential range,
$r_0$, has to be increased essentially linearly with the scattering
length. This does not seem to match the experimental conditions where
the scattering length is adjusted by tuning the atomic resonances in
an external magnetic field. The range of the inter-atomic interaction
is then left essentially unchanged.

Instead, an attractive finite-range potential might be a more
realistic interaction model for descriptions of trapped Bose gases
with Feshbach resonances.  Indeed, with an attractive potential an
arbitrary large positive scattering length can be achieved by
fine-tuning the energy of the bound two-body state around the
threshold, while maintaining the given realistic potential range.
However, attractive potentials with bound states bring in a major
complication for numerical calculations: A large number of many-body
self-bound negative-energy states appears in the system and the
condensate state in the trap becomes a highly excited state.

For a homogeneous Bose gas an approximate Jastrow-type wave-function
was employed where the pair-correlation function was essentially a
solution of the two-body equation \cite{cowell02,giorgini99}.  In
contrast we propose a direct numerical diagonalization of the
many-body Hamiltonian where the condensate state of trapped bosons
appears as a many-body excited state which is automatically orthogonal
to all the self-bound negative-energy states. This proposal opens up
the possibility to investigate the correlations which are completely
absent in standard zero-range mean-field models.

When correlations are allowed in the BEC wave-function these may induce
a depletion of the fully condensed state. The complete qualitative
connection between the correlations and the condensate fraction still
needs to be understood in detail.

The purpose of this chapter is to investigate the energy,
correlations, and condensate fraction of a system of trapped bosons.
The interaction is attractive and supports a single two-body state,
i.e. the scattering length is positive.  We will show that when the
scattering length is small compared to the trap length the system is
model-independent: All potential models -- attractive, repulsive and
zero-range -- provide similar results.  When the scattering length is
large the attractive model differs qualitatively from the repulsive
and zero-range models. In this regime the system with attractive
potential becomes independent of the scattering length, with both the
energy and the condensate fraction converging towards finite
constants. The pair-correlation function is shown to reflect the
two-body interaction at short distances and the mean-field solution at
larger distances.

\section{Condensation and Correlations}
Let us start by introducing the rigorous definitions of condensation
and correlations. This is done via the one-body density matrix and the
two-body correlation function. The results apply to the $T=0$
formalism used in this dissertation.  We also comment on some
subtleties for condensation in attractive systems.

\subsection{One-Body Density Matrix and Condensate Fraction}
Assume we have a pure state described by the symmetric many-body
wave-function $\Psi(\bm r_1,\dots,\bm r_N)$, which is normalized to
$N$.  The one-body density matrix (OBDM) is then defined as
\begin{equation}
  \label{eq:obdm}
  n^{(1)}(\bm r_1,\bm r_1')=\int 
  \Psi^*(\bm r_1,\bm r_2,\dots,\bm r_N)
  \Psi(\bm r_1',\bm r_2,\dots,\bm r_N)\ud\bm r_2\dots\ud\bm r_N.
\end{equation}
The diagonal part of the OBDM is the density, $n(\bm r)=n^{(1)}(\bm
r,\bm r)$, which ensures that the trace equals $N$. The OBDM is
hermitian so we can diagonalize it,
\begin{equation}
  \label{eq:obdm-eigeneq}
  \int n(\bm r,\bm r') \chi_i(\bm r')\ud \bm r'=N_i\chi_i(\bm r),
\end{equation}
obtaining a complete set of orthonormal eigenfunctions $\chi_i(\bm r)$
and corresponding eigenvalues $N_i$. The OBDM can then be expanded
uniquely as
\begin{equation}
  \label{eq:obdm-eigenfunction-expansion}
  n(\bm r,\bm r')=\sum_i N_i \chi_i^*(\bm r)\chi_i(\bm r').
\end{equation}
The eigenvalues $N_i$ are non-negative,
and taking the trace of the OBDM in the $\chi$-basis gives
\begin{equation}
  \label{eq:obdm-trace}
  \sum_i N_i=\int n(\bm r,\bm r)\ud\bm r=N.
\end{equation}
This allows us to interpret $N_i$ as the occupation numbers of the
single particle states or natural orbitals $\chi_i$.  We can now
define a system to be a Bose-Einstein condensate if the largest
eigenvalue, say $N_0$, is of order $N$, while all other eigenvalues
are of order $1$ \cite{penrose56,yang62}. This means that just one
single-particle state, $\chi_0$, is macroscopically occupied. The
ratio $\lambda_0=N_0/N$ is referred to as the condensate fraction.  

If more than one single-particle state is macroscopically occupied,
the system is called a fragmented condensate \cite{baym01,mueller06}.
The factorization of the OBDM into a single product $\chi^*(\bm
r)\chi(\bm r')$ is also referred to as perfect first order coherence
\cite{naraschewski99}.

From the OBDM we can get the expectation values of all one-body
observables, such as position or momentum density.
Equation~\eqref{eq:obdm} can be interpreted as the amplitude to stay
in the same state, if we destroy a particle at $\bm r_1'$ and create
one at $\bm r_1$. It can be related to the contrast of interference
fringes measured in spatial coherence experiments \cite{bloch00}.

In practice we calculate the condensate fraction by expressing the
OBDM in the harmonic oscillator basis and diagonalize numerically.
Details can be found in appendix~\ref{chap:appendix-obdm}.

\subsection{Condensation in Attractive and Bound Systems}
\label{sec:cond-self-bound}
The definition of condensation introduced above is the conventional
one.  However, \cite{wilkin98} considered a gas of weakly attractive
bosons in a harmonic trap with fixed angular momentum. They found that
the angular momentum is absorbed by the center-of-mass motion, and
that the system is fragmented, i.e.  non-condensed, according to the
definition used above.  This is quite surprising, because the system
is just a Bose-Einstein condensed ground state performing a
center-of-mass motion.  Ref.~\cite{pethick00} then proposed an
alternative definition in terms of an internal density matrix, where
the center-of-mass is integrated out. According to this definition the
system will be a pure condensate. However, \cite{gajda06} argued that
the internal density matrix is only useful when describing observables
that are independent of the center-of-mass. It is not useful when
measuring global properties of a gas such as condensation.

These subtleties are easily explained in terms of conditional
measurements. In a perfect BEC two subsequent measurements of
individual particles are completely uncorrelated. The spatial
probability distribution for the first particle is determined by the
mean-field density, and so is the second.  However if all the particles
move around in a correlated way, e.g. because the system is bound, a
measurement of the first particle will determine the systems
center-of-mass with high accuracy. A measurement of a second particle
will be close to the center-of-mass, and condensation is lost.

A concrete example on this is given in
appendix~\ref{chap:appendix-corr}, where we show that strong
hyper-radial correlation can deplete the condensate completely if the
internal length scale is very different from the center-of-mass scale.
Also see appendix~\ref{chap:appendix-corr} for a discussion on how
other types of correlation affects the condensate fraction.

In the light of all these subtleties we have chosen to use the
conventional criterion for Bose-Einstein condensation.

\subsection{Two-body Correlation Function}
The two-body correlation function is defined as
\begin{equation}\label{eq:tbcf}
  n^{(2)}(\bm r_1,\bm r_2)
  =\int\Psi^*(\bm r_1,\bm r_2,\dots,\bm r_N)
  \Psi(\bm r_1,\bm r_2,\dots,\bm r_N)\ud\bm r_3\dots\ud\bm r_N,
\end{equation}
and is interpreted as the joint probability of detecting two particles
at $\bm r_1$ and $\bm r_2$ simultaneously. If the particles are
totally uncorrelated the two-body correlation function factorizes as
$n^{(2)}(\bm r_1,\bm r_2) = n(\bm r_1)n(\bm r_2)$. The normalized
two-body correlation function\footnote{Simply referred to as the
  (two-body) correlation function below. Note that several different
  (but equivalent) definitions and notations are used in the
  literature.} is defined as
\begin{equation}\label{eq:normalized-tbcf}
  C^{(2)}(\bm r_1,\bm r_2)= \frac{n^{(2)}(\bm r_1,\bm r_2)}{n(\bm r_1)
    n(\bm r_2)}.
\end{equation}
It describes the tendency of particles to cluster ($C^{(2)}>1$)
or to stay separated ($0\le C^{(2)}<1$). If no two-body correlations are
present, $C^{(2)}$ equals $1$ identically.

In a homogeneous gas the correlation function
\eqref{eq:normalized-tbcf} is translational invariant and depends
only on $r_{12}=\lvert\bm r_1-\bm r_2\rvert$.  In a spherically
symmetric trap it also depends on the lengths $r_1$ and $r_2$.
However, for relative distances $r_{12}$ much smaller than the density
variations and the trap length, the correlation function will only
depend on $r_{12}$. Thus we will typically take $\bm r_2=0$.

The correlation function in quantum gases can be measured via noise
correlations \cite{altman04,bloch08}. The experimental procedure is to
release the gas from the trap and let it expand. The expanded density
profile is then measured by light absorption imaging, and correlations
in the noise are extracted. It is preferable that the gas is weakly
interacting during the expansion. For Feshbach resonance experiments
this is usually obtained by also ramping the scattering length down
to zero when the trap is turned off.

The technique of noise correlations have recently been used to
measure the correlations for fermionic gases and for bosons in optical
lattices \cite{altman04,bloch08}, but it also works for normal trapped
BECs.

\section{System and Numerical Techniques}

We consider a system of $N$ identical bosons with mass $m$ and
coordinates $\mathbf{r}_i$, $i=1,\dots,N$, in a spherical harmonic
trap with frequency $\omega$. The Hamiltonian of the system is given
by eq.~\eqref{eq:hamiltonian}, where the system parameters are taken
from \cite{blume01}: $m=86.909$amu ($^{87}$Rb) and
$\omega=2\pi\times77.87$Hz, giving the trap length
$b_t=\sqrt{\hbar/(m\omega)}=23095$au=$1.22\mu$m.

The zero-range model potential, eq.~\eqref{eq:zero-range-potential},
has the scattering length $a$ as the only length parameter. For dilute
bosonic systems this parameter is customarily chosen to be equal to
the inter-atomic scattering length.  The zero-range potential provides
the correct low-energy scattering amplitude in the first order Born
approximation.  The zero-range potential can only be used with an
appropriate non-correlated functional space, like the Hartree-Fock
product wave-functions. Using the zero-range potential in a correlated
space will lead to a complete Thomas collapse of the system
\cite{fedorov01a}.

For the finite-range potential model we use a Gaussian,
eq.~\eqref{eq:gauss-interaction}, with the range $r_0=11.65$au and a
varied negative strength $V_0$. The variation of the strength is
limited to the region where the potential provides exactly one
two-body bound state and has a positive scattering length. The range
of the interaction is much smaller than the trap length, i.e.
$r_0/b_t=5.04\times 10^{-4}$.

Our numerical procedure is the stochastic variational method described
in chapter~\ref{chap:theory}. The wave-function of the system is
represented as a linear combination of basis-functions taken in the
form of symmetrized correlated Gaussians, eq.~\eqref{eq:psi-full}. The
non-linear parameters $\alpha_{ij}^{(k)}$ in eq.~\eqref{eq:psi-full}
are optimized stochastically by random sampling from a region that
covers the distances from $r_0$ to $b_t$.  The center-of-mass motion
is assumed to be in the oscillator's ground state.

The zero-range potential eq.~\eqref{eq:zero-range-potential} requires
an uncorrelated variational wave-function, and we use the hyper-radial
wave-function $\Psi_\rho$ in eq.~\eqref{eq:psi-rho}.  The zero-range
potential with this variational space provides results similar to
the Gross-Pitaevskii equation \cite{sorensen05a}.

The calculation of a highly excited state with the fully correlated
basis $\Psi_\text{full}$, eq.~\eqref{eq:psi-full}, is a difficult
numerical task and is only possible for relatively small number of
particles.
However, for a typical system of trapped atoms even when the scattering
length is large the density of the system remains small, $n
r_0^3\ll1$, and one can assume that only binary collisions play a
significant role in the system's dynamics.  In this approximation the
variational wave-function can be simplified by only allowing two-body
correlations in the basis-functions as in $\Psi_{2B}$,
eq.~\eqref{eq:psi-2B}. The symmetrization of this function can be
done analytically \cite{sorensen05a} which greatly simplifies the
numerical calculations.

During the calculation of a given system the number of Gaussians in the
basis is increased and the stochastic optimization is carried out until
the number of negative energy states and the energy of the lowest state
with positive energy is converged.  The convergence within four digits
typically requires about $5\times 10^2$ Gaussian and about $10^5$
random trials per nonlinear parameter.
When the energies are converged, the relevant properties of the
wave-function can be extracted. We choose to consider densities,
condensate fractions, and two-body correlation functions.

\section{Results}

\subsection{Obtaining the BEC State}

For repulsive potential models it is simply the ground state of the
trapped many-boson system that is identified as the Bose-Einstein
condensate (BEC) state.
With deep attractive two-body potentials, however, the many-body
system in a trap has a large number of self-bound negative-energy
states and identification of the BEC-state is not obvious.

\begin{figure}[htb]
  \centering
  \includegraphics{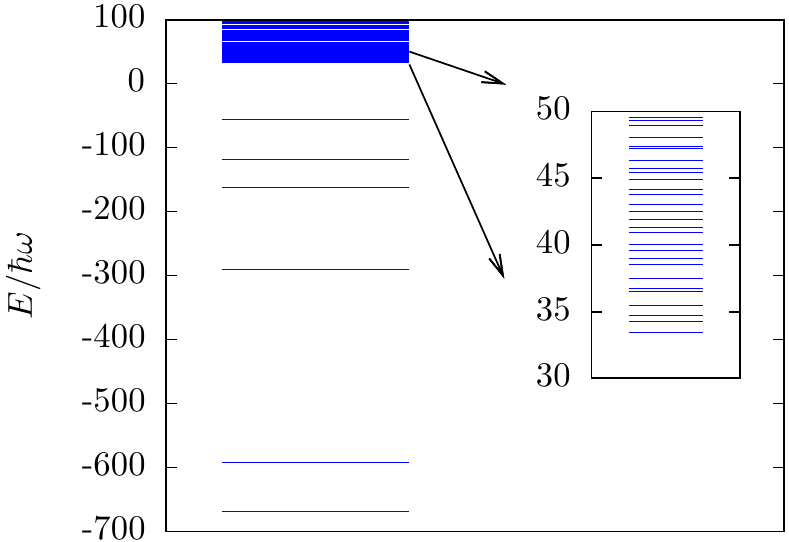}
  \caption{A typical spectrum (in the vicinity of zero energy) of a
    system of $N=20$ bosons in an oscillator trap, eq.~\eqref{eq:hamiltonian},
    interacting via an attractive two-body potential,
    eq.~\eqref{eq:gauss-interaction}, with one bound state and a
    positive scattering length. The inset shows the beginning of the
    so-called quasi-continuum spectrum.}
  \label{fig:spe}
\end{figure}

The typical spectrum of a trapped many-body system with attractive
potentials and including two-body correlations is shown in
fig.~\ref{fig:spe}. The system has a number of deeply bound states
with negative energies and then a positive quasi-continuum spectrum
which starts at about $(3N/2)\hbar\omega$ and has the
characteristic distance between levels of the order $\hbar\omega\ll
\hbar^2/(2m r_0)^2$.
Apparently the BEC-state should then be the lowest state of the
quasi-continuum spectrum or, equivalently, the lowest state with
positive energy.

\begin{figure}[htbp]
  \centering
  \includegraphics{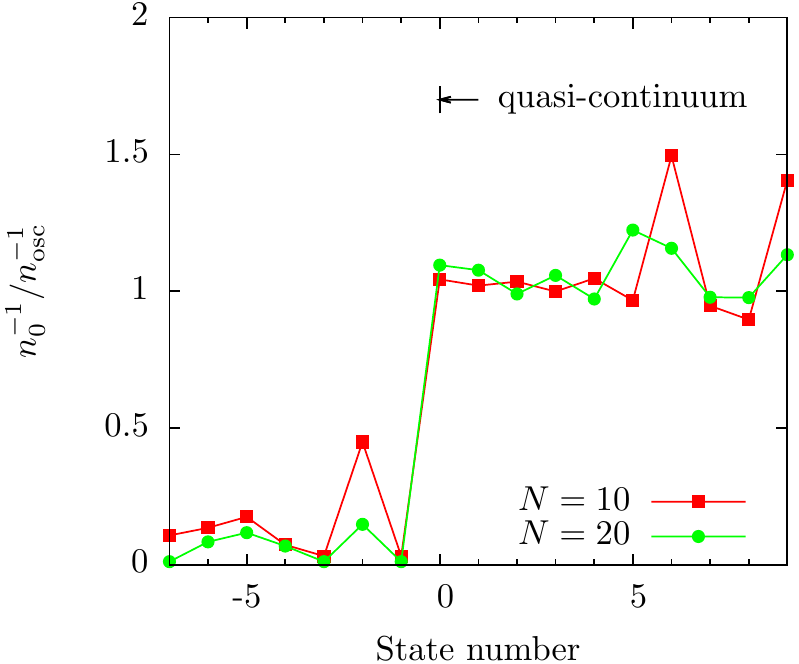}
  \caption{The inverse central density $n_0^{-1}$ (in oscillator units
    $n_\mathrm{osc}=\pi^{-3/2}Nb_t^{-3}$) for a system of $N$ bosons
    in an oscillator trap, eq.~\eqref{eq:hamiltonian}, interacting via an
    attractive two-body potential, eq.~\eqref{eq:gauss-interaction},
    with scattering length $a=119$au as function of the state number.
    The lowest state with positive energy is numbered zero.}
  \label{fig:v}
\end{figure}

To verify this conjecture we calculate the central density, $n_0$,
of the system for the negative- and positive-energy states around zero
energy. The results are shown in fig.~\ref{fig:v} in the form of the
inverse central density (the volume per particle) $n_0^{-1}$.  In the
BEC-state the atoms should occupy the whole volume of the trap and thus
the volume per particle should be close to one (in the correspondingly
scaled oscillator units). The states with negative energy are self-bound
states with much higher density and thus much smaller volume per
particle. And indeed that is what fig.~\ref{fig:v} shows -- a sharp increase
in the volume per particle from a small value to about unity exactly at
the lowest state with positive energy where the quasi-continuum starts.

\begin{figure}[htbp]
  \centering
  \includegraphics{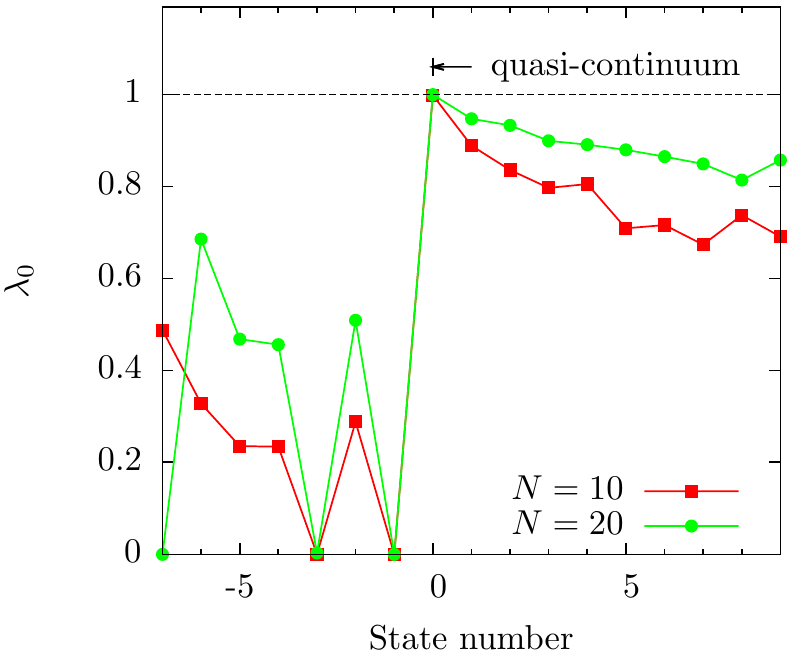}
  \caption{The condensate fraction of a system of $N$ trapped bosons
    from fig.~\ref{fig:v}.}
  \label{fig:cfs}
\end{figure}

Another test is the condensate fraction, shown for several states
around zero energy in fig.~\ref{fig:cfs}. The self-bound states with
negative energy must have smaller condensate fraction compared to the
BEC-state, and the excitations from the BEC-state must gradually
deplete the condensate fraction. Apparently this is what is seen in
fig.~\ref{fig:cfs} -- a sharp increase of the condensate fraction to
about 100\% at the lowest state with positive energy with the
subsequent gradual depletion.

We have thus verified that in the case of attractive potentials the
BEC-state of a system of trapped bosons is the lowest state with positive
energy.

\subsection{Accuracy of the Two-Body Correlated Basis}

In \cite{blume01} the energies of several low-density systems of
trapped bosons were calculated using the Gross-Pitaevskii equation as
well as repulsive hard-sphere models, both with the same scattering
length of 100~au.  In this regime the systems exhibit universality and
the energies calculated in both models were very close.

To test the accuracy of our two-body correlated basis,
eq.~\eqref{eq:psi-2B}, (which is expected to be a good approximation
in the low-density regime) as well as the identification of the
BEC-state for attractive potentials we consider the same systems with
the same scattering length but with the attractive Gaussian potential,
eq.~\eqref{eq:gauss-interaction}, and calculate the energy of the
BEC-state according to our prescription: The BEC-state is now an
excited state and is identified in the calculations as the lowest
state with positive energy.  We also calculate the energies for the
zero-range potential model, eq.~\eqref{eq:zero-range-potential}, with
the hyper-radial trial wave-function, eq.~\eqref{eq:psi-rho}.

The results are given in tab.~\ref{tab:blume}. As expected, these
low-density systems with relatively short scattering length seem to
exhibit universality as all the potential models give essentially the
same results. We conclude that our identification of the BEC-state is
correct and that the two-body correlated basis has an adequate
accuracy.

{\renewcommand{\arraystretch}{1.1}
\begin{table}[htbp]
  \centering
  \begin{threeparttable}
    \begin{tabular}{c|cccc}
      \hline \hline 
      $N$ & GP\tnote{$\dagger$} & HS\tnote{$\dagger$} & ZR & A\\
      \hline \hline 
      3   & 4.51032    & 4.51036(2)  & 4.5103        & 4.510\\
      \hline 
      5   & 7.53432    & 7.53443(4)  & 7.5342        & 7.534\\
      \hline 
      10  & 15.1534    & 15.1537(2)  & 15.1533       & 15.154\\
      \hline 
      20  & 30.638     & 30.640(1)   & 30.6394       & 30.640\\
      \hline \hline
    \end{tabular}
    \begin{tablenotes}
      \scriptsize
    \item[$\dagger$] Data taken from \cite{blume01}.
    \end{tablenotes}
  \end{threeparttable}
  \caption{ The energies in units of $\hbar\omega$ for the BEC-state
    of a system of $N$ bosons in a harmonic trap, eq.~\eqref{eq:hamiltonian},
    for different interaction models with the same scattering length
    $a=100$au. For the Gross-Pitaevskii (GP), hard-spheres (HS), and
    zero-range (ZR) models the BEC-state is the ground state. For
    the attractive model (A) the BEC-state is the lowest state with
    positive energy. The zero-range model employed the hyper-radial basis, 
    eq.~\eqref{eq:psi-rho}, while the attractive model employed the 
    two-body correlated basis, eq.~\eqref{eq:psi-2B}. }
    \label{tab:blume}
\end{table}
}

To check the accuracy of the two-body correlated basis also for large
scattering lengths we perform a test calculation for $N=4$ particles
with fully correlated and with two-body correlated basis for vastly
different scattering lengths. The results are given in
tab.~\ref{table:ft}.
Although the accuracy of the two-body correlated basis decreases
somewhat with the increase of the scattering length, the relative
accuracy is better than 1\% even for exceedingly large scattering
lengths.

{\renewcommand{\arraystretch}{1.1}
\renewcommand{\tabcolsep}{0.2cm}
\begin{table}[htbp]
  \centering
  \begin{tabular}{c c |cc}
    \hline \hline
    $V_0\times 10^7$, au & $a$, au & $E$(2B) & $E$(full)\\
    \hline\hline
    -1.400 &  119.4  &   6.025    &    6.025\\
    \hline
    -1.300 &  327.0  &   6.067    &    6.067\\
    \hline
    -1.290 &  402.4  &   6.083    &    6.083\\
    \hline
    -1.280 &  525.4  &   6.108    &    6.108\\
    \hline
    -1.270 &  761.0  &   6.155    &    6.156\\
    \hline
    -1.260 &  1400\phantom{00}   &   6.282    &    6.283\\
    \hline
    -1.255 &  2430\phantom{00}   &   6.478    &    6.481\\
    \hline
    -1.252 &  4370\phantom{00}   &   6.818    &    6.848\\
    \hline
    -1.251 &  5962\phantom{00}   &   7.059    &    7.112\\
    \hline \hline
  \end{tabular}
  \caption{The energies, in units of $\hbar\omega$, of the
    lowest state with positive energy for a system of $N=4$ bosons in a harmonic
    trap, eq.~\eqref{eq:hamiltonian}. The bosons interact via an attractive Gaussian
    potential, eq.~\eqref{eq:gauss-interaction}, with the strength $V_0$ and the scattering
    length $a$. The results from the fully correlated basis, eq.~\eqref{eq:psi-full},
    and from the two-body correlated basis, eq.~\eqref{eq:psi-2B}, are designated
    $E$(full) and $E$(2B), respectively.}
  \label{table:ft} 
\end{table}
}

\subsection{BEC Energies}

\begin{figure}[htbp]
  \centering
  \includegraphics{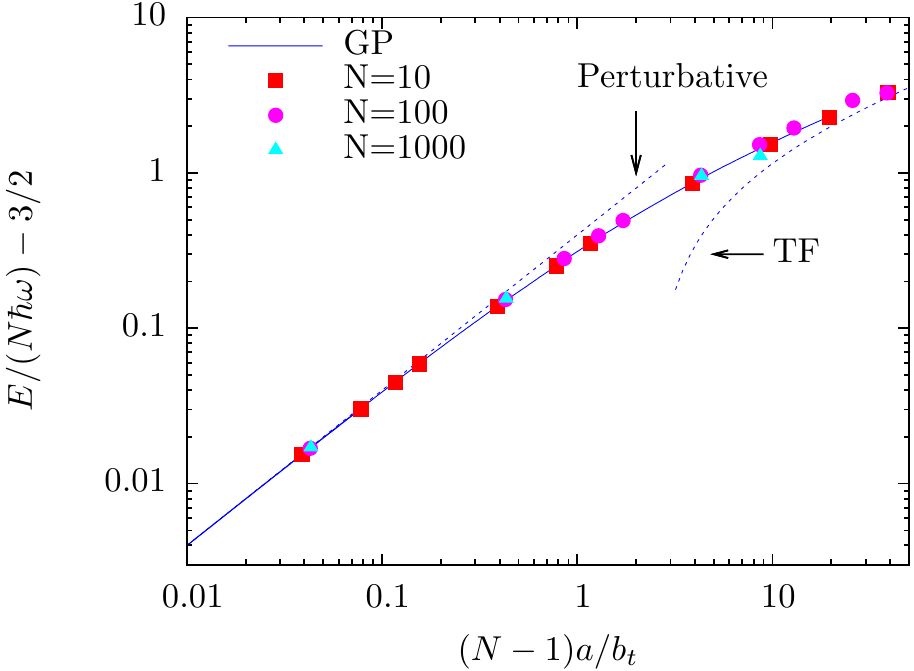}
  \caption{The energy per particle $E/N$ as function of the scattering
    length $a$ for a BEC-state of a system of $N$ identical bosons in
    a harmonic trap eq.~\eqref{eq:hamiltonian} for the zero-range
    model eq.~\eqref{eq:zero-range-potential} (points). The results
    are $N$-independent in this parametrization and in perfect
    agreement with the Gross-Pitaevskii (GP) equation. Also shown is the
    perturbative and Thomas-Fermi (TF) limits, see text.}
  \label{fig:corr-energy-zero}
\end{figure}

We first show the zero-range results for $N=10,100,1000$ in
fig.~\ref{fig:corr-energy-zero}. The energy depends on $N$ only
through the combination $(N-1)a/b_t$ and is in perfect agreement with
the Gross-Pitaevskii result. In the large $(N-1)a/b_t$ limit we
reproduce the energy of the Thomas-Fermi limit
\cite{pethick02}\footnote{See also chapter~\ref{chap:tf} for a
  complete discussion on the Thomas-Fermi approximation.},
\begin{equation}
  \frac{E_{TF}}{N}=\frac{5}{14}\left(\frac{15 Na}{b_t}\right)^{2/5}\hbar\omega.
\end{equation}
For small $(N-1)a/b_t$ the
interaction can be treated perturbatively: With the zero-range
perturbation $H'=\sum_{i<j}(4\pi \hbar^2 a/m)\delta(r_{ij})$ on the
non-interacting single-particle product ground state we find the
energy shift
\begin{equation}
  \label{eq:perturbation-energy}
  \frac{\Delta E}{N\hbar\omega}=\frac{1}{\sqrt{2\pi}}\frac{(N-1)a}{b_t}.
\end{equation}
Our results also reproduces this regime.
The agreement of our results with the GP equation was expected since
both models are based on the same zero-range interaction. Also, the
hyper-radial variational space for the zero-range interaction is
reminiscent of the Hartree-Fock product space used in the GP equation.

\begin{figure}[htbp]
  \centering
  \includegraphics[scale=0.95]{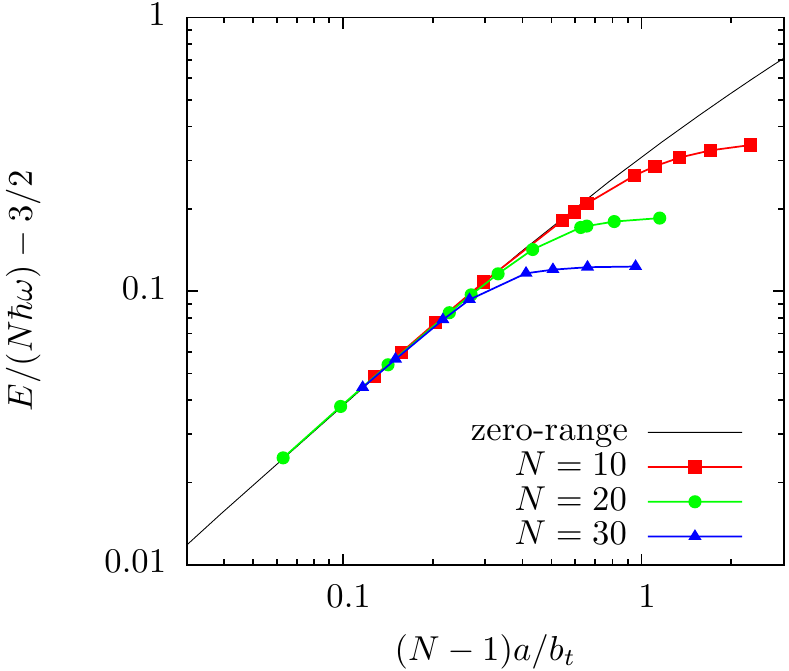}
  \caption{The energy per particle $E/N$ as function of the scattering
    length $a$ for a BEC-state of a system of $N$ trapped bosons.  The
    results for the attractive Gaussian interaction (points) is
    compared with the zero-range/Gross-Pitaevskii result from
    fig.~\ref{fig:corr-energy-zero}.}
  \label{fig:trap-e} 
\end{figure}

We now consider the energy per particle for the attractive potential
model in fig.~\ref{fig:trap-e} and compare with the zero-range model.
For small scattering lengths the different models give the same
universal results -- the system is model-independent. For larger
scattering lengths the energies from the attractive model are
systematically below the zero-range model, quite unlike the repulsive
model which goes above the zero-range model \cite{blume01,dubois01}.
For very large scattering lengths the attractive model, unlike the
zero-range and repulsive models, becomes insensitive to the scattering
length and the energies converge to a constant. This is consistent
with the Jastrow-type approximation used in \cite{cowell02}.

In the regime close to the two-body threshold, where the scattering
length is large, an arbitrarily large change in the scattering length
needs only an infinitesimally small change of the depth of the
attractive potential (see tab.~\ref{table:ft}). Therefore, when the
scattering length is larger than the trap length it ceases to be a
physical length scale for the system.
Since the external oscillator potential turns all continuum states
into discrete states all singularities due to various thresholds are
removed. Clearly an infinitesimally small change in the
potential, despite the large change in the scattering length, only
leads to an infinitesimal linear change in the energy which then
becomes independent of the scattering length.

On the contrary, for finite-range repulsive potential models an
increase of the scattering length needs an almost proportional increase
in the potential range. Thus the system never ceases
to depend on the scattering length.

\begin{figure}[htbp]
  \centering
  \includegraphics[scale=0.9]{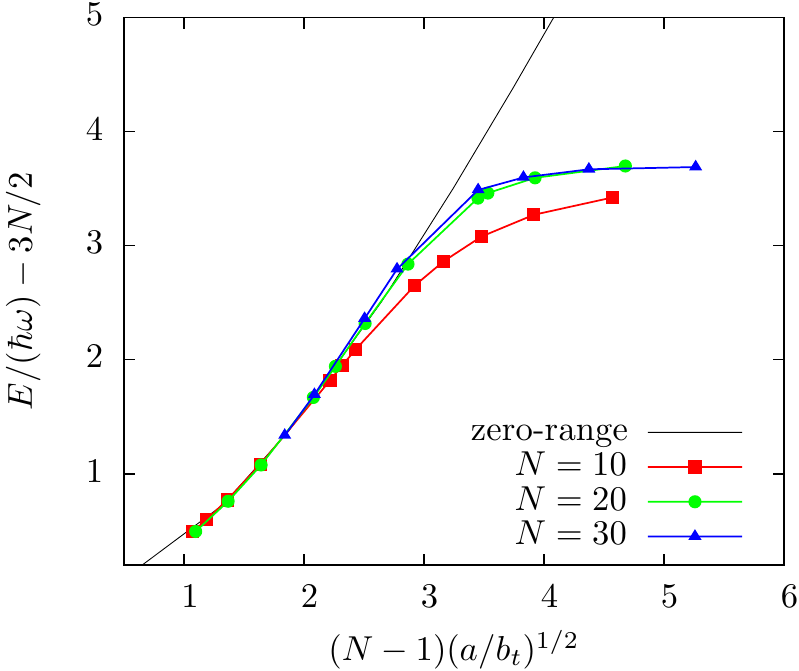}
  \caption{The same as is in fig.~\ref{fig:trap-e} but plotted with a
    different parametrization.}
    \label{fig:trap-e2}
\end{figure}

If we plot the data using a different parametrization, namely $E$ as
function of $(N-1)\left(a/b_t\right)^{1/2}$, the energy data points
seem to follow a universal curve as shown in fig.~\ref{fig:trap-e2}.  The
specific value to which the energy converges probably depends on an
independent energy scale, as for the Efimov effect in
chapter~\ref{chap:efimov} and \ref{chap:N-efimov}. We speculate that
other non-Gaussian interactions will reproduce the same universal
curve, by tuning only a single parameter.

\subsection{Condensate Fractions}

\begin{figure}[htbp]
  \centering
  \includegraphics[scale=0.9]{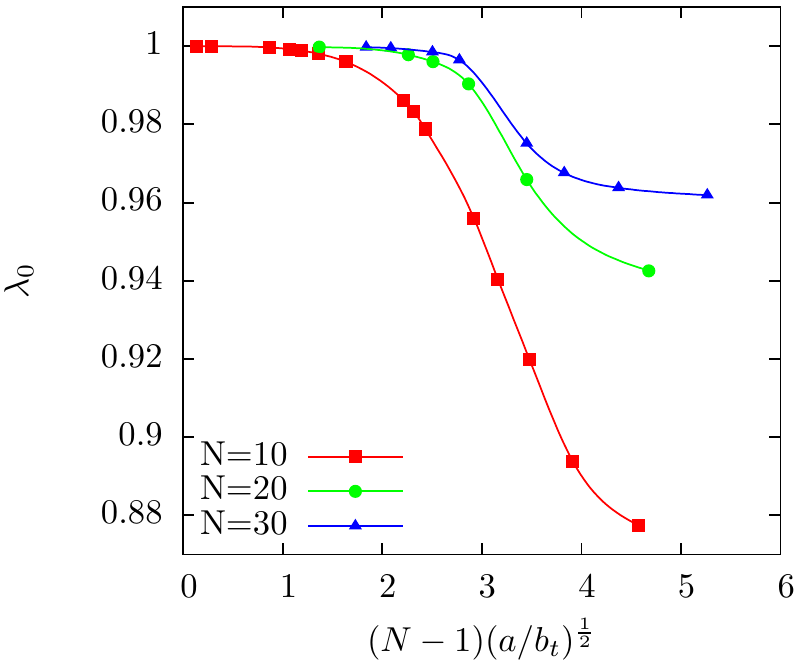}
  \caption{The condensate fraction of a BEC-state of a system of $N$
    identical bosons in a harmonic trap eq.~\eqref{eq:hamiltonian} with
    attractive inter-particle potential
    eq.~\eqref{eq:gauss-interaction} as function of $(N-1)(a/b_t)^{1/2}$.}
\label{fig:cf}
\end{figure}

Our results for the condensate fraction of a system of trapped
identical bosons in a BEC-state are shown in fig.~\ref{fig:cf}. For
small scattering lengths the system is 100\% condensate. When
$(N-1)\left(a/b_t\right)^{1/2}$ is about 3 the condensate fraction
rapidly drops a few percent before stabilizing again. Although we
could not reach further due to numerical convergence problems, we
expect that the condensate fraction will not appreciably change with
further increase of the scattering length.  This seems consistent with
the energies being stagnated when approaching the two-body threshold.

This behavior is qualitatively different from what happens within the
repulsive models. In the hard-sphere Monte-Carlo simulations
\cite{dubois01}, the condensate fraction starts to deviate from 100\%
much later when $a/b_t$ is of the order of 0.1, and with further
increase of the scattering length the condensate is completely
quenched.
Again our results are different from the Jastrow-type approximation
\cite{cowell02}, where the condensate also becomes fully quenched.
However, the authors of \cite{cowell02} note that their estimates for
the condensate fraction may be rather crude.

We have to note also that our wave-function includes only two-body
correlations which may lead to an over-estimate of the condensate
fraction.

\subsection{Two-Body Correlation Functions}

We now discuss the results for the two-body correlation function
$C^{(2)}(\bm r_1,\bm r_2)$, eq.~\eqref{eq:normalized-tbcf}.  For
simplicity one of the coordinates is set at the center of the trap,
$\bm r_2=0$.  This measures the probability to find a second particle
at radius $r$ when the first one is at the center. In this case the
two-body correlations depend on one coordinate only, and we can easily
investigate the properties over many length scales, including $r_0$
and $b_t$. However, the main conclusions will of course hold for
arbitrary $\bm r_2$.

\begin{figure}[htbp]
  \centering
  \includegraphics{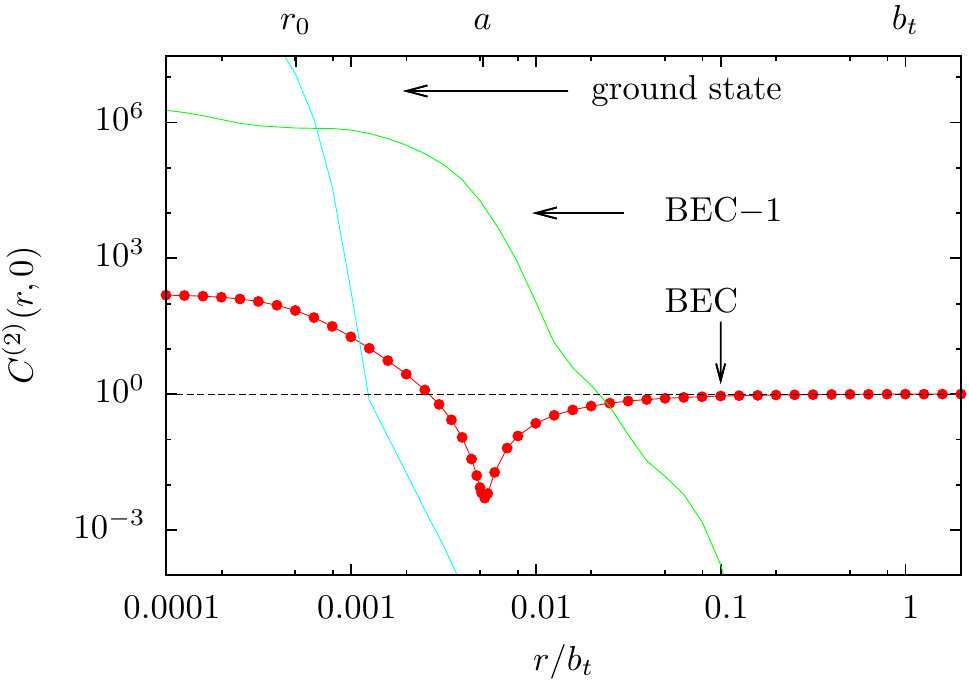}
  \caption{The two-body correlation function $C^{(2)}(r,0)$ for some
    of the different states found with the stochastic variational
    method.  The correlation for the BEC-state (lowest state in the
    quasi-continuum) reflects the two-body interaction at short
    distances and goes to unity at large distances. The first state
    below the BEC-state (BEC$-1$) is self-bound and the correlation
    function approaches zero at large distances. This is even more
    pronounced for the ground state (GS).}
  \label{fig:corr-func-states}
\end{figure}

Figure~\ref{fig:corr-func-states} shows the two-body correlation
function $C^{(2)}(r,0)$ for $N=20$ bosons in the case of an attractive
Gaussian interaction with one two-body bound state ($a>0$) and two-body
correlated functional space.
Three different states are shown: The BEC-state determined by the
lowest state in the quasi-continuum, the first state (BEC$-1$) right
below the BEC-state, and the ground state.

Let us first consider the asymptotics: The correlation function for
the BEC-state approaches unity for $r\to\infty$, thus the particles are
completely uncorrelated at very large distances. This is what we
expect, since we have shown earlier that the BEC-state has condensate
fraction close to unity.
The BEC$-1$ state is self-bound (bound without the trap) and hence the
particles are strongly correlated. The correlation function goes to
zero asymptotically, but is very large within the potential range
$r_0$. If one particle is measured at a particular point in space it
is very likely that all other particles are close to that point. Thus
the state is clearly non-condensed as discussed in
subsection~\ref{sec:cond-self-bound}.

Let us only consider the BEC-state in the rest of this chapter. We
observe that the two-body correlation function in
fig.~\ref{fig:corr-func-states} is large and positive within the range
$r_0$ of the potential. Outside $r_0$ we observe a large dip in the
correlation function. The value at the minimum of this dip is very
small but finite. The position of the dip occurs exactly at the
scattering length, $r=a$.

To understand this we consider the two-particle problem without a
trap. If the two-body potential supports a single two-body bound
state the relative wave-function $R(r)$ has a single node. If $a$ is
larger than $r_0$, the position of this node is approximately at
$a$.  So the joint probability, and hence the two-body correlation
function, is zero at the scattering length.

\begin{figure}[p]
  \centering
  \includegraphics{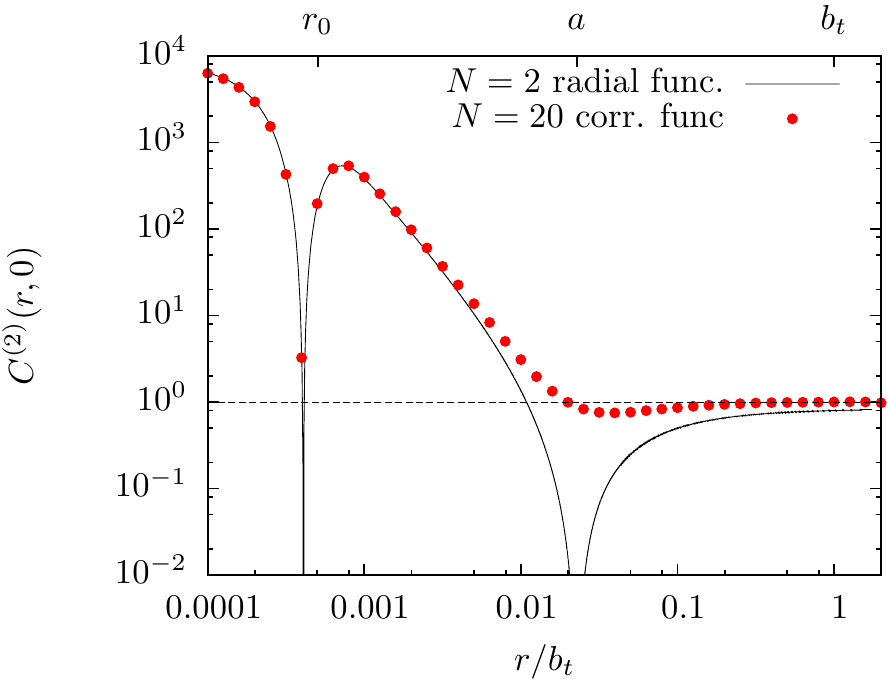}
  \caption{Correlation function (points) for the BEC-state of $N=20$
    trapped bosons with scattering length $a$ and two bound two-body
    states, compared with the squared two-body zero-energy radial
    function $R(r)^2$ (line). $R$ is normalized to to match the
    correlation function in $r=0$}
  \label{fig:corr-func-radial}
\end{figure}

To investigate this in more detail we take a new two-body potential
with two bound states and a large $a>0$. We calculate both the radial
function $R(r)$ for the two-body problem and the correlation function
for $N=20$ particles. The result is shown in
fig.~\ref{fig:corr-func-radial}.

\begin{figure}[p]
  \centering
  \includegraphics{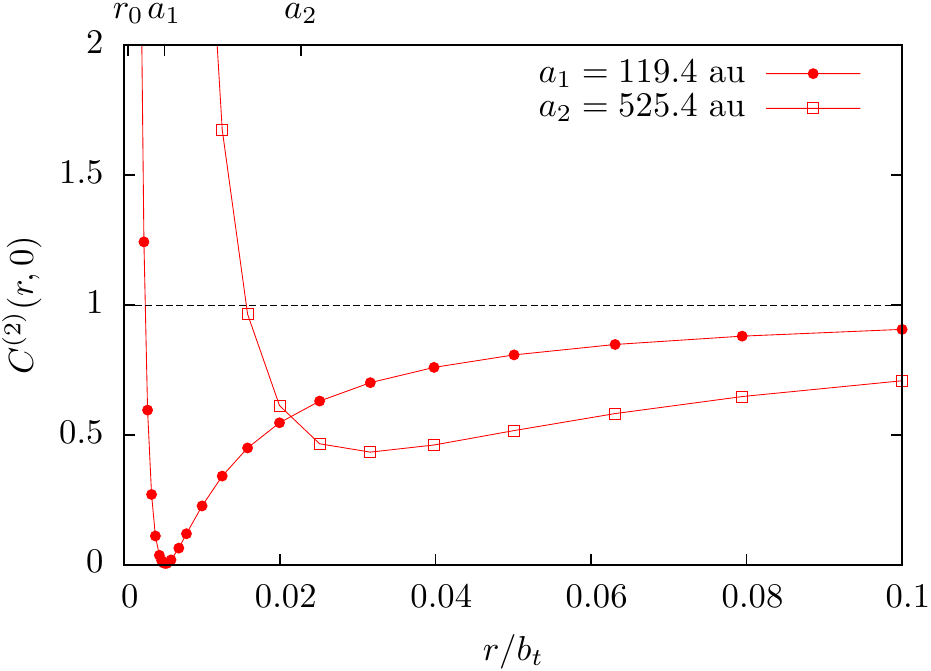}
  \caption{Correlation function for $N=20$ bosons in a trap. An
    attractive Gaussian interaction (range $r_0=11.65$au) was used in
    combination with the two-body correlated functional space. The
    bosons are in the BEC-state, identified as the lowest state of the
    quasi-continuum.  Two calculations with different scattering
    lengths ($a_1$ and $a_2$) are shown.  When the scattering length
    is small compared to the trap length $b_t$ a dip occurs at
    $a$. When $a$ increases the dip moves outwards and becomes
    shallower.}
  \label{fig:corr-func-a}
\end{figure}

At small distances the correlation function closely follows the
two-body wave-function while at large distances it converges to unity.
This behavior agree well with the Jastrow product approximation used in
\cite{cowell02}.  We conclude that the correlation function is
completely model-dependent for distances $\bm r_1 -\bm r_2$ comparable
to the potential range. However, for larger distances the scattering
length determines the properties.

Let us now investigate the large scattering length limit, where the
details are independent of the short-range potential. In
fig.~\ref{fig:corr-func-a} the correlation function is plotted for two
different scattering lengths. The $a_1$ data are the same as the
BEC-state in fig.~\ref{fig:corr-func-states}.  When the scattering
length is increased the dip moves correspondingly outwards but becomes
shallower. The decrease of the dip happens since the correlated pairs
starts to feel the mean-field of other particles between them. Thus the
probability to find two particles within a distance $a$ is increased
considerably. This is easily understood in terms of the universal
shallow (Feshbach) dimers with sizes of order $a$. We conclude that in
the large scattering length limit the two-body correlation function
has a universal behavior.  We propose to measure this effect via noise
correlations as discussed earlier in this chapter.

\section{Conclusions and Outlook}

We have calculated the energy and the condensate fraction of a system
of $N$ bosons in a harmonic trap as function of the number of bosons
and the scattering length $a$. Specifically, we considered the regime
where the scattering length is positive and comparable to the trap
length. The positive scattering length is modeled using an attractive
two-body potential with a bound two-body state. The many-body system
then has a large number of negative-energy self-bound states and the
condensate in the trap is identified as the lowest excited state with
positive energy.

When the scattering length is small compared to the trap length the
system shows model-independence (universality) -- the results from the
attractive potential model are very close to those from the zero-range
and the repulsive potential models.

In the limit of large scattering length the system properties become
independent of the scattering length, contrary to the zero-range and
the repulsive models.
For the attractive potentials the energy per particle of the system of
trapped bosons tends to follow a universal curve.
The condensate fraction decreases with the scattering length and reaches
a finite constant at large scattering lengths contrary to the repulsive
models where it reaches zero in this limit.

The two-body correlation function reflects the finite-range
interaction a short distances and the mean-field solution at larger
distances. When the scattering length becomes large the two-body
correlations become long-ranged and universal.

In general, the stochastic variational method allows for calculations
of correlations in BECs with attractive potentials. Such calculations
have traditionally not been possible because of large amount of
low-lying states. The obtained BEC states contain the short-range
correlations while the normal mean-field features are retained. Thus,
this is an important step towards understanding the non-universal
corrections in condensates.

%% file: mean-field.tex
\section{Introduction}

The stability of Bose-Einstein condensates (BECs) in ultra-cold alkali
gases is determined by the sign of the scattering length $a$
\cite{dalfovo99}. For $a<0$, one has effectively attractive
interactions and the condensate will collapse to a dense state when
the number of condensed particles, $N$, is larger than a critical
number $N_c$ \cite{ruprecht95,baym96,dalfovo99,gammal01}. This has
been beautifully demonstrated in experiments with $^{7}$Li
\cite{bradley97}, $^{87}$Rb \cite{roberts01,donley01}, and recently
with a dipolar $^{52}$Cr BEC \cite{lahaye08}. The findings indicate
that the theory based on the surprisingly simple Gross-Pitaevskii (GP)
equation can reproduce and describe most features of the experiments,
thus the scattering length is a universal parameter for these
many-particle systems.

The GP equation includes two-body terms through a contact interaction
which is parametrized by $a$. This is equivalent to a Born
approximation but with an effective coupling that is obtained by
replacing $a_{born}$ by the physical scattering length $a$. However,
higher-order terms in the expansion of the phase shifts at low
momenta, determined by the effective range $R_e$, the shape parameter
etc., give corrections to the simple GP equation. In this chapter we
explore the influence of the effective range term on the quantum
properties of a BEC. In particular, we show that the critical number
of condensed atoms depends strongly on the higher-order scattering
term when the scattering length approaches zero (zero-crossing). We
also show how the macroscopic quantum tunneling (MQT) rate, in which
the entire BEC tunnels as a coherent entity, can be modified for small
condensate samples.

The considered effects depend on a combination of $a$ and $R_e$ which
yield different behavior for wide and narrow Feshbach resonances.
Recent measurements on $^{39}$K found many both wide and narrow
resonances which allow for tuning of $a$ over many orders of magnitude
\cite{roati07,derrico07}. We therefore consider a selected example
from $^{39}$K in order to elucidate the general behavior for realistic
experimental parameters.

In this chapter we first introduce the modified GP equation with
effective range dependence. Using both a variational and numerical
approach we find a phase diagram describing the stability of the
condensate.  We then consider an extended Feshbach resonance model
including effective range variations.  The behavior of the critical
particle number near a scattering length zero-crossing is derived.  We
discuss MQT and show numerically how the rate is modified. We finally
discuss other possibilities for probing the higher-order interactions.

\section{Modified Gross-Pitaevskii Equation}
We assume that the condensate can be described by the GP equation and
we focus on the $a<0$ attractively interacting case.  Since we are
interested in the ultra-cold regime, where the temperature is much
smaller than the critical temperature for condensation, we adopt the
$T=0$ formalism. In order to include higher-order effects in the
two-body scattering dynamics, we use the modified GP equation derived
in \cite{collin07} for which the equivalent energy functional is
\begin{equation}
\label{efunc}
  E(\Psi)=\int \ud{\bm r} 
  \left[\frac{\hbar^2}{2m}|\nabla \Psi|^2+V_{ext}(\bm r)
  |\Psi|^2\right.
  \left.+\frac{U_0}{2}\left(|\Psi|^4 +g_2|\Psi|^2
      \nabla^2|\Psi|^2\right) \right],
\end{equation}
where $m$ is the atomic mass, $V_{ext}$ is the external trap,
$U_0=4\pi\hbar^2 a/m$, and $g_2=a^2/3-aR_e/2$ with $a$ and $R_e$ being
the $s$-wave scattering length and effective range, respectively.
The corresponding stationary GP equation found by variation of $\Psi$
is
\begin{equation}
  \left[-\frac{\hbar^2}{2m}\nabla^2
    +V_{ext}(r)+U_0\left(|\Psi|^2+g_2\nabla^2|\Psi|^2\right)
  \right]\Psi=  \mu\Psi,
\end{equation}
where the chemical potential, $\mu$, was introduced to fix the particle
number $N$.

\subsection{Effective Zero-Range Interaction}
\label{appeff}
To derive the modified GP equation we first need an effective two-body
zero-range interaction, which by construction will give the same
interaction energy as the real two-body potential. Partly following
\cite{roth01,collin07}, we first consider two non-interacting particles
and restrict ourselves to $s$-waves only. In order to count energy
levels and evaluate the corresponding energy shifts we choose the
radial wave-function to be zero some large but arbitrary radius $L$.
This gives the discrete energy levels $E_n=\hbar^2 k_n^2/m$, $k_n=\pi
n/L$ and normalized eigenstates $\psi_n(r)=\sqrt{2/(4\pi
  L)}\sin(k_nr)/r$.  Inclusion of the true two-body interaction gives
the new energy levels $\bar E_n=\hbar^2 \bar k_n^2/m$ and
corresponding small energy shift $\Delta E_n=\bar E_n-E_n$. The
criterion
\begin{equation}
  \label{eq:ZR-criterion}
  \Delta E_n=\langle \psi_n\vert \hat V_{ZR}\vert \psi_n\rangle
\end{equation}
then defines the coupling constant(s) of the effective zero-range
interaction $V_{ZR}$.

We first evaluate $\Delta E_n$. In case of interaction the asymptotic
wave-function is proportional to $\sin(\bar k_n r+\delta(\bar k_n))$,
where $\delta$ is the phase shift. Since the wave-function must still
vanish at $L$, the wave numbers are given by\footnote{If the two-body interaction
  supports $n_b$ bound states the phase shift must be reduced by
  $n_b\pi$ according to Levinsons theorem.} $\bar
k_nL=n\pi-\delta(\bar k_n)$.
Since $\Delta k_n=\bar k_n-k_n$ is small we have $\delta(\bar
k_n)\simeq\delta(k_n)$. Using $\Delta E_n\simeq 2\hbar^2 k_n\Delta
k_n/m$ we find
\begin{equation}
  \label{eq:energy-shift}
  \Delta E_n=-\frac{\delta(k_n)}{k_n}\ \frac{2E_n}{L}.
\end{equation}
Since we are only interested in terms up to order $k^2$, we use
the low-energy effective range expansion 
\begin{equation}
  \label{eq:phaseshift-exp}
  k\cot\delta=-\frac{1}{a}+\frac{R_e}{2}k^2+O(k^4),
\end{equation}
defining the scattering length $a$ and effective range $R_e$. Using
$\delta\simeq\tan\delta-(\tan\delta)^3/3$ we find
\begin{equation}
  \label{eq:phaseshift-exp2}
  -\frac{\delta(k)}{k}=a\left[1-(\frac{a^2}{3}-\frac{aR_e}{2})k^2+O(k^4)\right].
\end{equation}

We now construct the zero-range interaction. The simplest hermitian
ansatz up to order $k^2$ is $\hat V_{ZR}=\hat V_{0}+\hat V_{2}$
with
\begin{equation}
  \label{eq:V0-V2}
  \begin{split}
    \hat V_{0}({\bm r})&=U_0\delta({\bm r}),\\
  \hat V_{2}(\bm r)&=U_0 g_2\tfrac{1}{2}[
  \overleftarrow{\nabla}_{\bm r}^2\delta(\bm r)
  +\delta(\bm r)\overrightarrow{\nabla}_{\bm r}^2],
  \end{split}
\end{equation}
where $U_0$ and $g_2$ are unknown coupling strengths to be determined
from eq.~\eqref{eq:ZR-criterion}. Using the non-interacting
wave-functions $\psi_n$ we get
\begin{equation}
  \label{eq:ZR-matrixelement}
  \langle \psi_n\vert \hat V_{ZR}\vert \psi_n\rangle 
  =\frac{U_0m}{4\pi\hbar^2}\left(1-g_2 k_n^2\right)\frac{2E_n}{L}.
\end{equation}
Comparison of eqs.~\eqref{eq:energy-shift},
\eqref{eq:phaseshift-exp2}, and \eqref{eq:ZR-matrixelement} gives
$U_0=4\pi \hbar^2 a/m$ and $g_2=a^2/3-aR_e/2$. Thus we now have an
effective zero-range two-body interaction to order $k^2$. 
By constructing the many-body mean-field Hamiltonian from this interaction
(integrating out relative pair-distances), the last higher-order term
of eq.~\eqref{efunc} can be obtained \cite{fu03,collin07}.

\section{Variational Calculations}
\begin{figure}[htb]
  \centering
  \includegraphics{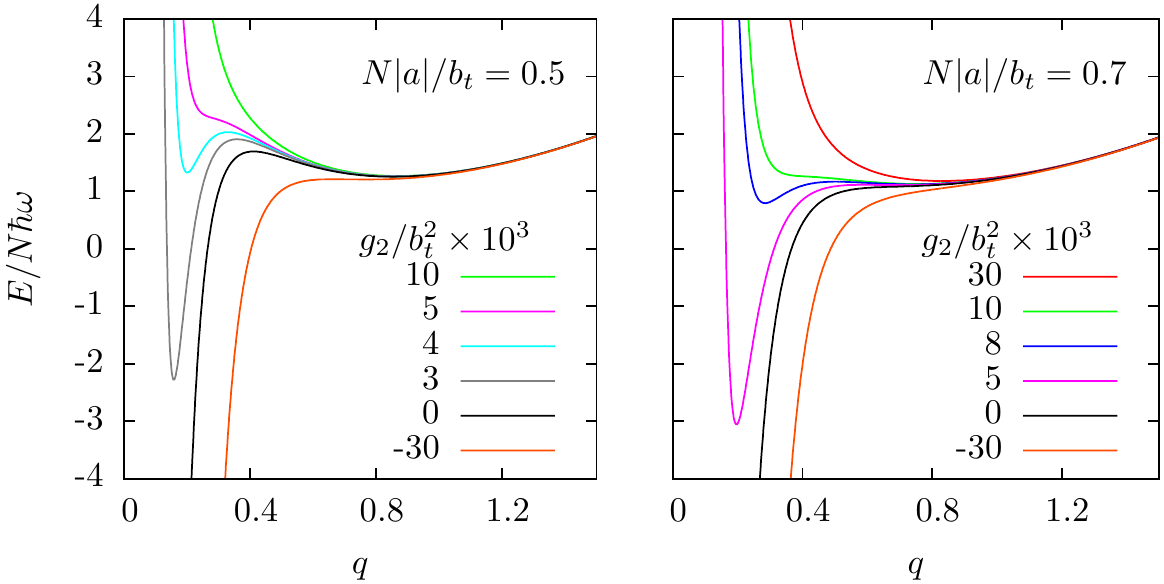}
  \caption{Energy of a BEC with fixed $N\lvert a\rvert/b_t$ as function
    of the variational parameter $q$, i.e. the size of the condensate.
    The higher-order interaction term $g_2$ modifies the height and
    shape of the barrier.}
  \label{fig1}
\end{figure}

We are interested in the stability properties of the ground-state and
we therefore perform a variational calculation on eq.~\eqref{efunc} using the mean-field
trial wave-function
\begin{equation}\label{trial}
  \Psi(r)
  =\frac{\sqrt{N}}{\pi^{3/4} \sqrt{(qb_t)^3}}\exp(-\frac{r^2}{(qb_t)^2}),
\end{equation}
where $q$ is the dimensionless variational parameter and
$b_t=\sqrt{\hbar/m\omega}$ is the trap length. The normalization is
\mbox{$N=\int \ud{\bm r}|\Psi(r)|^2$}. For simplicity we only consider
isotropic traps with $V_{ext}(r)=\tfrac{1}{2}m\omega^2r^2$.  However, the
effects found should hold for deformed traps as well (along the lines
of the analysis in \cite{ueda98}).  The variational energy is
\begin{equation}\label{energy}
  \frac{E(q)}{N\hbar\omega}
  =\frac{3}{4}q^2+\frac{3}{4}\frac{1}{q^2}+\frac{1}{\sqrt{2\pi}}\frac{N|a|}{b_t}
\left(-\frac{1}{q^3}
  +3\frac{g_2}{b_t^2}\frac{1}{q^5}\right).
\end{equation}
In fig.~\ref{fig1} we plot $E(q)$ for different parameters. As shown,
there are many possibilities for $g_2 \neq 0$, including stable,
unstable, and metastable systems. We see that the $g_2$ term modifies
the barrier for $N|a|/b_t=0.5$, implying that tunneling rates will be
altered.  For $N|a|/b_t=0.7$, the $g_2=0$ case has no barrier at all,
and here the $g_2$ term can in fact produce a small barrier on its
own. The $q^{-5}$ dependence of the term means that the effect is
small. However, the plot clearly shows that a new stability analysis
is needed.  In addition to the variational approach we have
numerically solved the full time-independent GP equation corresponding
to eq.~\eqref{efunc}.

\section{Phase Stability Diagram}
\begin{figure}[htb]
  \centering
  \includegraphics{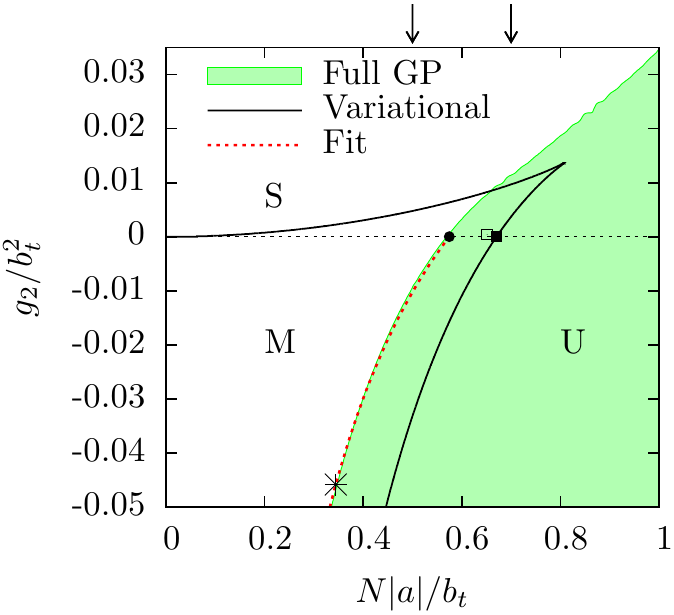}
  \caption{Phase stability diagram of a BEC with
    higher-order interactions. The solid line obtained from the
    variational ansatz, eq.~\eqref{trial}, divides the stable~(S),
    metastable~(M), and unstable~(U) regions.  The white region
    indicates where stationary solutions exists for the full GP
    equation. Filled points show the well-known $g_2=0$ results.
    Arrows correspond to the values in fig.~\ref{fig1}.  The dashed
    line is the fit in eq.~\eqref{eq:Nc-fit}. Two specific $^{39}$K
    values are chosen for zero-crossing (cross) and MQT rates (open
    square) calculations (see text).}
  \label{fig2}
\end{figure}

To determine the ground-state stability one looks for the vanishing of
the barrier towards $q=0$. For $g_2=0$ eq.~\eqref{energy} leads to
$N_c |a|/b_t\approx 0.671$ \cite{dalfovo99}. The full integration of
the GP equation gives $N_c |a|/b_t\approx k_0$, $k_0=0.5746$
\cite{ruprecht95,gammal01}. These values are indicated by filled
points in fig.~\ref{fig2}.  The stability coefficient has been
determined precisely to $0.547(58)$ using bound-state spectroscopy
near a $^{85}$Rb Feshbach resonance \cite{claussen03}. The deviation
is well understood in terms of the asymmetric trap used
\cite{gammal01}.

In the general $g_2\ne0$ case we first take the variational energy
eq.~\eqref{energy} and solve for multiple roots of $\ud E(q)/\ud q$.
The resulting ``phase diagram'' is plotted in fig.~\ref{fig2} (solid
line). In the upper left region (S) we have complete stability of the
condensate: only one minimum exists at large $q\sim 0.8$ (see
fig.~\ref{fig1}) and the potential goes to plus infinity at small $q$.
In the metastable region (M) a barrier and a minimum exists for large
$q$: for $g_2>0$ another minimum exists at small $q\lesssim 0.2$,
while for $g_2<0$ the potential goes to minus infinity.  In the
unstable region (U) the barrier has vanished: the potential either has
a minimum at small $q$ ($g_2>0$) or no minimum at all ($g_2<0$).  In
the variational approach the stable and unstable regions are connected
via the upper right part of fig.~\ref{fig2}. Going from (S) to (U)
corresponds to an adiabatic change, where the macroscopic ($q\sim 1$)
minimum is transferred to a microscopic ($q\ll 1$) high-density
minimum.

Next, we numerically solve the full the GP equation. The stationary
(white) and non-stationary (shaded) regions are shown in
fig.~\ref{fig2}. Since both stable and metastable solutions are
considered stationary, the white region covers both (S) and (M). Both
our variational and numerical results agrees with the known $g_2=0$
results, and have similar behavior for small and negative $g_2/b_t^2$.

\section{Feshbach Resonance Model}

In order to predict effects of the higher-order term, we need a
realistic model for $g_2$. Since $g_2/b_t^2$ is the relevant parameter,
and the trap length, $b_t$, is usually orders of magnitude larger than
atomic scales, it is necessary to look for divergences of $g_2$. Since
$g_2$ depends on $a$ and $R_e$, a Feshbach resonance is the obvious
mean. The standard single-channel models are inadequate since only $a$
is considered.  

We therefore use a multi-channel model \cite{bruun05},
which describes both $a$ and $R_e$ as a function of resonance position
$B_0$, width $\Delta B$, magnetic moment difference between the
channels $\Delta \mu$, and the background scattering length $a_{bg}$.
In \cite{bruun05} the on-shell $T$-matrix in the open-open channel is
given by
\begin{equation}
  \label{eq:Too-bruun}
  \begin{split}
    T_{oo}&=\frac{4\pi \hbar^2}{m}\cdot\frac{a_{bg}}{\left(
        1+\frac{\Delta\mu \Delta B}{\hbar^2 k^2/m-\Delta\mu(B-B_0)
        }\right)^{-1}+i a_{bg}k}\\
    &=\frac{4\pi\hbar^2}{m}\cdot\frac{1}{\frac{1}{a_{bg}}
      \left(1-\frac{1}{R_{e0}a_{bg}k^2/2+\eta}\right)^{-1}+ik},
  \end{split}
\end{equation}
where $k$ is the relative momentum of the atoms and we introduced
$R_{e0}=-2\hbar^2/ma_{bg}\Delta \mu\Delta B<0$ from
eq.~\eqref{eq:reff-fesh-const} and $\eta=(B-B_0)/\Delta B$.  Expanding
the denominator of eq.~\eqref{eq:Too-bruun} to second order in $k$
we get
\begin{equation}
  T_{oo}\simeq\frac{4\pi \hbar^2}{m}\cdot
  \frac{1}{\frac{1}{a_{bg}}(1-\frac{1}{x})-\frac{R_{e0}}{2(1-\eta)^2}k^2+ik}.
\end{equation}
By comparing this to the effective range expansion of the vacuum $T$-matrix,
on-shell $T$-matrix \cite{pethick02}
\begin{equation}
  T=-\frac{4\pi\hbar^2}{m}\cdot\frac{1}{k\cot\delta(k)-ik}
  \simeq\frac{4\pi\hbar^2}{m}\cdot\frac{1}{\frac{1}{a}-\frac{R_e}{2}k^2+ik},
\end{equation}
the scattering length and effective range becomes%
\footnote{We note that eq.~\eqref{eq:reff-fesh} can also be obtained
  from \cite[eqs.~(16) and (44)]{jonsell04}.}
\begin{equation}
  a(B)=a_{bg}\left(1-\frac{\Delta B}{B-B_0}\right),
\end{equation}
\begin{equation}
  \label{eq:reff-fesh}
  R_e(B)=R_{e0}\left( 1-\frac{B-B_0}{\Delta B}\right)^{-2} 
  = R_{e0}\left(1-\frac{a_{bg}}{a(B)}\right)^2.
\end{equation}
In \cite{bruun05} only the regime $|B-B_0|\ll \Delta B$ was considered
where $R_e\simeq R_{e0}$ is constant. The modification above
includes the field dependence of $R_e$ and since we later want to
consider a broad region around $B_0$ we retain this explicit field
dependence.

We note that $R_e$ can become large (and negative) in two different
ways. The first way is to consider narrow resonances where $\Delta B$
is small, and hence $R_{e0}$ is large. The other option is to tune the
scattering length near a zero-crossing to small values, $a\ll a_0$.
The effective range goes to zero far away from the resonance where
$a\simeq a_{bg}$.

We now have a field-dependent $g_2(B)$ for given values of $B_0$,
$\Delta B$, $\Delta \mu$, and $a_{bg}$. Notice that with this Feshbach model
\begin{equation}
  \label{eq:g2(a)}
  g_2(a)=\frac{a^2}{3}-\frac{a R_{e0}}{2}\left(1-\frac{a_{bg}}{a}\right)^2,
\end{equation}
hence $g_2$ diverges when $a\to 0$ (referred to as zero-crossing) or
$a\to\infty$ (on resonance).  These are well-known features for model
potentials like the square-well and van der Waals interactions.

\begin{figure}[htb]
  \centering
  \includegraphics{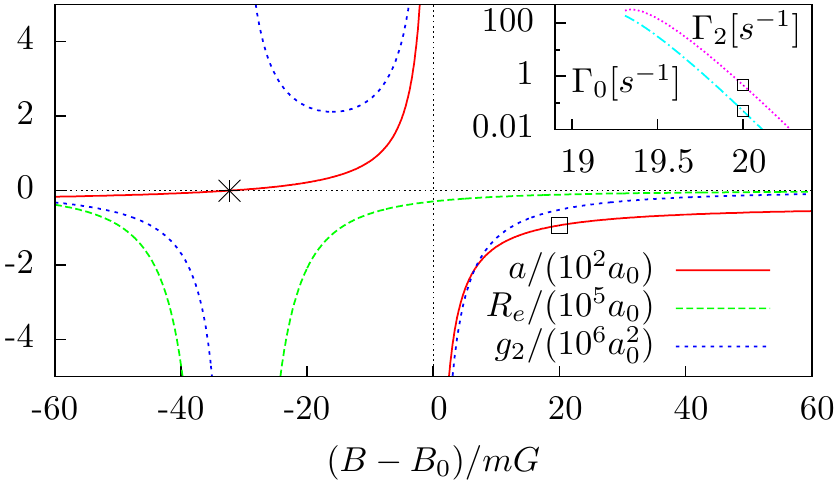}
  \caption{Scattering length $a$, effective range
    $R_e$, and coupling constant $g_2$ as function of $B$-field for
    the narrow $^{39}$K Feshbach resonance at $B_0=825$~G. The inset
    shows the MQT rate $\Gamma_2$ (and $\Gamma_0$ for $g_2=0$) for
    $N=242$ and $b_t=1.84\mu$m. The cross and open square are as in
    fig.~\ref{fig2}.}
  \label{fig3}
\end{figure}

As a concrete example we use the extremely narrow Feshbach resonance
in $^{39}$K found at $B_0=825$G, with $\Delta B=-0.032$G,
$\Delta\mu=3.92\mu_B$, and $a_{bg}=-36a_0$ \cite{derrico07}. For this
resonance we find a large $R_{e0}=-2.93\times10^4a_0$.  The variation
of $a$, $R_e$ and $g_2$ as function of $B$ is shown in
fig.~\ref{fig3}.  We use trap length $b_t=1.84\mu\textrm{m} =3.48\times
10^4a_0$ in all calculations unless indicated otherwise.

\section{Critical Particle Number Near Zero-Crossings}
As noted above, large values of $g_2$ are possible at zero-crossings
($a=0$) since $R_e$ diverges here, see e.g. fig.~\ref{fig3} around
$B-B_0\sim-32$mG. The scattering length is currently being tuned with
extreme accuracy near such zero-crossings in $^7$Li
\cite{pollack09}.

The advantage is that many particles $N_c\propto 1/|a|$ can be
accommodated in the condensate.  However, for $a\rightarrow 0$ we have
$ag_2\rightarrow -R_{e0}(a_{bg})^2/2$, i.e. a finite limit.
Remembering that $R_{e0}$ and $\nabla^2|\Psi|^2$ are negative, the
last term in the energy functional eq.~\eqref{efunc} also becomes
negative. Thus larger densities or density fluctuations gives lower
total energy. This implies less stability and smaller $N_c$ near
$a=0$.

\begin{figure}[htb]
  \centering
  \includegraphics[scale=1.0]{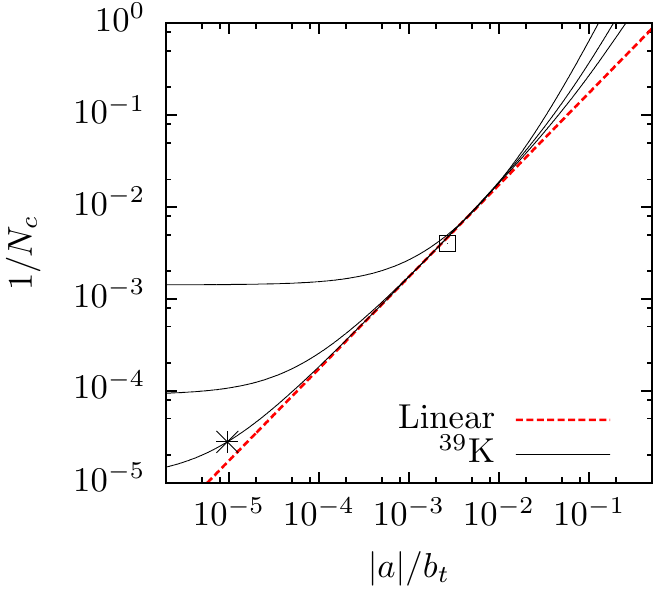}
  \caption{Critical particle number $N_c$ as function
    of scattering length $a$. The linear dashed (red) line is the
    well-known $g_2=0$ result $N|a|/b_t=k_0$. The solid (black) lines
    are for a narrow $^{39}$K resonance and trap length $b_t=1.84\mu$m,
    (bottom), $0.5b_t$ (middle), and $0.2b_t$ (top).  The cross and open
    square are as in fig.~\ref{fig2}.}
    \label{fig4}
\end{figure}

To calculate quantitative effects on $N_c$ we focus on the critical
line between the unstable and metastable regions in fig.~\ref{fig2}
for $g_2\le 0$.  We fit the dependency as
\begin{equation}
  \label{eq:Nc-fit}
  N_c=k_0 \frac{b_t}{|a|} \times \left(1-k_1 \frac{g_2(a)}{b_t^2}\right)^{-1},
\end{equation}
where $k_1=14.5$, as shown in fig.~\ref{fig2}. With $g_2(a)$ given by
eq.~\eqref{eq:g2(a)}, $N_c$ becomes a function of $|a|/b_t$ for fixed
values of $a_{bg}/b_t$ and $R_{e0}/b_t$.  In fig.~\ref{fig4} we plot
$1/N_c$ as function of $|a|/b_t$ for the $^{39}$K resonance with various
trap lengths.  The curve deviates from the linear $g_2=0$ result both
at $|a|\sim 0$ and $|a|\sim b_t$.  Squeezing the trap makes the
effect even more pronounced.  In the limit $|a|/b_t\to 0$ the critical
particle number approaches a finite value, $N_c \to 2 k_0 b_t^3 /(k_1
|R_{e0}| a_{bg}^2)$.  Large effects also occur near $|a|\sim b_t$,
however, here the allowed particle number is very small.

To exemplify with the $^{39}$K resonance we use $B-B_0=-32.3$mG (cross
in fig.~\ref{fig3}) giving $a=-0.334a_0$ and $g_2/b_t^2=-0.0461$.  The
critical particle number 
is then $N_c=3.58\times 10^4$. The values are indicated by crosses in
fig.~\ref{fig2} and~\ref{fig4}. The $g_2=0$ prediction is
$N_c=5.975\times 10^4$. In conclusion, we predict a 40\% reduction in
$N_c$ for macroscopic particle number $N_c\sim 10^4$.

\section{Macroscopic Quantum Tunneling}

Close to $a=\infty$, $g_2/b_t^2$ is still a small term in typical
traps.  As the MQT rate \cite{mozyrsky07} is exponentially dependent
on the integral under the barrier, we expect changes due to non-zero
$g_2$ to be amplified. A similar point was raised in \cite{ueda98}
where the MQT rate was shown to increase dramatically close to $N_c$.
The MQT rate, $\Gamma_2$, can be obtained from field theory
\cite{stoof97} by considering the bounce solution of the effective
action in variational $q$-space with potential $E(q)/N\hbar\omega$.
For comparison, we denote by $\Gamma_0$ the rate with $g_2=0$.
All MQT calculations are done within the variational approach.

We are interested in metastable states with small barriers and large
MQT rates, thus we have to work close to $N_c$ \cite{ueda98}.  Below
we find that $g_2/b_t^2$ is only important for the rate when $N$ is of
order $10^2$ or less.  Since the three-body loss depends on the
density to the third power \cite{adhikari02}, we expect it to be
small in the outer metastable minimum and to be large in the inner
one. Thus, the physical picture is that of a metastable BEC in the
outer minimum that knocks on the barrier as a coherent state with a
common quantum tunneling probability. When it tunnels to the inner
minimum it swiftly decays as the large increase in density amplifies
the three-body loss.  Thermal fluctuations can of course also induce
MQT and we therefore need to operate at very low temperature.  An
estimate of the thermal tunneling rate is given in \cite{stoof97},
and leads to the following criterion for thermal fluctuations to be
suppressed: $E(q_m)-E(q_0)\gg k_B T$, where $q_m$ and $q_0$ are the
positions of the barrier maximum and the outer minimum, respectively.

For the $^{39}$K resonance we now consider $B-B_0= 20$mG (see
fig.~\ref{fig3}) where $g_2/b_t^2=-4.3\times 10^{-4}$ and $a=-93.6a_0$.
The critical particle number becomes $N_c\simeq249$ (see
fig.~\ref{fig4}).  We now pick $N=242<N_c$ to obtain metastability
(see fig.~\ref{fig2}).  The rates are found to be $\Gamma_2=0.49$
s$^{-1}$ and $\Gamma_0=0.05$ s$^{-1}$ (shown by the open squares in
the inset of fig.~\ref{fig3}), i.e. a tenfold enhancement of the MQT
rate. The temperature must be below 8~nK for thermal fluctuations to
be small.  It is possible to obtain larger $N$ with the cost of a
smaller effect on the MQT rate.  However, going to $N\sim 10^{3}$ does
not seem possible.

We propose to start from the $a>0$ side where the condensate is
stable. Preparing a sample with low enough $N$ and temperature of a
few nK for the effect to be observable is the very difficult challenge
for the MQT scenario. A sudden ramp of the magnetic field to the
appropriate $a<0$ is then performed. The system has to retain a small
barrier so that the MQT rate is considerable and to avoid collapse.
Then one monitors $N$ as a function of time, as was done in e.g.
\cite{donley01}. The low $N$ is also challenging for optical imaging
of the atomic cloud, but should be possible with current techniques.
We expect not only to see modified MQT rates, but also changes in the
decay after the condensate tunnels to a high density state: The
three-body recombination is very sensitive to the density, and the
inner barrier caused by the $g_2$ term should affect this process as
well.

\section{Other Signatures}

From the structure of the $g_2$ term it is clear that density
variations are needed to see effects. As we have demonstrated already,
tighter traps amplify the contribution, and we expect optical lattices
to do the same.  Solitons and vortices are other features with density
variation. For a simple one vortex state, the
$|\Psi|^2\nabla^2|\Psi|^2$ term will leave the core and the asymptotic
regions unchanged but change the profile in between. Work is in
progress \cite{zinner09b}.

\section{Conclusion and Outlook}
We have explored the effect of higher-order terms in the
Gross-Pitaevskii description of a Bose-Einstein condensate,
particularly the effective range correction near a Feshbach resonance.
Using both a variational and numerical approach we find an interesting
new phase diagram for the stability of a condensate with negative
scattering length.

The critical particle number is strongly affected near zero-crossings
of the scattering length for narrow resonances. Effects of 40\% are
found for particle number of order $10^4$.  These deviations increase
for more narrow resonances or tighter traps. We find that the critical
number will be reduced as the zero-crossing is approached from either
side since the higher-order term is attractive here.  
Macroscopic quantum tunneling is also modified by these higher-order
corrections and we have discussed some experimental conditions for
exploring the physics.  Narrow Feshbach resonances are the best way to
isolate the effect of tunneling. However, one needs small samples of
order $10^2$ particles or low temperatures. Other possible
experimental signatures are tight traps, optical lattices, solitons,
and vortices in rotating BECs.

In a more general sense, the fact that higher-order interactions can
become large and dominant in determining the stability properties
means that it is no longer a correction and that still higher terms
might become important. This is the subject of future work. For the
moment we have demonstrated that experiments targeting the regions
discussed above could find interesting new stability properties beyond
those that were already understood about a decade ago.

%% file: tf.tex
\section{Introduction}
The Gross-Pitaevskii (GP) equation
\cite{dalfovo99,pethick02,pitaevskii03} has been extremely successful
in describing a wide range of mean-field features for experiments with
Bose-Einstein condensates (BECs).  In particular, the Thomas-Fermi
(TF) approximation \cite{baym96,dalfovo96,lundh97}, where the kinetic
energy is neglected, has been very rewarding \cite{hau98}. This
approximation holds for repulsive condensates with positive scattering
length $a$ and large particle numbers.  In the regime of validity of
the TF approximation, the total energy is distributed between
interaction energy and potential energy from the confining trap, while
the kinetic energy becomes negligible.

Because of the non-linear nature of the GP equation, it is only solved
analytically in a few cases, e.g. vortices
and solitons in homogeneous condensates \cite{pethick02,pitaevskii03}. 
The TF solution is also analytical, although it only holds in
the bulk of the condensate. At the surface the approximation breaks
down and is usually patched by including the kinetic energy at the surface 
\cite{dalfovo96,lundh97}.

The interactions of the ordinary GP equation is based on the lowest
order zero-range potential, which is governed by the scattering length
alone. Although this approximation is usually very good, the
higher-order corrections to the scattering dynamics
\cite{roth01,fu03,collin07} can be crucial in certain cases, e.g.  for
Rydberg molecules embedded in BECs \cite{collin07} and for narrow
Feshbach resonances \cite{zinner09a}. Inclusion of higher-order terms
are well known and applied in Skyrme-Hartree-Fock calculations in
nuclear physics \cite{brack85}. Here they often play a crucial role in
order to get bulk nuclear properties right \cite{skyrme56,siemens87}.
However, the effects of similar higher-order terms in the GP equation
have been less investigated.

In this chapter we solve the modified GP equation with higher-order
interactions analytically in the TF approximation. The chapter is
organized as follows: We introduce the modified GP equation and its
parameters and show how it is derived from an appropriate energy
density functional with careful treatment of boundary terms. We
present the analytical solution in the TF approximation and discuss
the condensate size and chemical potential as function of the
interaction parameters. The density profiles and energies are
discussed and we address the consistency of the TF approximation by
considering the kinetic energy of the solutions. We compare to some
relevant atomic systems and finally present our conclusions.

\section{Modified Gross-Pitaevskii Equation}\label{MGPeq}
We assume that the condensate can be described by the GP equation.
Since we are interested in the ultra-cold regime, where the
temperature is much smaller than the critical temperature for
condensation, we adopt the \mbox{$T=0$} formalism. In order to include
higher-order effects in the two-body scattering dynamics we use the
modified GP equation derived in \cite{collin07}, which in the
stationary form reads
\begin{equation}
  \left[-\frac{\hbar^2}{2m}\nabla^2
    +V_{ext}(r)+U_0\left(|\Psi|^2+g_2\nabla^2|\Psi|^2\right)
  \right]\Psi=  \mu\Psi,
\label{GPE}
\end{equation}
where $m$ is the atomic mass, $V_{ext}$ is the external trap,
$U_0=4\pi\hbar^2 a/m$, and $g_2=a^2/3-aR_e/2$, with $a$ and $R_e$
being respectively the $s$-wave scattering length and effective range
\cite{collin07}. We assume an isotropic trap, $V_{ext}(r)=m\omega^2
r^2/2$, and introduce the trap length $b_t=\sqrt{\hbar/m\omega}$. The
single-particle density, $\rho(r)=|\Psi(r)|^2$, is normalized to the
particle number, $N=\int\ud{\bm r}\rho(r)$, and $\mu$ is the chemical
potential.

As the boundary conditions are important for the TF 
approximation applied below we now discuss the procedure for obtaining
the modified GP equation from the corresponding energy functional which is
\begin{equation}\label{efunc-tf}
  E(\Psi)=\int \ud{\bm r} (
  \epsilon_K +\epsilon_V +\epsilon_I +\epsilon_{I2} ),
\end{equation}
with kinetic, potential and interaction energy densities 
\begin{align}
  \epsilon_K&=\dfrac{\hbar^2}{2m}|\nabla\Psi|^2,&
  \epsilon_V&=V_{ext}(\bm r)|\Psi|^2,\\
  \epsilon_I&=\dfrac{1}{2}U_0 |\Psi|^4,&
  \epsilon_{I2}&=\dfrac{1}{2}U_0 g_2|\Psi|^2 \nabla^2|\Psi|^2.
\end{align}
The corresponding integrated energy contributions are denoted $E_K$,
$E_V$, $E_I$, and $E_{I2}$ respectively.  To obtain eq.~\eqref{GPE},
we vary eq.~\eqref{efunc-tf} with respect to $\Psi^*$ for fixed
$\Psi$.  To first order in $\delta\Psi^*$ we have
\begin{equation}\label{GPvar}
  \begin{split}
    \delta E
    &=E[\Psi^*+\delta\Psi^*]-E[\Psi^*]\\
    &=\int\ud{\bm r} \bigg[ -\frac{\hbar^2}{2m}\nabla^2\Psi+ V_{ext}(\bm r)\Psi
    +U_0\left( |\Psi|^2+g_2\nabla^2|\Psi|^2\right)\Psi \bigg] \delta\Psi^*\\
    &+\int\ud{\bm S} |\Psi|^2\nabla\left(\Psi\delta\Psi^*\right)
    -\int\ud{\bm S} \Psi\delta\Psi^*\nabla|\Psi|^2
    +\int\ud{\bm S} \delta\Psi^*\nabla\Psi.
  \end{split}
\end{equation}
Here $\bm S$ is the outward-pointing surface normal.  In the usual
analysis one assumes that $\Psi$ and $\nabla\Psi$ vanishes at
infinity, drops the boundary terms, and eq.~\eqref{GPE} is obtained by
varying $E-\mu N$.  However, the existence of these surface terms are
essential for the inclusion of higher-order interactions as discussed below.

In the rest of this chapter we use trap units, $\hbar\omega=b_t=1$,
i.e.  energies ($E$, $V_{ext}$, $\mu$, etc.) are measured in units of
$\hbar\omega$ and lengths ($a$, $R_e$, $r$, etc.) in units of
$b_t$. Note that $g_2$ has dimension of length squared.

\section{Thomas-Fermi Approximation}\label{TFA}
Let us briefly review the standard Thomas-Fermi approximation
\cite{baym96,dalfovo99,pethick02,pitaevskii03}. Neglecting the kinetic
energy term, as compared to the trap and interaction energies, the GP
equation has the solution
\begin{equation}
  \label{eq:density-TF}
  \rho_{TF}=\frac{1}{4\pi a}(\mu_{TF}-\frac{1}{2}r^2),
\end{equation}
with chemical potential $\mu_{TF}$. This solution is used out to the
surface, $R_{TF}$, while outside $\rho_{TF}=0$. The
normalization and surface condition $\rho_{TF}(R_{TF})=0$ gives
\begin{equation}
  \label{eq:mu-rmax-TF}
  \mu_{TF}=\frac{1}{2}R_{TF}^2,\qquad
  R_{TF}=(15Na)^{1/5}.
\end{equation}
The total energy becomes
\begin{equation}
  \frac{E_{TF}}{N}=\frac{5}{7}\frac{R_{TF}^2}{2}.
\end{equation}
The trap and interaction energies are $E_V=3E/5$ and $E_I=2E/5$,
respectively. Since $R_{TF}>0$ in eq.~\eqref{eq:mu-rmax-TF}, these
results only hold for $a>0$.
The TF approximation is good for $Na\gg 1$, except at the surface
region where the kinetic energy density diverges. Here the solution
can be corrected as in \cite{dalfovo96,lundh97,pethick02,pitaevskii03}, essentially
giving a small exponential tail.

\subsection{Inclusion of Higher-Order Interactions}
We now consider the TF approximation with the
higher-order interaction term, $\epsilon_{I2}$. Ignoring the boundary
terms in eq.~\eqref{GPvar}, the modified GP equation can then be
written in terms of the density $\rho(r)=|\Psi(r)|^2$ as
\begin{equation}
  \label{eq:gpe-dimless}
  {\mu} =\frac{1}{2} r^2+4\pi a \left(\rho+ g_2
    \nabla^2 \rho\right).
\end{equation}
With scaled coordinate $x=r/\sqrt{g_2}$ (assuming $g_2>0$ for the
moment) and density $f(r)=4\pi a x\rho(r)/g_2$, this becomes
\begin{equation}
 \frac{\ud^2 f}{\ud x^2}+f= \frac{\mu}{g_2} x -\frac{1}{2} x^3,
\end{equation}
The inhomogeneous and homogeneous solutions with boundary condition
$f(0)=0$ are
\begin{equation}
  \label{eq:f-solution}
  f_i(x)=(\frac{\mu}{g_2}-\frac{1}{2}x^2 +3) x, \quad
  f_h(x)=\frac{A}{g_2} \sin x,
\end{equation}
where $A$ is a constant (with dimensions of length squared) to be
determined later. The full solution is
\begin{equation}
  \label{eq:density-solution}
   \rho( x)=\frac{g_2}{4\pi a}\left[
    \frac{\mu}{g_2} -\frac{1}{2} x^2 +3
    +\frac{A}{g_2}\frac{\sin x}{x}
  \right].
\end{equation}
For a given $A$, the chemical potential $\mu$ and the condensate
radius $R$ are determined by the normalization and the surface
condition,
\begin{equation}
  \label{eq:norm-surf-cond}
  \int_{0}^{x_0} 4\pi x^2 \rho(x)\ud r=N
  \quad\textrm{and}\quad
  \rho(x_0)=0,
\end{equation}
where $x_0=R/\sqrt{g_2}$.
The solution $\rho$ should be positive for $x<x_0$ which must be
explicitly checked. Outside $x_0$ we use $\rho=0$.

We now consider the boundary terms in eq.~\eqref{GPvar}. Above we
assumed that $\rho(x_0)=0$ at some finite radius $x_0$ which we
identify as the condensate size. However, only the first two boundary
terms in eq.~\eqref{GPvar} vanish on account of this condition.  For
the last term in eq.~\eqref{GPvar} to vanish we need
$\nabla_x\Psi(x_0)=0$, which implies that
\begin{equation}
  \label{eq:surface-derivative}
  \frac{d\rho}{dx}(x_0)=0.
\end{equation}
Notice that this latter derivative is in fact non-zero in the $g_2=0$
case, which is the root of the divergence of the kinetic energy at the
condensate surface as we discuss later.
Equation~(\ref{eq:surface-derivative}) gives a closed expression for
the remaining free parameter $A$,
\begin{equation}
  \label{eq:A-vs-z}
  \frac{A}{g_2}=\frac{x_{0}^{3}}
  {x_0\cos x_0-\sin x_0}.
\end{equation}

This additional requirement on the derivative at the edge of the
condensate implies that higher-order terms require a smoothing at the
surface of the cloud. 
In addition, the
discussion of which kinetic operator structure to use
($|\nabla\Psi|^2$ or $\Psi^*\nabla^2\Psi$ \cite{lundh97}) is
obsolete in our treatment since the boundary term $\delta\Psi^*\nabla\Psi$
vanishes. In this sense the
inclusion of a higher-order term neatly removes some of the
difficulties of the traditional TF treatment.

The solutions with a finite boundary $R$ of the modified GP equation
only minimize the energy functional if
eq.~\eqref{eq:surface-derivative} holds. We note that extremal states
of the energy functional always satisfy the virial theorem. Thus,
enforcing the virial theorem on the GP solutions is equivalent to
eq.~\eqref{eq:surface-derivative}. For completeness, we show in
appendix~\ref{chap:appendix-tf} that the virial theorem approach also
leads to eq.~\eqref{eq:A-vs-z}.

\section{Size and Chemical Potential}\label{sizechem}
We now determine the condensate size $R$ and chemical potential $\mu$.
The normalization condition is
\begin{equation}
  \label{eq:normalization-condition2}
  \frac{Na}{g_2^{5/2}}=x_0^3\left(\frac{\mu}{3g_2}-\frac{x_0^2}{10}\right),
\end{equation}
while the surface condition reads
\begin{equation}
  \label{eq:surface-condition2}
  \frac{\mu}{g_2} -x_0^2/2+3+\frac{A}{g_2}\frac{\sin x_0}{x_0} =0.
\end{equation}
Combining eq.~\eqref{eq:A-vs-z},
eq.~\eqref{eq:normalization-condition2}, and
eq.~\eqref{eq:surface-condition2} gives
\begin{equation}
  \label{eq:rmax-equation}
  \frac{Na}{g_2^{5/2}}=x_0^3\left(\frac{x_0^2}{15}-1+\frac{x_0^2/3}{1-x_0\cot x_0}\right),
\end{equation}
which determines $R$ for given $Na$ and $g_2$, and upon 
back-substitution also $\mu$.

\begin{figure}[tbhp]
  \centering
  \includegraphics[scale=1.0]{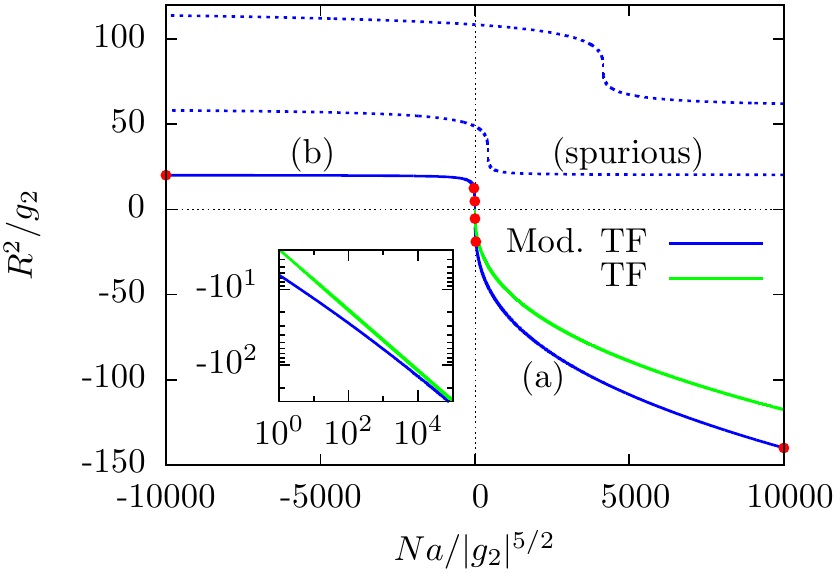}
  \caption{ Condensate size ($R$) as function of $Na$ and $g_2$ as
    found in the modified TF approximation,
    eq.~\eqref{eq:rmax-equation2}.  The solutions (a) and (b)
    correspond to the sign combinations ($a>0$, $g_2<0$) and ($a<0$,
    $g_2>0$), respectively.  No solutions exist for $a g_2>0$. The
    spurious solutions (dotted) have negative densities for one or
    more intervals inside $R$. The branch (a) approaches the normal TF
    result eq.~\eqref{eq:mu-rmax-TF} when $Na\to+\infty$ or $g_2\to
    -0$. Note that the convergence is only relative, see
    eq.~\eqref{eq:rmax-equation2}, and the TF limit is better
    represented in the logarithmic inset. Points indicate the data
    from tab.~(\ref{tab:data}). All values are in trap units.}
  \label{fig-rmax}
\end{figure}

The $g_2<0$ case can be worked out analogously by replacing trigonometric 
functions with hyperbolics and keeping track of signs. The two cases
can in fact be combined into one equation
\begin{equation}
  \label{eq:rmax-equation2}
  \frac{Na}{|g_2|^{5/2}} 
  =|x_0|x_0^2\left(\frac{x_0^2}{15}-1+\frac{x_0^2/3}{1-|x_0\cot x_0|}\right).
\end{equation}
This equation determines $x_0^2=R^2/g_2$ implicitly as function of $N
a/|g_2|^{5/2}$. The result is shown in fig.~(\ref{fig-rmax}). We
notice that in principle $R$ becomes a multi-valued function. However,
all the higher solutions for $g_2>0$ (dotted in fig.~(\ref{fig-rmax}))
are spurious, since the density becomes negative on one or more
intervals inside $R$.  The non-spurious solutions (solid line in
fig.~(\ref{fig-rmax})) defines $R$ as a single-valued function of $a$
and $g_2$, which was not guaranteed a priori.  The four quadrants in
fig.~(\ref{fig-rmax}) correspond to the different sign combinations of
$a$ and $g_2$.  The sign of the extra interaction energy, $E_{I2}$, is
determined by $a g_2\nabla^2\rho$. For a typical concave density the
Laplacian term will be negative. We therefore see that for $ag_2>0$
the higher-order interaction is attractive, whereas for $ag_2<0$ it is
repulsive.  The TF solution only exists for $a g_2<0$. We discuss both
cases separately below.

\subsection{The Attractive Regime: $ag_2>0$}

For $a<0$, $g_2<0$ (third quadrant in fig.~(\ref{fig-rmax})) there are
no solutions, which is expected since the normal TF approximation has
no solutions for $a<0$ as the interaction energy $E_I$ is negative
and the kinetic energy that could prevent collapse is neglected.

The $g_2>0$, $a>0$ case in the first quadrant has only spurious
solutions. Here the $g_2$ term is attractive for the typical concave
density and a collapse towards a high-density state is possible in
complete analogy to the usual discussion of attractively interacting
condensates within the standard GP theory. Whereas there can be
metastable states at large values of $Na/g_{2}^{5/2}$, these are
stabilized by kinetic energy and thus are not present in our TF
approach.  Thus, even when the the total kinetic energy is small, it
is still needed to prevent the attractive higher-order term from
amplifying local density variations.

This important point can also be established by considering the
stability of the homogeneous condensate through linearization of the
GP equation.  By repeating the analysis of \cite{pethick02} with the
higher-order term, we find that for $g_2>0$ and $a>0$ the kinetic
energy term is crucial for the stability of the excitation modes. In
fact, exponentially growing modes will always be present if the
kinetic energy is neglected.  This will be discussed elsewhere in
relation to the numerical solution of the full GP equation
\cite{zinner09b}.

\subsection{The Repulsive Regime: $ag_2<0$}
For $g_2<0$, $a>0$ a single solution (a) exists.  This was expected
since $E_{I2}>0$ gives extra stability.  The solution approaches the
normal TF result in eq.~\eqref{eq:mu-rmax-TF} when
$Na/|g_2|^{5/2}\to+\infty$, as can also be seen from
eq.~\eqref{eq:rmax-equation2}. Of course in this limit $E_{I2}\ll
E_I$.  However, the convergence in terms of $Na/|g_2|^{5/2}$ is only
on a relative scale, see inset in fig.~(\ref{fig-rmax}) and
eq.~\eqref{eq:rmax-equation2}.

For $g_2>0$,
$a<0$ there is a single solution (b) which connects smoothly to the
(a) solution.  In the limit $Na/|g_2|^{5/2}\to -\infty$, which is
determined by $x_0\cot x_0=1$, we find $R^2/g_2=20.1907$. This solution
is possible when the $g_2$ term provides just enough repulsion to
cancel the usual $a<0$ collapse behavior.

\subsection{Chemical Potential}
\begin{figure}[tbhp]
  \centering
  \includegraphics[scale=1.0]{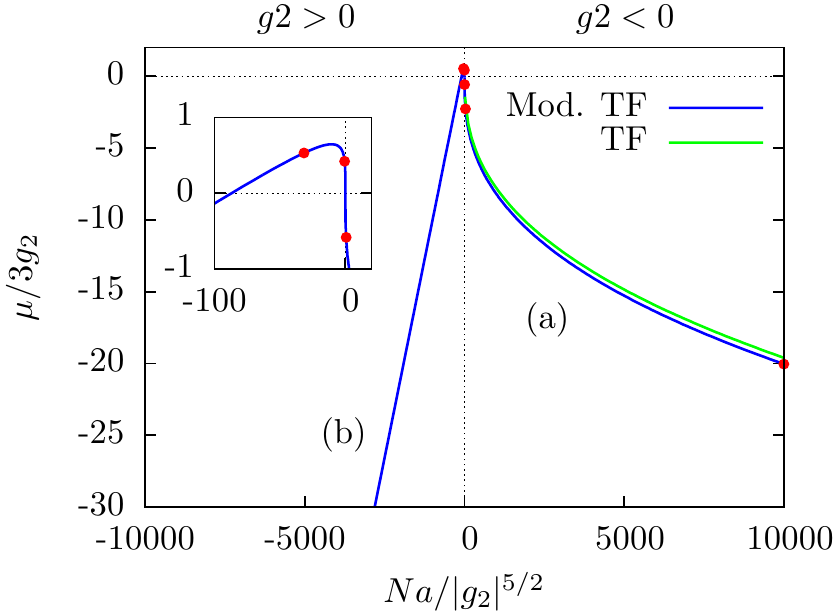}
  \caption{ Chemical potential $\mu$ as function of $Na$
    and $g_2$ as found in the modified TF approximation, using the
    solutions (a) and (b) from fig.~(\ref{fig-rmax}).  For branch (a)
    and the upper part of branch (b) (see inset) we have $\mu>0$. The
    lower part of (b) has $\mu<0$. Points indicate the data from
    tab.~(\ref{tab:data}). All values are in trap units.}
  \label{fig-mu}
\end{figure}

In fig.~(\ref{fig-mu}) we show the chemical potential for the smoothly
connecting solutions (a) and (b). Again we see that (a) approaches the
normal TF limit for large $Na/|g_{2}|^{5/2}$. Here it is interesting
to note how $\mu$ turns around near the origin (amplified in the inset
in Fig~(\ref{fig-mu})) and maintains a positive value. This occurs in
the region where the lowest-order interaction gives a large negative
energy contribution which the $g_2$ term is still able to balance
yielding a well-defined TF solution. This behavior is analogues to the
balancing of attraction by the kinetic term in the usual $a<0,g_2=0$
case \cite{baym96,dalfovo99}. As $a$ becomes increasingly negative so
too does $\mu$ and collapse is inevitable (and likewise when
$g_2\rightarrow 0^+$).

\section{Densities and Energies}\label{densene}
With $R$ and $\mu$ determined we can find the density profile, energy
densities and integrated energy contributions. With 
eq.~\eqref{eq:density-solution} the energy densities are given by
\begin{equation}
  \epsilon_V=\frac{ x^2}{2} \rho,\qquad
  \epsilon_I=2\pi  a \rho^2,\qquad
  \epsilon_{I2}  =-\frac{1}{2} \rho(3
  +\frac{A}{ g_2 } \frac{\sin x}{x}).
\end{equation}
Using eq.~\eqref{eq:gpe-dimless} the total energy density (without
$\epsilon_K$) becomes
\begin{equation}
  \label{eq:energy-dens}
  \epsilon \equiv  \epsilon_V+\epsilon_I+\epsilon_{I2}
  =\frac{1}{2} \rho(x) (V_{ext}(x)+\frac{\mu}{g_2}).
\end{equation}

\begin{figure}[tbhp]
  \centering
  \includegraphics[scale=1.0]{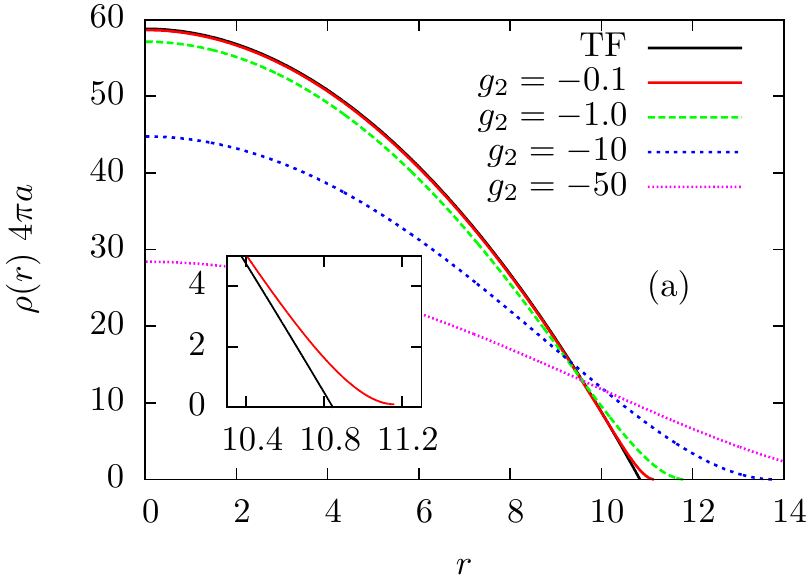}
  \caption{ Densities for branch (a) in
    fig.~(\ref{fig-rmax}) ($g_2<0$ and $Na=10^4$). The $g_2=-0.1$ curve
    is on top of the normal TF result. The inset shows the smooth behavior at the surface
    for $g_2<0$.
    All values are in trap units.}
  \label{fig-dens-a}
\end{figure}

In fig.~(\ref{fig-dens-a}) we show the density profile of the (a)
solutions for $Na=10^4$ and selected $g_2<0$. We clearly see that the
higher-order term tends to expand the condensate through its
repulsion. Importantly, at the boundary there is a smoothing caused by
the condition in eq.~(\ref{eq:surface-derivative}), see inset in
fig.~(\ref{fig-dens-a}). We will discuss how this affects the
estimated kinetic energy in the next section.  As $|g_2|$ grows we see
the condensate flatten and in the limit of very large $|g_2|$ it
becomes a constant density.

\begin{figure}[tbhp]
  \centering
  \includegraphics[scale=1.0]{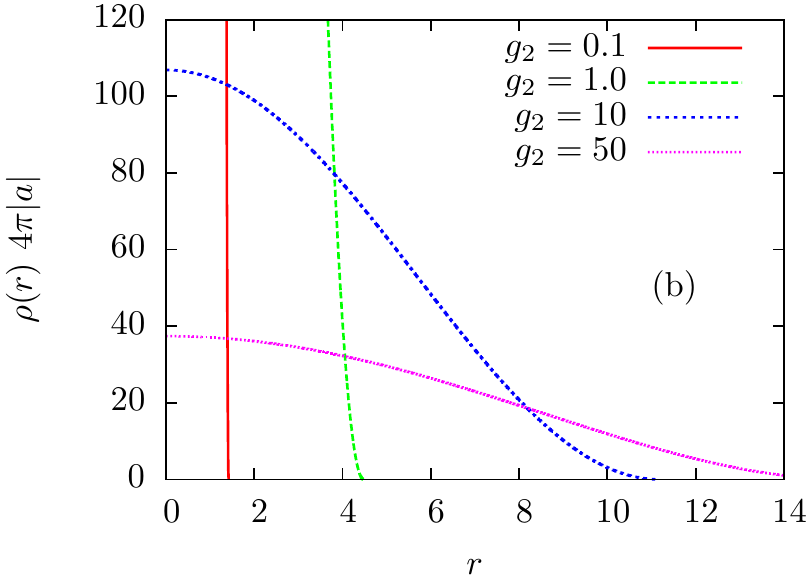}
  \caption{ Same as fig.~(\ref{fig-rmax}) but for
    solutions (b), i.e. opposite signs $g_2<0$ and $Na=-10^4$. }
  \label{fig-dens-b}
\end{figure}

Figure~(\ref{fig-dens-b}) displays the density profile for the (b)
solutions with $a<0$ for selected $g_2>0$. Here we see the profile
collapse towards the expected delta-function with decreasing $g_2$.
It is interesting to follow the (a) solution through the origin in
fig.~(\ref{fig-rmax}) and onto branch (b), passing from
$g_2=-\infty$ to $g_2=\infty$. On the (a) branch the solution flattens
as $g_2$ decreases and eventually becomes effectively constant in
space. This is also true for the (b) branch at $g_2=\infty$, and as
$g_2$ is decreased the solution proceed to shrink as the $g_2$ term
becomes unable to provide the repulsion needed to prevent the $a<0$
collapse induced by the lowest-order term.

From the figures we see that large $|g_2|$ induces large changes in cloud size. As the condensate
can be imaged with very good resolution \cite{hau98}, this should be measurable if the 
regime of large $|g_2|$ can be accessed.

{\renewcommand{\tabcolsep}{0.3cm}
\begin{sidewaystable}[p]
\begin{threeparttable}
  \centering
  \begin{tabular}{cr|cccccccc}
    \hline \hline
   &$g_{2}$\phantom{ \tnote{$\dagger$}}&$R$&$\mu$&$E_{V}/N$&$E_{I}/N$&$E_{I2}/N$&$E/N$&$E_R/N$&$E_{K}/|E|$\\
    \hline
    TF&---\phantom{ \tnote{$\dagger$}}&10.8447&58.8040&25.2017&16.8011&---&42.0028&16.8012&3.135$\times10^{-3}$ \tnote{$\star$} \\
    \hline 
       &$-0.01$\phantom{ \tnote{$\dagger$}}&10.9447&58.8188&25.2164&16.7865&0.01465&42.0176&16.8012& 1.8$\times10^{-3}$\\
       &$-0.1$\phantom{ \tnote{$\dagger$}}&11.1607&58.9481&25.3430&16.6635&0.13909&42.1456&16.8026& 1.4$\times10^{-3}$\\
    (a)&$-1.0$ \tnote{$\dagger$}&11.8364&60.1210&26.4309&15.6818&1.16330&43.2760&16.8451& 1.0$\times10^{-3}$\\
       &$-10$ \tnote{$\dagger$} &13.7835&68.4515&32.9856&11.3469&6.38609&50.7186&17.7330& 0.57$\times10^{-3}$\\
       &$-50$ \tnote{$\dagger$} &16.439 &87.8248&45.6836&6.99293&14.0777&66.7542&21.0706&0.30$\times10^{-3}$\\
    \hline
       &50 \tnote{$\dagger$} &15.407&63.0102&38.9723 &$-8.92496$&20.9439&50.9912 &12.0189&0.43$\times10^{-3}$\\
       &10 \tnote{$\dagger$} &11.170&15.9128&19.8375 &$-24.7434$&22.7810&17.8751 &$-1.9623$&2.3$\times10^{-3}$\\
    (b)&5.14 \tnote{$\ddagger$}&9.1999&$-13.1384$&13.1579&$-46.0283$&32.8801&0.0097  &$-13.148$&6.098\\
       &1.0 \tnote{$\dagger$}&4.4801&$-327.612$&3.04199&$-416.359$&251.032&$-162.285$&$-165.33$&1.5$\times10^{-3}$\\
       &0.1\phantom{ \tnote{$\dagger$}}&1.4204 &$-10456.4$&0.30571&$-13071.1$&7842.81&$-5227.98$ &$-5228.4$&0.47$\times10^{-3}$\\
    \hline \hline
  \end{tabular}
  \begin{tablenotes}
    \scriptsize
  \item[$\star$] The kinetic energy estimated by surface corrections as in \cite{pethick02}.
  \item[$\dagger$] Values are indicated by points in fig.~(\ref{fig-rmax}) and (\ref{fig-mu}).
  \item[$\ddagger$] The total energy $|E|$ is zero near $g_2=5.14$, hence the TF approximation is
    invalid here.
    \end{tablenotes}
    \caption{Condensate size $R$ and chemical potential $\mu$ for
      different $g_2$ and fixed $N|a|=10^4$. Region (a) has $a>0$ and
      region (b) has $a<0$.  The integrated energies are trap ($E_V$),
      interaction ($E_I,E_{I2}$), total ($E=E_V+E_I+E_{I2}$), and
      release energy ($E_R=E-E_V$).  The TF limit is approached for
      $g_2\to -0$. The ratio of kinetic energy $E_K$ to total energy
      $E$ indicates where the TF approximation is valid. The
      corresponding density distributions are shown in
      fig.~(\ref{fig-dens-a}) and (\ref{fig-dens-b}). All values are
      in trap units.}
  \label{tab:data}
  \end{threeparttable}
\end{sidewaystable}
}

\begin{figure}[tbhp]
  \centering
  \includegraphics[scale=1.0]{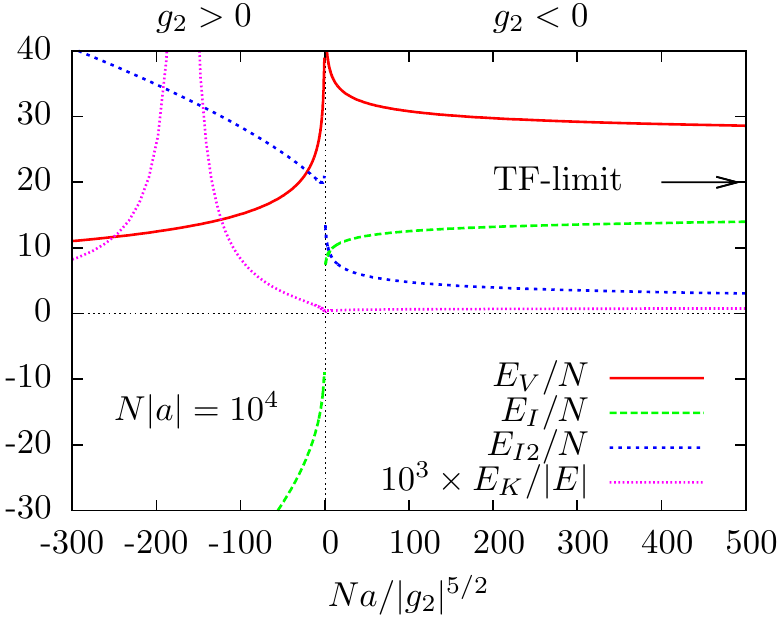}
  \caption{Different total energy contributions.
    $N|a|=10^4$. Values are in trap units.}
  \label{fig-energy}
\end{figure}

We now discuss the energy contributions which are interesting since the 
release energy are in fact measurable quantities \cite{pitaevskii03}. Since we neglect
the kinetic term in the TF approximation, the release energy is simply $E_R=E_I+E_{I2}=E-E_V$.
In tab.~(\ref{tab:data}) we give the integrated energy contributions for some relevant values of $g_2$
calculated for $N|a|=10^4$, whereas fig.~(\ref{fig-energy}) gives the energies as function of $Na/|g_2|^{5/2}$.
We note that for smaller values of $N|a|$ the same overall behavior is found,
however, the kinetic term is more important and the TF approximation becomes worse.

We observe that $E/N$ grows towards the $|g_2|=\infty$ point. This is
due to the trap energy increasing as the density flattens ($E_V$
diverges around the origin in fig.~(\ref{fig-energy})).  Furthermore,
as $g_2\rightarrow 0^+$ the energy diverges towards $-\infty$ as the
collapse sets in ($E_I$ diverges on the $g_2>0$ side in
fig.~(\ref{fig-energy})).  The boundary where the energy vanishes is
around $g_2=5.14$ for $N|a|=10^4$, but this depends on the choice of
$N|a|$. With respect to the release energy, we find that somewhere in
the region $10<g_2<50$, $E_R$ becomes negative. This is a result of
the unavoidable collapse, and also indicates that kinetic energy
cannot be ignored at this point. Notice, however, that the release
energy changes considerably and could provide a way to measure the
influence of the $g_2$ term.

\section{Thomas-Fermi Approximation: Consistency}
\label{cons}
We now address the validity of the TF approximation with the $g_2$ term included. In 
order to do so we must consider the contribution of the kinetic energy.
The kinetic energy density can be written
\begin{equation}
  \label{eq:eK}
  \epsilon_K
  =\frac{g_2}{8 \rho(4\pi a)^2}\left(
     x+\frac{A}{g_2}
    \frac{x
    \sin x
    -\cos x}{x}
  \right)^2.
\end{equation}
Strictly speaking, this is not the true kinetic energy, since the
kinetic term were neglected from the start. However
eq.~\eqref{eq:energy-dens} and eq.~\eqref{eq:eK} can be used to test
whether the TF approximation holds locally, i.e. $\epsilon_K \ll
\epsilon$ should hold for the solution $\rho$ to be consistent. In
tab.~(\ref{tab:data}) we calculate the integrated contribution of the
kinetic energy relative to the total TF energy, and we find that the
contribution is small everywhere except the point where $E=0$ on the
$g_2>0$ side of fig.~(\ref{fig-energy}). Here the kinetic energy is of
course the most important term and the TF approximation is poor.

In the standard TF, the kinetic energy causes trouble at the boundary of the cloud.
Here $\nabla\Psi\propto \nabla\rho/\sqrt{\rho}$ and since the density vanishes 
and the derivative is finite (see eq.~(\ref{eq:density-TF})) this diverges at $R_{TF}$.
When including the higher-order term we need to use the additional boundary 
condition $\nabla\Psi=0$ at $R$, so the kinetic energy will be strictly zero
at $R$. However, as one approaches the boundary the kinetic energy density grows 
rapidly before it descends towards zero within a very small interval 
at $R$. The total energy density in eq.~(\ref{eq:energy-dens}) goes to zero at this point 
and we find that $\epsilon_K/\epsilon$ is very large near the boundary as in 
the usual $g_2=0$ case.

We conclude that the inclusion of the higher-order term does not alleviate the 
difficulties with kinetic energy at the boundary. The techniques for 
addressing this problem described in \cite{dalfovo96,lundh97} should therefore
be generalized to include the higher-order interaction term in order to 
improve the description at the boundary of the cloud.

\section{Comparison to Atomic Systems}\label{compare}
The considerations above show that deviations from the usual TF approximation
can be strong when $g_2$ is large. In the following we reintroduce explicit units
for comparison with real systems. We have to consider $g_2/b_t^2$.
Of course the $b_t^2$ factor means
that this quantity is generally very small since $g_2$ is of order $a_{0}^{2}$
and $b_t$ is of order $10^4 a_0$.

We first consider some typical 
background values for bosonic alkali atoms away from Feshbach resonances. We estimate
the effective range to be of order the potential range, and assuming a 
van der Waals interaction we have $R_e\sim 50-200a_0$.
For typical one-component gases we have 
$-450a_0\lesssim a \lesssim 2500a_0$ \cite{chin09}. Since $g_2=a^2/3-aR_e/2$,
we see that the $a^2$ term will dominate and in all cases
$0<g_2\lesssim 10^6a_0$. In trap units this becomes
$g_2/b_t^2\lesssim 5\times10^{-3}(1\mu\text{m}/b_t)^2$.
In typical traps of $b_t\sim 1-10\mu\text{m}$ 
the higher-order term is therefore very small. 
These values also predominantly lie in the first quadrant of 
fig.~(\ref{fig-rmax}) and thus no TF solution exists. 

We now consider Feshbach resonances in order to increase the influence of
the $g_2$ term.
We use a multi-channel Feshbach model \cite{bruun05,zinner09a},
see chapter~\ref{chap:mean-field},
which describes both $a$ and $R_e$ as a function of resonance position
$B_0$, width $\Delta B$, magnetic moment difference between the
channels $\Delta \mu$, and the background scattering length $a_{bg}$.
We have $a=a_{bg}(1-\Delta B/(B-B_0))$ and
$R_e=R_{e0}/(1-(B-B_0)/\Delta B)^2$, where
$R_{e0}=-2\hbar^2/ma_{bg}\Delta \mu\Delta B<0$. 
Notice that $R_e=R_{e0}(1-a_{bg}/a)^2$ and
\begin{equation}
  g_2(a)=\frac{a^2}{3}-\frac{a R_{e0}}{2}(1-\frac{a_{bg}}{a})^2.
\end{equation}
Hence $g_2$ diverges when $a\to 0$ (referred to as zero-crossing) or
$a\to\infty$ (on resonance). Near a zero-crossing $g_2$ behaves as $a
g_2\simeq |R_{e0}|a_{bg}^2/2$, where $R_{e0}<0$.

As a concrete example, we consider the alkali isotope $^{39}$K where
several Feshbach resonances of vastly different widths were found
recently \cite{derrico07}.  First we focus on zero-crossing and
consider the very narrow resonance at $B_0=28.85$G with $\Delta
B=-0.47$G, $\Delta \mu=1.5\mu_B$ and $a_{bg}=-33a_0$.  We obtain
$R_{e0}=-5687a_0$ and $ag_2\rightarrow 93.8\times 10^3a_{0}^{3}$ for
$a\rightarrow 0$.  It is important to notice that $ag_2>0$ around
$a=0$. This means that we are looking for solutions in the first and
third quadrant of fig.~(\ref{fig-rmax}), and again we have to conclude
that no TF solutions can be found when higher-order terms are taken
into account.

Another case of interest is around resonance where $|a|=\infty$. Here
we have $R_e\sim R_{e0}$ and $g_2\propto a^2>0$ on both sides of the
resonance. Thus the $a>0$ side will be in the first and the $a<0$ in
the second quadrant of fig.~(\ref{fig-rmax}).  This makes it difficult
to imagine sweeping the resonance from either side to probe the
solutions on branch (b) in fig.~(\ref{fig-rmax}). One could imagine
starting on the $a>0$ side with small $g_2>0$. The full GP equation
will have perfectly sensible solution here, however, when one
approaches the resonance the $g_2$ term will diverge and induce
collapse already on the $a>0$ side. If we approach from the $a<0$ side
then we face the problem that the critical number of particles
decreases dramatically before $g_2$ grows sufficiently, and one
therefore needs a very small condensate since $Na/b_t\sim 0.5$
\cite{zinner09a}. At this point the TF approximation is no longer
valid.

From the examples above we see considerable problems in accessing the
TF solutions presented above in current experiments with ultra-cold
alkali gases. In particular, we notice that realistic systems which
have been used for creation of BECs in alkali gases for the last
decades have parameters that predominantly lie in the first quadrant
of fig.~(\ref{fig-rmax}). As we have discussed there are no
well-defined TF solutions in that region. Therefore we see that the
kinetic energy plays a decisive role and we are forced to consider it
in principle, even if it is small for all practical purposes. The
physical reason is that for $a>0$ and $g_2>0$ the higher-order
interaction is effectively attractive and induces collapse which will
have to be balanced by a barrier from the kinetic term, similar to the
$a<0$, $g_2=0$ case \cite{dalfovo99}. Since we neglect the kinetic
term in the TF approximation we should not expect to find solutions in
the $ag_2>0$ case.

\section{Conclusion and Outlook}\label{conc}
We have considered the effect of higher-order interactions in
Bose-Einstein condensates within the Gross-Pitaevskii theory. We
derived the GP equation with effective range corrections included and
solved it analytically in the Thomas-Fermi approximation.
Higher-order interaction terms act as derivatives on the condensate
wave function which means that the boundary conditions on the
solutions of the GP equation must be carefully considered.  We then
discussed the solutions for various parameters, presented the chemical
potential, density profiles, and the energy contributions.

We find that no TF solutions are possible when the higher-order term
is attractive. This conclusion holds both in the trapped system and in
the homogeneous case.  An estimate of the relevant
parameters for alkali atoms showed, however, that they typically lie
in the region where the effective range correction is effectively
attractive. In order for this to be consistent with current
experiments that have found the standard TF approximation to be quite
good we conclude that the kinetic energy, even if very small, is
crucial in order to stabilize collapse due to higher-order interaction
terms.
A full numerical analysis is in progress \cite{zinner09b}.

%% file: summary.tex
In this dissertation we presented theoretical investigations of
universality and finite-range corrections in few- and many-boson
systems. The main focus was on ultra-cold atomic gases seen both from
the three-body, many-body, and mean-field perspective.

In the introductory chapter~\ref{chap:intro} we gave a soft
introduction to the concept of universality in few- and many-body
systems, in particular the wide range of effects in atomic gases and
Bose-Einstein condensates. We also focused on the search for the
elusive Efimov effect and its relation to nuclei and atomic gases.
In chapter~\ref{chap:theory} we presented the relevant theoretical and
numerical background for universal few- and many-body systems.

The purpose of chapter~\ref{chap:efimov} was to go beyond the
scattering length approximation for three-boson Efimov physics and
express the corrections in terms of model-independent parameters.  We
found universal scaling with corrections determined by the effective
range.  We showed that for negative scattering lengths the effective
range corrections to Efimov physics and Borromean binding are two
aspects of the same quantitative effect.  This leads to a linear shift
in critical scattering lengths at the trimer threshold.  For positive
scattering lengths near the atom-dimer threshold the effective range
corrections to the trimer energies are mainly determined by the dimer
corrections.  Results agree quantitatively with the newest effective
field theory predictions. This has consequences for three-body
recombination rates in atomic gases when the scattering length is
comparable to the effective range.  The main effect is that the
universal scaling factor becomes smaller for the lowest Efimov states
(smallest scattering length) on the negative scattering length side,
corresponding to a smaller Borromean window.  We also described the
effects of putting the system in a finite trap.

Chapter~\ref{chap:efimov2} continued this line by analytically
investigating the conditions for Efimov physics, in particular for
large effective range.  We used the adiabatic hyper-spherical
approximation to derive rigorously a transcendental equation for the
adiabatic potentials for general finite-range potentials. Solutions
agreed with exact numerical results.  We concluded that in the
hyper-spherical adiabatic approximation it is insufficient to include
effective range only, as another term of the same order appears.
However, non-adiabatic corrections restores model-independence.  For
large negative effective range the window for Efimov physics is
precisely open between the effective range (not the potential range)
and the scattering length.

In chapter~\ref{chap:N-efimov} we provided theoretical support for the
existence of long lived meta-stable $N$-body Efimov states in atomic
Bose-gases.  The inclusion of two-body correlations (or equivalently a
Faddeev-Yakubovski decomposition in two-body amplitudes) gave an
effective hyper-radial potential as for the three-body Efimov effect.
This in turn led to universal scaling, but with new scaling factors.
Results were confirmed both numerically and analytically.  The
experimental signatures in atomic recombination experiments were
discussed including possible connections to other universal
four-body effects.  The four-body Efimov states would be harder to
detect due to the low density of experiments, but also easier to
probe because of the smaller universal scaling factor.  The effect
could be applicable to Borromean systems too.

In chapter~\ref{chap:correlations} we considered trapped few-boson
systems with large positive scattering length. This was modeled using
attractive two-body potentials, giving the many-body system a large
amount of low-lying bound states.  We presented a novel numerical
technique to identify the BEC state as a highly excited many-body
state.  The obtained state contained short-range correlations
determined by the interaction, while the normal mean-field features
were retained at long distances. We found that for scattering lengths
smaller than the trap the system shows universality. For larger
scattering lengths the system properties becomes independent of the
scattering length, contrary to the zero-range and the repulsive
models.  The correlations become long-ranged and universal in the
large scattering length limit.  This is an important step towards
understanding the non-universal corrections in condensates.

Chapter~\ref{chap:mean-field} approached the question of effective
range corrections in condensates from the mean-field point of view. A
modified GP equation with effective range dependence was introduced.
Using variational and numerical approaches we found a phase diagram
describing the condensate stability.  We then considered an extended
Feshbach resonance model including effective range variations.
Effects on macroscopic quantum tunneling were small for realistic
systems.  The behavior of the critical particle number was modified
near a scattering length zero-crossing with observable consequences.

In chapter~\ref{chap:tf} we continued the mean-field analysis of the
condensates using the modified Gross-Pitaevskii equation.  We solved
this equation analytically in Thomas-Fermi approximation where the
kinetic energy can be neglected, but keeping higher-order interaction
terms.  Boundary conditions of the solutions were carefully considered.
We presented the chemical potential, density profiles, and the energy
contributions.  No Thomas-Fermi solutions were possible when the
higher-order term was attractive. This holds both in the trapped and
homogeneous system.  However, realistic atomic systems typically lie
in the region where the term is attractive.  In order for this to be
consistent with current experiments, the kinetic energy, even if very
small, is crucial in order to stabilize collapse due to higher-order
interactions.

In conclusion, we have carried out new theoretical investigations of
universality and its limits in few- and many-boson systems.  The focus
was on ultra-cold trapped atomic gases, but results were presented in
universal model-independent terms. Thus, much of the work may
hopefully be used or continued in other areas of physics.
The subject of universality in atomic gases will face many new
challenges during the next decades.  Interesting new physics is
clearly within reach in ultra-cold gases and indicate a promising
future.

\vspace{1cm}
\paragraph{Note added.}
The observation of an Efimov spectrum in an ultra-cold gas has
recently been published \cite{zaccanti09} giving the first definite
proof of the universal scaling factors. We note that this experimental
group is independent of the group measuring the first isolated Efimov
state \cite{kraemer06}.

In \cite{zaccanti09} five experimental features were found. For
positive scattering length two interference minima were observed in the
recombination loss rate, corresponding to two Efimov states. Also on
this side two small peaks were found, corresponding to a secondary
loss mechanism. For negative scattering length a single peak was found
in the loss rate (the second Efimov state being out of the
experimental range).

All five features agree well with the theoretical universal scaling
factors except for systematic shifts. They conclude that the scaling
factors are reduced on the negative scattering length side and
increased on the positive side. They attribute this to finite range
corrections and compare with the effects in \cite{thogersen08-3} (i.e.
chapter~\ref{chap:efimov}) and the identical conclusions of effective
field theory \cite{platter09}. The theory and experiment agrees.  A
more precise analysis is needed at this point, but the need for
effective range corrections to Efimov physics is more evident now than
ever before.

%% file: appendix-obdm.tex
To calculate the condensate fraction of the states found with the
stochastic variational method in chapter~\ref{chap:correlations} we
first need to calculate the one-body density matrix (OBDM),
eq.~\eqref{eq:obdm}, for the two-body correlated wave-functions,
eq.~\eqref{eq:psi-2B}. We choose to change basis to the harmonic
oscillator eigenfunctions before diagonalization. This basis turns out
to be good, since the OBDM is almost diagonal and numerically only a
few basis states are needed to ensure that the occupation numbers sum
up to $N$.

Here we give an outline on how the matrix elements are calculated and
the condensate fraction is obtained.
First we expand the eigenfunctions of eq.~\eqref{eq:obdm-eigeneq} in
the harmonic oscillator basis,
\begin{equation}
 \chi_i(\bm{r})=\sum_{l m}a_{nl m}^{(i)}R_{nl}(r) Y_{l m}(\theta,\varphi)  ,
\end{equation}
\begin{equation}
  R_{nl}(r)=\sqrt{\frac{2g^3n!}{\Gamma(n+l+\frac{3}{2})}}
  e^{-\frac{1}{2}g^2r^2}(gr)^lL_n^{l+\frac{1}{2}}(g^2r^2),
\end{equation}
where $Y_{lm}$ are the spherical harmonics, $L_n^{l+\frac{1}{2}}$ are
the associated (or generalized) Laguerre polynomials
\cite{abramowitz95}, and $g$ is an arbitrary oscillator scale
(typically chosen as $b_t^{-1}$). In this basis the eigenvalue
equation eq.~\eqref{eq:obdm-eigeneq} takes the form
\begin{equation}
\sum_{n'l'm'} A_{nl m,n'l 'm'} a_{n'l' m'}^{(i)} = 
 N_i a_{nl m}^{(i)},
\end{equation}
where the matrix elements are given by
\begin{equation}
  \label{eq:obdm-ho-basis}
   A_{nl m,n'l 'm'}=\int  n(\bm{r},\bm{r}') R_{n'l'}(r') R^*_{nl}(r)
 Y_{l 'm'}(\Omega') Y^*_{l m}(\Omega)
 \ud\tau\ud\tau',
\end{equation}
with $\ud\tau=r^2\ud r\ud \Omega$.  Most work lies in the evaluation
of these matrix elements. It can be shown that the OBDM (in the
$\bm r$-basis) for the two-body correlated wave-function
eq.~\eqref{eq:psi-2B} has the form
\begin{equation}
  \label{eq:obdm-svm}
 n(\bm r,\bm r')= \sum_{kk'}\sum_{t=1}^9 a^{kk'}_t
  \exp\left(- b^{kk'}_t \bm r^2 - c^{kk'}_t 
 \bm r'^2 + d^{kk'}_t \bm r 
 \cdot \bm r' \right).
\end{equation}
The parameters $\{a_t^{kk'},b_t^{kk'},c_t^{kk'},d_t^{kk'}\}$ are
functions of $N$ and $\{C_k,\alpha^{(k)},\beta^{(k)}\}$.%
\footnote{The exact forms of the functions are analytical but very
  complicated. It is more convenient to do the transformation partly
  numerically.}
The summation over $k$ and $k'$ corresponds to the linear-combination
in eq.~\eqref{eq:psi-2B}. The $9$ different terms labeled by $t$ comes
from the symmetrization, which was done analytically. By expanding the
factor $\exp(d_t^{kk'}\bm{r\cdot r'})$ in eq.~\eqref{eq:obdm-svm} in
spherical harmonics and inserting it into eq.~\eqref{eq:obdm-ho-basis},
it is straightforward, although tedious, to calculate the matrix
elements. The result is

\begin{equation}
  \begin{split}
    A_{nl m,n'l'm'}=&\delta_{ll'}\delta_{mm'}
 \binom{k}{n}^\frac{1}{2}\binom{k}{n'}^\frac{1}{2}
  \sum_{kk't} a_t^{kk'}
  \frac{(g^2)^{3/2}x^ly_{+-}^ny_{-+}^{n'}}{y_{++}^{3/2+l+n+n'}}\\
  &\times {}_2F_1(-n,-n';-k;1-\frac{x^2}{y_{+-}y_{-+}}),\\
  \end{split}
\end{equation}
where
\begin{equation}
  \begin{split}
      y_{++}=&(b_t^{kk'}+g^2/2)(c_t^{kk'}+g^2/2)-(d_t^{kk'}/2)^2,\\
      y_{+-}=&(b_t^{kk'}+g^2/2)(c_t^{kk'}-g^2/2)-(d_t^{kk'}/2)^2,\\
      y_{-+}=&(b_t^{kk'}-g^2/2)(c_t^{kk'}+g^2/2)-(d_t^{kk'}/2)^2,\\
  x=&g^2 d_t^{kk'}/2,\qquad k=n+n'+l+\frac{1}{2}.
  \end{split}
\end{equation}
The hypergeometric function ${}_2F_1$ is here reducing to a Jacobi
polynomium in the last argument of degree min$(n,n')$
\cite{abramowitz95}.

We see that $A$ is diagonal in $l$ and $m$, and also independent of
$m$. This is because eq.~\eqref{eq:psi-2B} is rotationally invariant. We
can therefore diagonalize one block at a time. The eigenvalues and
eigenfunctions are re-labelled as $N_j^{(lm)}$ and
$\chi_j^{(lm)}(\bm r)$, with $j$ labeling the eigenvalues in
decreasing order. The condensate fraction will usually be the first eigenvalue
$\lambda_0=N_0^{(00)}/N$.

The matrix $A$ can now be diagonalized numerically and the condensate
fraction obtained as the largest eigenvalue.

%% file: appendix-corr.tex
The correlated wave-functions used in the stochastic variational method
opens up the possibility for the condensate to be depleted
considerably.  It is therefore useful to know how the different types
of correlations are able to lower the condensate fraction. In this
appendix we give some examples of strong hyper-radial and two-body
correlations.

\section{Strong Hyper-Radial Correlation}
If only one term ($k=1$) is included in the hyper-radial variational
function, eq.~\eqref{eq:psi-rho}, it reads
\begin{equation}\label{eq:psi-rho-single}
  \Psi_\rho=c_1 \exp \left(-\frac{N R^2}{2 b_t^2}\right)
  \exp\left(-\frac{1}{2}N\alpha^{(1)}\rho^2\right).
\end{equation}
In this case the occupation numbers $N_i$ defined in
eq.~\eqref{eq:obdm-eigeneq} can be obtained analytically. Since we are
considering a spherically symmetric trap, the total angular momentum
$l$ and its projection $m$ are good quantum numbers, and we can
re-label the occupation numbers as $N_j^{(lm)}$. The subscript $j$ labels
the eigenvalues in decreasing order for given $l,m$. By expanding in
the harmonic oscillator basis as in appendix~\ref{chap:appendix-obdm},
the result is
\begin{equation}
  N_j^{(lm)}=\frac{8N\xi^{l+2j}}{[1+\sqrt{1+\xi}]^{2l+4j+3}},
\end{equation}
with
\begin{equation}
    \xi=\frac{(\gamma-1)^2}{\gamma}\frac{(N-1)}{N^2},\qquad
    \gamma=N\alpha^{(1)}b_t^2.
\end{equation}
Here $\alpha^{(1)}$ is the non-linear parameter in
eq.~\eqref{eq:psi-rho-single}. Note that $\gamma=1$ corresponds to the
fully condensed gas where
$N_j^{(lm)}=\delta_{j0}\delta_{l0}\delta_{m0}N$.  When $\gamma$ is far
away from unity the condensate fraction can be made arbitrarily small.
This means that the condensate can be destroyed just because the
internal length scale (measured e.g. by $\sqrt{<\rho^2>}$) is
different from the center-of-mass length scale.  This conclusion still
holds for more $k$-terms in eq.~\eqref{eq:psi-rho}.  The same
conclusion is also found in \cite{gajda06}.

\section{Strong Two-Body Correlation}
To investigate the impact of two-body correlations on the condensate
fraction we choose one term ($k=1$) in
eq.~\eqref{eq:psi-2B} with
$N\alpha^{(1)} b_t^2=1$. The wave-function then has the form
\begin{equation}
  \label{eq:simple-2b}
  \Psi_{2B}=\prod_{i=1}^N \exp\left({-\frac{r_i^2}{2b_t^2}}\right)
  \times \hat S \exp\left(-\frac{1}{2}\beta^{(1)} r_{12}^2\right).
\end{equation}
This is a special case of the more general form
\begin{equation}
  \label{eq:general-2b}
  \Phi=\prod_{i=1}^N \phi(\bm r_i)\times \hat S f(\bm r_{12}),
\end{equation}
which we will investigate instead. It is a mean-field product state
modified by a correlation function. The real correlation function $f$
only includes one pair $\bm r_{12}$, and the function is subsequently
symmetrized. We can assume without loss of generality that $\phi$ and
$f$ are normalized to unity,
\begin{equation}
  \int|\phi(\bm r)|^2\ud\bm r=\int|f(\bm r)|^2\ud \bm r=1.
\end{equation}
We now consider the strong two-body correlated limit, where $f$
goes to zero over a range much smaller than the range over which $\phi$ varies, say
\begin{equation}
  \begin{split}
    f(\bm r)&\simeq0 \text{ for } \lvert\bm r\rvert>\epsilon,\\
    \phi(\bm r+\delta \bm r)&\simeq\phi(\bm r) \text{ for }
    |\delta \bm r|<\epsilon.
  \end{split}
\end{equation}
Here $\epsilon$ is some small length going to zero.  In this limit we
are able to calculate the eigenvalues and eigenfunctions of the OBDM,
eq.~\eqref{eq:obdm}. Since eq.~\eqref{eq:general-2b} is not normalized
we must first calculate the overlap integral. Because of the
symmetrization in eq.~\eqref{eq:general-2b} we get
$[N(N-1)/2]^2$ terms. Many of these terms are identical,
however, and only $3$ different types occur, namely
\begin{equation}
  \label{eq:olap1}
  \langle \Phi\vert\Phi\rangle=½N(N-1)[\langle12|12\rangle+2(N-2)\langle12|13\rangle+½(N-2)(N-3)\langle12|34\rangle],
\end{equation}
where
\begin{equation}
  \begin{split}
  \langle12|12\rangle
  &=\int|\phi(\bm r_1)|^2|\phi(\bm r_2)|^2 f(\bm r_{12})^2\ud \bm r_1\ud \bm r_2
  \simeq \int|\phi(\bm r)|^4\ud \bm r,\\
      \langle12|13\rangle
  &=\int|\phi(\bm r_1)|^2|\phi(\bm r_2)|^2|\phi(\bm r_3)|^2f(\bm r_{12})f(\bm r_{13})\ud \bm r_1\ud \bm r_2\ud \bm r_3\\
  &\simeq \int|\phi(\bm r)|^6\ud \bm r(\int f(\bm r)\ud \bm r)^2,\\
    \langle12|34\rangle
    &=\int|\phi(\bm r_1)|^2|\phi(\bm r_2)|^2|\phi(\bm r_3)|^2\phi(\bm r_4)|^2f(\bm r_{12})f(\bm r_{34})\ud \bm r_1\ud \bm r_2\ud \bm r_3\ud\bm r_4\\
    &\simeq (\int|\phi(\bm r)|^4\ud \bm r)^2(\int f(\bm r)\ud \bm r)^2.
  \end{split}
\end{equation}
The terms $\int f(\bm r)\ud \bm r$ are of order $\epsilon$.
To lowest order in $\epsilon$ we find
\begin{equation}
  \label{eq:olap3}
  \langle \Phi\vert\Phi\rangle\simeq N(N-1)\langle12|12\rangle.
\end{equation}
The symmetrization of the non-normalized OBDM (denoted $\tilde n(\bm
r,\bm r')$) can be done
in a similar manner. It consists of $9$ different types of terms, but
to lowest order in $\epsilon$ only two terms remain,
\begin{equation}
  \label{eq:obdm1}
  \begin{split}
    \tilde n(\bm r,\bm r')=(N-1)(N-2)n_{2323}(\bm r,\bm r')+(N-1)n_{1212}(\bm r,\bm r').
  \end{split}
\end{equation}
These two terms are given by
\begin{equation}
  \label{eq:obdm2}
  \begin{split}
    n_{2323}(\bm r,\bm r')
    &=\int\phi^*(\bm r)\phi(\bm r')|\phi(\bm r_2)|^2|\phi(\bm r_3)|^2f(\bm r_{23})^2\ud \bm r_2\ud \bm r_3\\
    &=\phi^*(\bm r)\phi(\bm r')\langle12|12\rangle,\\
    n_{1212}(\bm r,\bm r')
    &=\int\phi^*(\bm r)\phi(\bm r')|\phi(\bm r_2)|^2f(\bm r-\bm r_2)f(\bm r'-\bm r_2)\ud \bm r_2\\
    &\simeq \int|\phi(\bm s)|^4f(\bm r-\bm s)f(\bm r'-\bm s)\ud \bm s.
  \end{split}
\end{equation}
The normalized OBDM $n(\bm r,\bm r')=N\tilde n(\bm r,\bm r') /
\langle\Phi|\Phi\rangle$ finally becomes

\begin{equation}
  \label{eq:obdm-expans-pointstates}
  n(\bm r,\bm r')\simeq (N-2)\phi^*(\bm r)\phi(\bm r)
  +\int \!\ud \bm s\, N_{\bm s} \,
  \chi^*_{\bm s}(\bm r)\, \chi_{\bm s}(\bm r'),
\end{equation}
where we have defined
\begin{equation}
  \label{eq:pointstates}
  \chi_{\bm s}(\bm r)=f(\bm r-\bm s),\qquad
  N_{\bm s}=2\frac{n(\bm s)^2} {\int n(\bm r)^2\ud
    \bm r}.
\end{equation}
Here $n(\bm r)=\lvert\phi(\bm r)\rvert^2$ is the mean-field density.
Notice that eq.~\eqref{eq:obdm-expans-pointstates} has trace equal to
$N$.  The set $\{\phi,\chi_{\bm s}|\bm s\in \mathbb R^3\}$ is
orthonormal in the limit considered. Since the expansion of the OBDM
in eigenfunctions, eq.~\eqref{eq:obdm-eigenfunction-expansion}, is
unique, we can compare eq.~\eqref{eq:obdm-eigenfunction-expansion}
with eq.~\eqref{eq:obdm-expans-pointstates} and read off
eigenfunctions and corresponding eigenvalues directly. The condensate
fraction is reduced to $1-2/N$ with $\phi$ still being the
corresponding single-particle state. The remaining $2$ ``particles''
are distributed over infinitely many states $\chi_{\bm s}(\bm r)$,
which are peaked around $\bm s$.  The occupation numbers $\ud \bm s\,
N_{\bm s}$ for $\chi_{\bm s}(\bm r)$ are proportional to the
square-density, $n(\bm s)^2$, of the mean-field.
%
%
%

%
%
%

%
%

%

%
%
%
%
%
%

%% file: appendix-tf.tex
In this appendix we derive the free parameter $A$ of the Thomas-Fermi
solution eq.~\eqref{eq:density-solution} in
chapter~\ref{chap:tf}. This is done explicitly by enforcing the virial
equation. We show that this is equivalent to the boundary condition
eq.~\eqref{eq:surface-derivative}.

Even though eq.~\eqref{eq:density-solution} is a solution to the
modified GP equation eq.~\eqref{eq:gpe-dimless} for all $A$, it does
not necessarily minimize the energy functional as discussed in
the section~\ref{TFA}. This can also be seen from
the virial equation (with neglected kinetic energy),
\begin{eqnarray}
  \label{eq:mgp-virial}
  -2E_V+3E_I+5E_{I2}=0,
\end{eqnarray}
which holds for all extremal points of the energy
functional. Equation~\eqref{eq:mgp-virial} is derived from the energy
functional using scaling arguments as in \cite{pitaevskii03}.

As an example consider the $A=0$ solution in
eq.~\eqref{eq:density-solution}. This solution has a chemical
potential shifted by $3 g_2$ compared to the $g_2=0$ TF result. But
the density is unchanged and so is $E_V$ and $E_I$. Hence the usual
virial equation $-2E_V+3E_I=0$ for $ g_2=0$ also holds for $ g_2\ne
0$. Since $E_{I2}=-3 g_2/2 \ne 0$ the virial equation
eq.~\eqref{eq:mgp-virial} is not fulfilled, and hence the $A=0$
solution is not extremal. Below we use the virial equation to calculate 
the value of $A$ that minimizes the energy functional and the 
corresponding $R$ and $\mu$. We will also prove that this condition
is in fact equivalent to assume $\rho(x_0)=\nabla_x\rho(x_0)=0$ at the 
boundary.

The general results for $A$, $R$, and $\mu$ can be derived using the
normalization and surface conditions eq.~\eqref{eq:norm-surf-cond},
and the virial equation eq.~\eqref{eq:mgp-virial}.  For convenience we
introduce the variables $\bar\mu = \mu/(3 g_2)+1$, $ \bar A =A/(
3g_2)$, and $c=N a/g_2^{5/2}$. 
The different energy contributions are
\begin{equation}
  \label{eq:energy-integrals}
  \begin{split}
    E_V&=3s\int_0^{x_0} \ud x x^4(\bar\mu -\frac{x^2}{6}
    +\bar A \frac{\sin x}{x}),\\
    E_I&=9s\int_0^{x_0} \ud x x^2(\bar\mu -\frac{x^2}{6}
    +\bar A \frac{\sin x}{x})^2,\\
    E_{I2}&=-9s\int_0^{x_0} \ud x
    x^2(\bar\mu -\frac{x^2}{6}+\bar A \frac{\sin x}{x})
    (1+\bar A \frac{\sin x}{x}),
  \end{split}
\end{equation}
where $s= g_2^{7/2}/(2 a)$. Direct integration of
eq.~\eqref{eq:energy-integrals}, insertion of $\bar\mu$ from the
normalization eq.~\eqref{eq:surface-condition2}, and some algebra
gives the virial equation
\begin{equation}
  \label{eq:Evir}
  \begin{split}
    0&= -2E_V+3E_I+5E_{I2}\\
    &=-\frac{s}{x_0}(x_{0}^{3}-3\bar A (x_0\cos x_0-\sin x_0))^2.
  \end{split}
\end{equation}

We immediately see that this is in fact equivalent to
eq.~\eqref{eq:A-vs-z}. Therefore, the solution we have explicitly
found above minimizes the energy functional with boundary conditions
$\rho(x_0)=\nabla_x\rho(x_0)=0$. More generally, when we solved the
MGP without considering the boundary terms in section~(\ref{TFA}), we
found a one-parameter family of solutions (parametrized by $A$). The
virial theorem is merely a constraint on $A$ for obtaining a minimum
of $E$.

%% file: all.bbl
\newcommand{\etalchar}[1]{$^{#1}$}